\documentclass[a4paper,11pt]{article}
\usepackage{jheppub} 
\usepackage{xcolor}
\definecolor{ForestGreen}{RGB}{34,139,34}
\usepackage[T1]{fontenc}
\usepackage{esint}
\usepackage{physics}

\def\a{{\sf a}}
\def\b{{\sf b}}
\newcommand{\eg}{{e.g.,}\ }
\newcommand{\ie}{{i.e.,}\ }

\usepackage{cancel}
\usepackage{mdframed,framed}

\newcommand{\beq}{\begin{eqnarray}}
\newcommand{\eeq}{\end{eqnarray}}
\newcommand{\beqn}{\begin{eqnarray}}
\newcommand{\eeqn}{\end{eqnarray}}
\newcommand{\bee}{\begin{equation} \begin{aligned}}
\newcommand{\eee}{ \end{aligned} \end{equation}}

\newcommand{\cO}{{\cal O}}

\usepackage{tikz-cd}
\usepackage{pict2e}
\makeatletter
\newcommand{\variable@rule}[1]{%
  \fontdimen8  
  \ifx#1\displaystyle\textfont3\else
    \ifx#1\textstyle\textfont3\else
      \ifx#1\scriptstyle\scriptfont3\else
        \scriptscriptfont3\relax
  \fi\fi\fi
}
\usepackage{mathtools}

\usepackage{tocloft}
\newcommand{\cD}{{\cal{D}}}

\newcommand{\rd}{\text{d}}

\newcommand{\chkM}{{\color{red} \,\checkmark\kern-5pt{}_{M}}}

\newcommand{\be}{\begin{equation}}
\newcommand{\ee}{\end{equation}}
\newcommand{\bea}{\begin{eqnarray}}
\newcommand{\eea}{\end{eqnarray}}

\usepackage{ragged2e}
\usepackage{blindtext}
\usepackage{ulem}

\newcommand{\perimeter}[1]{
	\centerline{
		\begin{minipage}[c]{0.7\textwidth}
			\begin{center}
			$^a$ Perimeter Institute for Theoretical Physics,\\
			 31 Caroline Street North, Waterloo, Ontario N2L 2Y5, Canada
			\end{center}
		\end{minipage}
		}
	}

\newcommand{\reef}[1]{(\ref{#1})}

\newcommand{\beqa}{\begin{eqnarray}}
\newcommand{\eeqa}{\end{eqnarray}}

\renewcommand{\(}{\left(}
\renewcommand{\)}{\right)}
\renewcommand{\[}{\left[}
\renewcommand{\]}{\right]}

\title{\texorpdfstring{Geometric modular flows in 2d CFT and beyond}{Geometric modular flows in 2d CFT and beyond}}

\author[a,b]{Jacqueline Caminiti,}
\author[a,c]{Federico Capeccia,}
\author[a]{Luca Ciambelli}
\author[a]{and Robert C. Myers}
\affiliation[a]{Perimeter Institute for Theoretical Physics,\\
31 Caroline Street North, Waterloo, Ontario N2L 2Y5, Canada}
\affiliation[b]{Department of Physics \& Astronomy, University of Waterloo,\\
200 University Avenue West, Waterloo, Ontario N2L 3G1, Canada}
\affiliation[c]{Dipartimento di Fisica, Università degli Studi di Milano,\\ Via Celoria 16, Milano, Italy,  I-20133}

\emailAdd{jcaminiti@perimeterinstitute.ca}
\emailAdd{rmyers.perimeter@gmail.com}
\emailAdd{federico.capeccia@studenti.unimi.it}
\emailAdd{ciambelli.luca@gmail.com}

\abstract{We study geometric modular flows in two-dimensional conformal field theories.
We explore which states exhibit a geometric modular flow with respect to a causally complete subregion and, conversely, how to construct a state from a given geometric modular flow.
Given suitable boundary conditions, we find that generic geometric modular flows in the Rindler wedge are conformally equivalent.
Based on this insight, we show how conformal unitaries can be used to explicitly construct a state for each flow. We analyze these states, deriving general formulas for the energy density and entanglement entropy.
We also consider geometric flows beyond the Rindler wedge setting, and in higher dimensions.}

\begin{document}

\maketitle
\flushbottom

\section{Introduction}
\label{sec:intro}

The study of quantum subsystems lies at the heart of modern theoretical physics and quantum information theory.
In the absence of knowledge about the entire universe --- \eg omitting the interior of a black hole, the exterior of our cosmological horizon, or the environment which decoheres a superconducting qubit --- we rely on the ability to analyze subsystems in a precise and environment-independent way.

In QFT, a universal and elegant object characterizing the structure of a subsystem is the modular flow, which was first introduced in the context of
Tomita-Takesaki theory \cite{Takesaki:1970aki} (see also \cite{Witten:2018zxz}), a rigorous algebraic framework for continuum QFTs.
Like density matrices in finite dimensional quantum mechanics, modular flows can be used to study quantum information problems such as the distinguishability and entanglement of states with respect to a given subsystem.
In continuum QFTs, density matrices are ill-defined, while modular flows are well-defined.
Modular flows have played a pivotal role in a variety of developments in high energy physics such as understanding the origin of Hawking radiation \cite{Hawking:1975vcx, Hartle:1976tp, SEWELL198023, kay1991theorems}, 
reconstructing bulk from boundary physics in AdS/CFT \cite{Dong:2016eik, Harlow:2016vwg, Faulkner:2017vdd, Cotler:2017erl, Faulkner:2018faa, Kamal:2019skn, Czech:2019vih, Engelhardt:2021mue, Parrikar:2024zbb}, formulating entropy conjectures in quantum gravity \cite{Susskind:1994sm, Bousso:1999xy, Wall:2011hj, Bousso:2015mna, Bousso:2015wca, Gesteau:2023hbq}, revealing energy conditions \cite{Koeller:2017njr, Balakrishnan:2017bjg, Ceyhan:2018zfg} and entropy bounds \cite{Longo:2018zib} in general QFTs, and calculating entanglement entropy and anomalies in CFTs \cite{Casini:2006es, Casini:2011kv, Cardy:2016fqc}.

In finite-dimensional quantum systems, the modular flow is simply a repackaging of the information in the density matrix $\rho$ into a one-parameter family of automorphisms. 
Given an element $\a$ of the subsystem algebra, we have\footnote{The modular flow in eq.~\eqref{eq:rho} is  well-defined only when the density matrix has no zero eigenvalues. This roughly means the subsystem must be sufficiently entangled with a suitable purifying system. \label{foot:foot1}}
\begin{equation}
    \mathsf{a} \mapsto \mathsf{a}_s= \rho^{-is}\, \a\, \rho^{is}=: U(s)^{\dag}\, \a \,U(s) \,,
    \label{eq:rho}
\end{equation}
and we can write
\begin{equation}
    U(s) = e^{-ihs}\,,
    \label{eq:modularunitary}
\end{equation}
where the Hermitian operator $h$ is the modular Hamiltonian.
A key property of the modular flow, both in finite dimensions and in the continuum, is that it satisfies the Kubo–Martin–Schwinger (KMS) \cite{Kubo:1957mj, Martin:1959jp} condition
\begin{equation}
    \langle \a_{s} \,\b\rangle = \langle \b\,\a_{s+i} \rangle \,,
\label{eq:definingeq}
\end{equation}
which, in finite dimensions, follows from the cyclicity of the trace used to compute expectation values.
Here $\b$, like $\a$ and $\rho$, is an element of the subsystem algebra.

The KMS condition encodes a generalized notion of thermality \cite{Kubo:1957mj, Martin:1959jp,Haag:1996hvx}.
Indeed, when the modular flow aligns with temporal evolution, the KMS condition implies physical observers may experience thermal effects.
Consider the Gibbs thermal state, $\rho = e^{-\beta H}/Z$, where the modular flow corresponds to time-evolution in the Heisenberg picture:
\begin{equation}
    \a \mapsto e^{i \beta s H} \a\, e^{-i \beta s H}\,.
\end{equation}
In this case, the KMS periodicity $s \sim s+ i$ corresponds to periodicity in imaginary time, $t \sim t+i\beta$, the signature of thermality in QFT \cite{Matsubara, Martin:1959jp}.

This indicates a rich link between thermality and entanglement. 
This link is particularly poignant in the context of QFT states with geometric modular flows.
A modular flow is \textit{geometric} when $h$ can be expressed as a local integral of the QFT stress tensor $T^{\mu}_{\;\;\nu}(\mathbf{x})$ on any Cauchy slice $\Sigma$,
\begin{equation}
    h:= 2\pi \int_{\Sigma} \, \rd \Sigma_{\mu}\, T^{\mu}{}_{\nu}(\mathbf{x})\,\xi^{\nu}(\mathbf{x}) \,,
    \label{eq:localHam}
\end{equation}
where $\xi(\mathbf{x})$ generates an infinitesimal local flow. 
In this case, $U(s)$ implements the flow along integral curves of $2\pi\xi(\mathbf{x})$.\footnote{A more precise definition of geometric modular flows is the following \cite{Sorce:2020ama}: Let $U$ be a unitary operator, and let $\varphi$ be a finite diffeomorphism.
In particular, $\varphi=\varphi_s$ implements the flow with affine parameter $s$ along integral curves of $\xi$.
Then, $U$ acts geometrically, implementing $\varphi$, if for any smearing function $f$ of any local operator $\phi(x)$, the operator $U^{\dag} \phi[f] U$ can be written as $U^{\dag} \phi[f] U =  \phi[\tilde{f}]$, where $\mathrm{supp}(\tilde{f}) = \varphi(\mathrm{supp} (f))$. 
The function $\tilde{f}$ need not equal the pullback $f(\varphi^{-1}(x))$; it can carry an extra conformal weight.
In a CFT, if $\xi$ is a conformal Killing vector, the family of unitaries \eqref{eq:modularunitary} built from eq.~\eqref{eq:localHam} satisfies these properties. Thus, we use eq.~\eqref{eq:localHam} as the defining property of a geometric modular flow.
\label{foot:GUF}}
When these curves are future-directed, describing  observer worldlines,
the KMS condition \eqref{eq:definingeq} implies that observers may experience thermal effects.\footnote{Note, we use $s$ to parameterize integral curves of $\xi$ rather than $2\pi \xi$. 
For this reason, we must replace the standard $i$-periodicity in the KMS condition \eqref{eq:definingeq} with $2\pi i$ going forward.
The Canadian authors apologize for this unconventional choice.
In most equations, standard conventions can be restored via $\xi \mapsto \xi/2\pi$.
}

\begin{figure}[ht]
    \centering
    \includegraphics[width=0.99\linewidth]{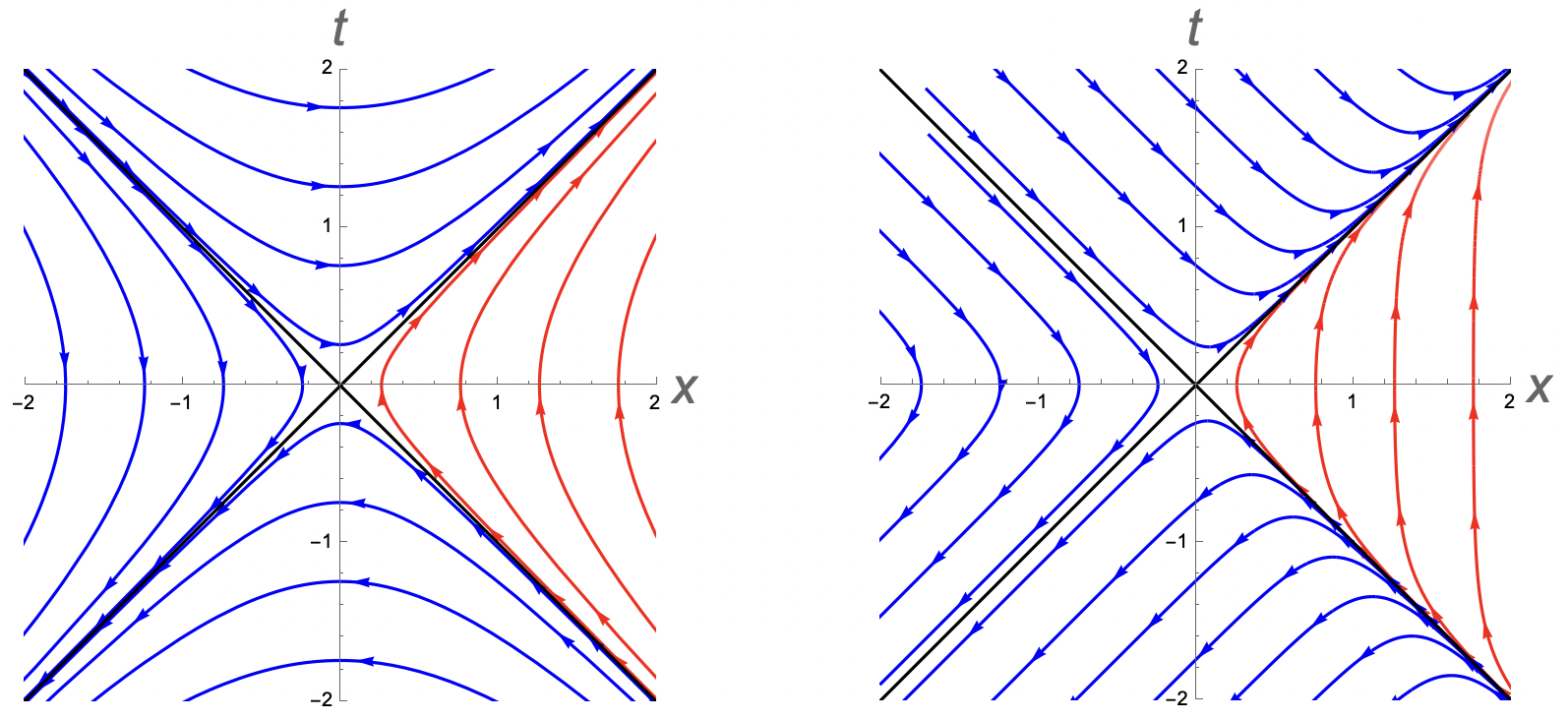} 
    \caption{Two examples of geometric modular flows. Left: in the Minkowski vacuum state, observers along the red trajectories in the right Rindler wedge are uniformly accelerated, and they experience a temperature proportional to their acceleration. Right: in the Minkowksi thermal state in a CFT$_2$, certain non-uniformly accelerated observers also experience Unruh-like effects.
    This flow is discussed around eq.~\eqref{eq:borchers1} below.}
    \label{fig:flows}
\end{figure}

For example, in the two-dimensional Minkowski vacuum state, where $\rd s^2 = -\rd t^2 + \rd x^2$, the modular flow associated with the right Rindler wedge is a boost,
\begin{equation}
    \xi(t, x) = x\, \partial_t + t\, \partial_x\,,
    \label{eq:boost}
\end{equation}
as shown in figure \ref{fig:flows}.
Integral curves of this geometric modular flow correspond to uniformly accelerated observers, and the KMS condition \eqref{eq:definingeq} implies that each observer experiences a constant temperature $\frac{a}{2\pi}$ proportional to their acceleration, $a$.
This is the celebrated Unruh effect \cite{Unruh:1976db, Unruh:1983ms}, which 
holds in any spacetime dimension and was established using algebraic QFT methods by Bisognano and Wichmann \cite{Bisognano:1976za}.

A similar effect occurs when considering, instead of the vacuum state, a thermal state at inverse temperature $\beta$ in Minkowski spacetime. In the context of a $2d$ CFT, when this state is further reduced to the Rindler wedge, the resulting modular flow is geometric \cite{Borchers:1998ye}, such that non-uniformly accelerated observers along the trajectories shown in figure \ref{fig:flows} can experience thermal effects. 
In the Rindler wedge, the flow is \cite{Borchers:1998ye}
\begin{equation}
    \xi = \frac{\beta}{2\pi}\[\left(1-e^{-\frac{2\pi x}{\beta}}\cosh\(\frac{2\pi t}{\beta}\)\right)\partial_t
    +e^{-\frac{2\pi x}{\beta}}\sinh\(\frac{2\pi t}{\beta}\)\partial_x \]\,.
    \label{eq:thermalflow}
\end{equation}
For small $x$ and $t$, it behaves as a boost: $\xi\sim x\partial_t+t\partial_x$. In contrast for $x\to\infty$ and finite $t$, it approaches the constant vector $\xi\sim\frac{\beta}{2\pi} \partial_t$, reflecting the thermality of the underlying state.
Unlike in the vacuum case, an observer traveling along this modular flow does not experience a constant temperature; at best, they experience a slowly-varying temperature.\footnote{ 
Interestingly, we can also consider a family of accelerated observers in the thermal state which experience zero temperature. See \cite{Borchers:1998ye}, figure 4, where this is referred to as the ``reverse Unruh effect.''}

Generic modular flows in QFT are not geometric; the modular Hamiltonian for a typical subregion, even in the vacuum state, is highly nonlocal, as discussed in \cite{Holzhey:1994we, Calabrese:2004eu, Wong:2013gua}.
Further, as shown in \cite{Arias:2016nip,Sorce:2020ama}, geometric modular flows are necessarily future-directed and described by a conformal Killing vector. That is, within the open and causally complete subregion $\mathcal{V}$ of interest, $\xi^{\mu}$ obeys
\begin{equation}
    \xi^\mu\xi_\mu < 0 \qquad \textrm{and} \qquad \nabla_{\mu} \xi_{\nu} + \nabla_{\mu}\xi_{\nu} =  \frac{2}{d}\nabla_{\rho}\,\xi^{\rho}g_{\mu \nu}\,.
    \label{eq:ckv}
\end{equation}
Therefore, the number of states with geometric modular flow in any QFT is limited by the number of vector fields satisfying eq.~\eqref{eq:ckv}, and depends on the conformal properties of the theory \cite{Sorce:2020ama}. However, as noted in \cite{Jensen:2023yxy}, it is often useful to consider a broader class of states whose modular flows are only instantaneously geometric, such that the modular Hamiltonian can be expressed by eq.~\eqref{eq:localHam} only on one Cauchy slice $\Sigma$.\footnote{In the context of CFT$_2$,
it suffices to study globally geometric flows, because in $d=2$ any vector field on an interval $A$ admits a unique extension to a conformal Killing vector on the causal development of $A$.}
In this case, the condition \eqref{eq:ckv} no longer applies.

Our study of geometric modular flows and their abundance in CFTs is partially motivated by certain results in semiclassical gravity.
It has recently been appreciated that gravitational constraints can be implemented to reduce the severity of ultra-local divergences in QFTs \cite{Witten:2021unn}. 
In the language of von Neumann algebras, the type of the von Neumann algebra for a subregion is reduced from type III in continuum QFT to type II once the gravitational constraints are imposed \cite{Chandrasekaran:2022eqq, Penington:2023dql, AliAhmad:2023etg, Klinger:2023tgi, Kudler-Flam:2023qfl, Jensen:2023yxy, AliAhmad:2024eun, Chen:2024rpx, Kudler-Flam:2024psh, Kaplan:2024xyk, DeVuyst:2024grw, Kolchmeyer:2024fly, Penington:2024sum}. 
This type reduction implies the emergence of a well-defined trace that is unique up to rescaling, enabling precise definitions of density matrices and von Neumann entropies, which are unavailable in the type III setting.
The outcomes of this research program have been striking, including a new semiclassical definition of the generalized entropy \cite{Witten:2021unn, Jensen:2023yxy,Chandrasekaran:2022cip}.
In this type-reduction procedure, it is important to consider states admitting geometric and instantaneously-geometric modular flows, partially motivating our analysis.

In this manuscript, we begin by studying states with geometric modular flows in the Rindler wedge $\mathcal{R}$ of a $2d$ CFT. 
We find that there exists a state $|\Psi\rangle$ for every flow preserving $\mathcal{R}$ which obeys eq.~\eqref{eq:ckv}, suitable boundary conditions, and a condition on the purity of the underlying state (see eq.~\eqref{eq:purity} below).
We then construct these states by acting with conformal unitaries on either the vacuum or the thermal state in Minkowski spacetime.

Building on \cite{Norma:1981,  Hislop:1981uh, Birrell:1982ix,  Casini:2011kv, Wong:2013gua, Cardy:2016fqc}, a key observation is that \textit{local} conformal symmetries act non-trivially on the CFT vacuum state, even though the \textit{global} conformal group acts trivially.
Consider exciting the vacuum with an infinitesimal Virasoro transformation
\begin{equation}
    |\Psi\rangle := e^{i \epsilon (L_n+L_{-n})} |\Omega \rangle = \left(1+i \epsilon L_{-n} + \cO(\epsilon^2) \right) |\Omega \rangle \,,
\end{equation}
with $n\ge 2$. This state has positive energy owing to the failure of $L_0$ to commute with $L_{-n}$.
Further, since conformal transformations are geometric, they can be used to map geometric modular flows in the vacuum to new geometric modular flows, with the vector field $\xi$ of the original modular flow simply transforming via push-forward under the conformal map.

While we focus on geometric modular flows which are defined on the entire spacetime, note that one is often interested in studying a subregion $\mathcal{V}$, such as $\mathcal{V}=\mathcal{R}$, without specifying the behavior of the modular flow outside $\mathcal{V}$.
In this context, our results guarantee the existence of a state for every geometric flow in $\mathcal{R}$ which obeys eq.~\eqref{eq:ckv} and suitable boundary conditions (see section \ref{sec:uniqueness}).

We then study the energy and entanglement entropy of the states we have constructed, and relate them to the modular flow. The von Neumann entropy $S(\rho_A) =- \Tr_A(\rho_A\log \rho_A)$
is invariant under $\rho_A=\Tr_{\bar{A}}(|\Psi\rangle \langle \Psi |)\to  \Tr_{\bar{A}}(U|\Psi\rangle \langle \Psi |U^{\dag})$ if $U$ splits as $U=U_{A}\otimes U_{\bar{A}}$.\footnote{One can write $\rho_A=U_A\Tr_{\bar{A}}(|\Psi\rangle \langle \Psi |)U_A^{\dag}$ and use the  cyclicity of $\Tr_A$.}
However, our excited states $|\Psi\rangle$ and the vacuum $|\Omega\rangle$ have different entanglement structure between the left and right Rindler wedges, because the conformal unitaries utilized do not factorize. Indeed, we argue that conformal unitaries can be used to disentangle the state. This marks a key difference from \cite{Casini:2011kv}, where the central ingredient in the derivation is the use of unitary transformations which preserve the von Neumann entropy.

After studying conformally excited states in the Rindler wedge, we discuss extensions to other causally complete subregions and geometries, leading to two insights for CFT states with geometric modular flows in a subregion $\mathcal{V}$ in higher dimensions. 

First, we find their entanglement structure is quite simple.
The entanglement entropy can be evaluated with a local integral, encapsulated by the general formula
\begin{equation}
    S(\rho_{\Psi}^{\mathcal{V}}) = - \alpha \int_A \rd \Sigma_{\mu}\, \frac{\xi^{\mu}_{\Psi}}{\norm{\xi_{\Psi}}^d}\,,
    \label{eq:TheEntropyIntro}
\end{equation} 
where $\alpha$ is a state-independent thermal entropy density coefficient.
This result is the covariant extension to higher dimensions of the $2d$ CFT expression in \cite{Wong:2013gua}.
We give various consistency checks of our formula in $d\ge 2$, including matching to several known examples.
Notably, the physical reasoning which led us to propose eq.~\eqref{eq:TheEntropyIntro} is based on a local inverse temperature $\beta = 2\pi \norm{\xi}$ associated instantaneously with observers along the flow, and thus we speculate it applies also to states with only instantaneously geometric modular flows.

Secondly, we define a procedure in $d=2$ for removing the entanglement between the left and right Rindler wedges in some state $|\Psi\rangle$.
When $|\Psi\rangle=|\Omega\rangle$, this procedure yields the Rindler vacuum $|\Omega\rangle_{\mathcal{R}}$\footnote{This is the vacuum state with respect to one-sided boosts, rather than time-translations. It is disentangled, such that we can write $|\Omega_{\mathcal{R}}\rangle=|0\rangle_{\mathrm{left}}\otimes|0\rangle_{\mathrm{right}}$.
By contrast, the usual Minkowski vacuum is entangled: $|\Omega\rangle = \sum_{n} e^{-\pi E_n}|n\rangle_{\mathrm{left}}\otimes|n\rangle_{\mathrm{right}}/\sqrt{Z}$, where $E_n$ is the eigenvalue of $|n\rangle_{\mathrm{right}}$ under right-sided boosts.\label{foot:rindlervac}}. 
We then propose that in $d=2$, and for analogous disentangled states in higher $d$,
\begin{equation}
    \langle T_{\mu \nu}\rangle_{\Psi} - \langle T_{\mu \nu}\rangle_{\Psi_{\mathrm{dis}}} =
    \frac{\alpha}{2\pi d\norm{\xi}^{d+2}} \left(d \,\xi_{\mu} \xi_{\nu} + \norm{\xi}^2 g_{\mu\nu}\right)\,.
    \label{eq:EnergyIntro}
\end{equation}
In particular, this implies a general proposal for the Rindler vacuum energy-momentum tensor in CFT$_d$. 
Again, the physical reasoning which leads to this formula is based on local temperatures associated instantaneously with observers along the flow.

In section \ref{sec:discuss}, we explore the meaning of the local inverse temperature $\beta = 2\pi \norm{\xi}$ used to develop eqs.~\eqref{eq:TheEntropyIntro} and \eqref{eq:EnergyIntro}.
This temperature is often discussed in the study of the modular flow in the Unruh effect \cite{SEWELL1982201,Martinetti:2002sz,Arias:2016nip}.
We find that while $T=\frac{1}{2\pi \norm{\xi}}$ can be interpreted as a temperature using information theoretic arguments \cite{Arias:2016nip}, it may in general differ from the physical temperature $T_{\mathrm{phys}}$ experienced by an observer along the modular flow.
One has that $T=T_{\mathrm{phys}}$ whenever $\xi$ is a Killing vector, as occurs in the Unruh effect.

The paper is organized as follows: In section \ref{sec:AStateForEveryUnruhFlow}, we construct states $|\Psi\rangle$ for every flow obeying eq.~\eqref{eq:ckv} with suitable purity and boundary conditions.
In section \ref{sec:stateProperties}, we explore the physical properties of the corresponding states, and apply our results to a family of examples, including cases where there is no entanglement between the left and right wedges.
In section \ref{sec:beyondRindler}, we
discuss analogous insights for a finite diamond, and for subregions of the cylinder. Section \ref{sec:entanglementcurrents} summarizes and explores our proposals about entanglement entropy and disentangled states in higher dimensions.
We close with a discussion in section \ref{sec:discuss}.

Various technical details are provided in the appendices.
In appendix \ref{app:pushforward}, we demonstrate that if two states with geometric modular flows are related by a conformal unitary, the flows simply transform as the push-forward under the corresponding conformal map.
In appendix \ref{app:connesappendix}, we show that states with geometric modular flows in a subregion can be perturbed outside this subregion, without changing their flow inside, which is important for understanding the uniqueness of our various constructions.
In appendix \ref{app:areabalance}, we ensure that the conformal mappings discussed in sections \ref{sec:vacuumsector} and \ref{sec:uniqueness} are well-defined.
In appendix \ref{app:annulus}, we use the entanglement entropy formula discussed in \cite{Cardy:2016fqc} to independently check the formula \eqref{eq: EE Wong} in \cite{Wong:2013gua}, which is a special case of our proposal eq. \eqref{eq:TheEntropyIntro}. Appendix \ref{app:nonlocal} deals with nonlocal transformation rules for primary operators and the subtleties arising in the study of special conformal transformations in Minkowski spacetime. Finally, Appendix \ref{App:multiple} shows that in the case of multiple intervals in CFT$_2$, our proposed entropy formula \eqref{eq:TheEntropyIntro} reproduces the correct entanglement entropy even when the modular Hamiltonian contains nonlocal terms. Readers seeking an introductory reference to modular flows and the theory of algebraic QFT are encouraged to consult
\cite{Sorce:2023fdx, Sorce:2024pte, Sorce:2023gio, Witten:2018zxz, Hollands:2018wxj, Hollands:2022dem, Fewster:2019ixc, Hollands:2014eia}.

\section{\texorpdfstring{Excited states in $d=2$ with geometric modular flows}{Excited states in d=2 with geometric modular flows}}
\label{sec:AStateForEveryUnruhFlow}

In this section, we explore the space of states with geometric modular flows in $2d$ CFTs.

For a vector field to give rise to a geometric modular flow, it must be conformal and future-directed in the subregion of interest. 
Moreover, generic modular flows are expected to be boost-like near the entangling surface \cite{Fredenhagen:1984dc, Haag:1996hvx, Sorce:2020ama}.
We therefore introduce the concept of \textit{Unruh flows}, defined as vector fields $\xi$ preserving an open, connected, causally complete region $\mathcal{V}=D(A)$\footnote{Here and later on, $D(\cdot)$ denotes the causal development, and $A$ denotes a codimension-1 subregion.} which
\begin{itemize}
    \item satisfy eq.~\eqref{eq:ckv} everywhere in $\mathcal{V}$ and its causal complement, $D(\bar{A})$,
    \item are future-directed in $\mathcal{V}$ and past-directed in  $D(\bar{A})$, and timelike in both,\footnote{When Haag duality holds, we have $U_{\Psi}^{D(A)}(s)=U_{\Psi}^{D(\bar{A})}(-s)$ \cite{Haag:1996hvx}.
    Thus, if the modular flow with respect to $D(\bar{A})$ is future-directed in $D(\bar{A})$, then the modular flow with respect to $\mathcal{V}$ is past-directed in $D(\bar{A})$.}
    \item are boost-like at $\partial A$, \ie there exists a regular coordinate system centered at $\partial A$ such that $\xi = t \partial_x + x\partial_t + \cO(t^2, x^2, tx)$.\footnote{This condition pertains to $d=2$ spacetimes. A higher-dimensional analog is the constant surface gravity condition of \cite{Jensen:2023yxy}. Note that any continuous vector field preserving $\mathcal{V}$ is boost-like at $\partial A$, up to rescaling by an overall coefficient. Our condition fixes the coefficient to $1$.}
\end{itemize}
In the case of an infinite subregion $\mathcal{V}$, we also demand $\xi$ cannot have exotic asymptotic behavior; that is, it must asymptote to the modular flows of either the vacuum state (becoming a \textit{vacuum-sector flow}) or the thermal state (becoming a \textit{thermal-sector flow}).\footnote{We are glad to thank Jon Sorce for discussions and explanations on this point and its relation to Hilbert space sectors.
We refer readers to a forthcoming work by Jon Sorce and Gautam Satishchandran on the topic of boundary conditions for modular flow in free theories \cite{Sorce:2025Upcoming}.\label{foot:jon}}

In this section we demonstrate that every vacuum- or thermal-sector Unruh flow $\xi$ with respect to the Rindler wedge $\mathcal{R}$, obeying a condition on the state's purity, coincides with the modular flow for some state in a $2d$ CFT on Minkowski spacetime. We construct states with Unruh flows in the vacuum and thermal sectors in subsections \ref{sec:vacuumsector} and \ref{sec:thermalsector}, respectively. 

A note on notation:
We employ both Cartesian $(t,x)$ and double-null $(u,v)$ coordinates. 
The Minkowski line element is $\rd s^2 = -\rd t^2 + \rd x^2=-\rd u\, \rd v$, where $u=t-x$ and $v=t+x$. Therefore, the Rindler wedge $\mathcal{R}$ is the causal development of $A := \{t=0, x>0\}$, which in null coordinates reads $\mathcal{R}=\{u<0,v>0\}$, and $\partial A$ consists of the single point $(0, 0)$ in either coordinate system.
Functions of $u=t-x$ ($v=t+x$) are left-movers (right-movers).

\subsection{Excited states in the vacuum sector}
\label{sec:vacuumsector}

The conformal Killing equation \eqref{eq:ckv} in $d=2$ has general solution $\xi=c(u)\partial_{u} + b(v)\partial_{v}$.
We accordingly define a vacuum-sector Unruh flow with respect to $\mathcal{V}=\mathcal{R}$ as an Unruh flow of this form, which approaches the vacuum modular flow \cite{Bisognano:1976za}
\begin{equation}
    \xi_{\Omega} := c(u)\, \partial_u + b(v)\, \partial_v = - u\, \partial_u + v\, \partial_v\,,
    \label{eq:vacflow}
\end{equation}
at large $|u|$ and $|v|$. 
Specifically,
\begin{equation}
    c(u) =- u + \cO(1)\qquad
    b(v) = v + \cO(1)\,.
    \label{eq:vacFalloffsRindler}
\end{equation}
This definition is motivated by the Gelfand-Naimark-Segal (GNS) construction of sectors in QFT \cite{zbMATH03097323, bams/1183510397}.
The GNS Hilbert space
$\mathcal{H}_{\omega_1}$ for the algebra $\mathcal{A}_{QFT}$ is generated by acting on a privileged algebraic\footnote{A state $\omega$ on a von Neumann algebra $\mathcal{A}$ is a positive, normalized linear functional $\omega: \mathcal{A} \rightarrow \mathbb{C}$, \ie an assignment of expectation values on the algebra. For an introduction to the GNS construction, see \cite{Fewster:2019ixc, Witten:2018zxz}.} state $\omega_1$ with all locally smeared quantum fields in the theory.
Since GNS Hilbert spaces are generated by smearing fields on \textit{finite} regions, states in a given sector share the same asymptotic behavior; see appendix \ref{app:connesappendix} and footnote \ref{foot:jon}. The precise definition of asymptotic behavior is subtle, and we address this issue in section \ref{sec:exDiamond}.

In short, we are demanding that states in the vacuum sector look like the CFT vacuum $|\Omega\rangle$ in the deep IR, \ie at large distances.
This is an additional requirement on top of the standard notion that physical states look like the vacuum at short (UV) distance scales \cite{Fredenhagen:1984dc, Haag:1996hvx, Sorce:2020ama}.
We assume $|\Omega\rangle$ is the unique state invariant under global conformal symmetries \cite{Haag:1996hvx}. Furthermore, from the GNS construction, we remark that vacuum sector states cannot have arbitrarily large excitations at asymptotic infinity.

We now show that every vacuum-sector Unruh flow $\xi_{\Psi}$ coincides with the modular flow of some state in the vacuum sector Hilbert space, based on the key observation that conformal transformations can be used to map $|\Omega\rangle$ to other states with geometric modular flows. 
At the classical level, $2d$ conformal transformations treat left- and right-movers independently, i.e.~$\varphi: (u, v) \rightarrow (\tilde{u}(u)$, $\tilde{v}(v))$.
This map is orientation-preserving if the corresponding Jacobian $\tilde{u}'(u) \tilde{v}'(v)$ is positive everywhere.
This implies that the line element
\begin{equation}
    \rd s^2 = - \rd\tilde{u} \,\rd\tilde{v} = - \tilde{u}'(u) \, \tilde{v}'(v)\, \rd u\, \rd v =: -\Omega^2(u,v) \,\rd u\,  \rd v
    \label{eq:metric}
\end{equation}
carries a positive conformal factor $\Omega^2(u,v)$.
Since conformal mappings are invertible, this means that $\tilde{u}'>0$ and $\tilde{v}'>0$ everywhere; we exclude the alternative case where both are negative because we focus on conformal mappings which preserve $\mathcal{R}$.

Let such a conformal transformation $\varphi$ be implemented at the quantum level by a conformal unitary $U_{\varphi}$.
For a precise discussion of such unitaries, see e.g.~\cite{Fewster:2004nj,Oblak:2017ect}.
Under $U_{\varphi}$, conformal primary operators transform according to the first derivatives of $\tilde{u}(u)$, $\tilde{v}(v)$:
\begin{equation}
    U_{\varphi}\, O(u,v) \,U_{\varphi}^{\dag} = \Omega(u,v)^{-\Delta}\,O(\tilde u,\tilde v) = (\tilde{u}'\,\tilde{v}')^{-\frac{\Delta}{2}}\,O(\tilde u,\tilde v)\,.
    \label{eq:primary}
\end{equation}
Moreover, the Schwarzian appearing in the transformation of the stress tensor contains a third derivative, so will require that $\tilde{u}(u)$ and $\tilde{v}(v)$ are at least three times differentiable, \ie $\tilde{u}(u), \tilde{v}(v) \in C^3(\mathbb{R})$.
This guarantees that the state $|\Psi\rangle := U_{\varphi}|\Omega\rangle$ has well-defined expectation values for most primary and quasi-primary operators of interest.\footnote{Since higher derivatives will appear in transformation laws for descendants, infinite differentiability would be ideal, but it is difficult to explicitly check this in the cases of interest. Some of the conformal maps considered in this paper are not smooth, notably the disentangling maps of section \ref{sec:disentangling}; see further discussion in section \ref{sec:uniqueness}.
Smoothness, together with the falloff prescriptions \eqref{eq:vacFalloffsRindler}, additionally helps ensure good behavior of the Bott-Thurston cocycle which defines the Virasoro central extension at the group level \cite{bott1977characteristic}, \eg see also \cite{Oblak:2017ect}.}

In appendix \ref{app:pushforward}, we demonstrate that $|\Psi\rangle$ has modular flow
given by
\begin{equation}
U_{\Psi}(s)=U_{\varphi}\, U_{\Omega}(s)\, U_{\varphi}^\dag\,,
\label{eq:PsiModularUnitary}
\end{equation}
because $U_{\Psi}(s)$ yields the relevant KMS condition. 
Note, eq.~\eqref{eq:PsiModularUnitary} holds also when the subregion is not preserved; see appendix \ref{app:pushforward}.
The main point is that the modular flow for $|\Psi\rangle$ in $\mathcal{R}$ is a composition of geometric unitaries, and hence is geometric.\footnote{Had we considered a highest-weight state $|h\rangle$ instead of the vacuum, the modular flow unitary
\begin{equation}
U_{\Psi}(s):=U_{\varphi} \,U_{h}(s)\,U_{\varphi}^\dag
\label{eq:PsiModularUnitaryNonGeometric}
\end{equation}
for $|\Psi\rangle := U_{\varphi}|h\rangle$  has a geometric piece $U_{\varphi}$ and a generically non-geometric piece $U_{h}(s)$ (see, \eg \cite{Caputa:2023vyr}).}

Specifically, if $U_{\Omega}(s)$ implements the diffeomorphism $\xi_{\Omega}$, then $U_{\Psi}(s)$ implements $\xi_{\Psi}$, where $\xi_{\Psi}$ is the pushforward of $\xi_{\Omega}$ under the conformal map $\varphi$.
Further, given suitable asymptotic boundary conditions, any $\xi_{\Psi}$ constructed with this procedure is an Unruh flow with respect to $\mathcal{R}=D(A)$. We prove these statements in appendix \ref{app:pushforward}.

We now illustrate that every vacuum-sector Unruh flow in $\mathcal{R}$ 
obeying a purity condition
is just the pushforward of the vacuum modular flow $\xi_{\Omega}$ under a conformal mapping $\varphi$.
From the previous discussion, this implies that for every such flow, there exists a state $|\Psi\rangle = U_{\varphi}|\Omega\rangle$ whose modular flow implements $\xi$.
Schematically, we arrive to the statement
\begin{equation}
\boxed{
\textrm{(Pure, vacuum-sector) Unruh flow  w.r.t. }\mathcal{R}  \quad \iff \quad |\Psi\rangle=U_{\varphi} |\Omega\rangle \textrm{ for some }\varphi}\,.
\end{equation}
While the $\impliedby$ direction is straightforward, we wish to demonstrate the $\implies$ direction. Note that distinct states may have the same geometric modular flow in $\mathcal{R}$, just as distinct quantum mechanical states may have the same reduced density matrix for a subsystem.

We define  the \textit{purity condition}
\begin{equation}
\begin{aligned}
  \fint_{-\infty}^{\infty}\frac{\dd{\tilde{u}}}{\tilde{c}(\tilde{u})}=0
   \qquad
   \fint_{-\infty}^{\infty}\frac{\dd{\tilde{v}}}{\tilde{b}(\tilde{v})}=0\,,
\end{aligned}
\label{eq:purity}
\end{equation}
where $\fint$ indicates a principal value integral, designed to handle the vanishing of $\tilde{c}$ and $\tilde{b}$ at $\tilde{u}=0$ and $\tilde{v}=0$, respectively.\footnote{One might be concerned that this expression depends on the choice of UV/IR regulators.
Indeed, we will see that
eq.~\eqref{eq:purity} can be derived from eqs.~\eqref{eq:alphapfromalpha} and \eqref{eq:alphafrom2} below by imposing a symmetric UV/IR cutoff scheme (e.g., $\tilde{u}_{\textrm{IR}}=\pm L$, with $L\to \infty$).
Eqs.~\eqref{eq:alphapfromalpha} and \eqref{eq:alphafrom2} themselves involve a regularization-independent comparison of finite quantities.
}
For a pure state in quantum mechanics, $S(\rho^{D(A)})=S(\rho^{D(\bar A)})$, and so we refer to eq.~\eqref{eq:purity} as the purity condition since it ensures that the entanglement entropies of the left and right Rindler wedges coincide --- see eq.~\eqref{eq:TheEntropy2d} below. 
However, note that the two conditions in eq.~\eqref{eq:purity} are actually imposing a stronger constraint, namely that the contribution to the entanglement entropy from the right-movers alone coincides for both wedges, and the same for the left-movers.

Consider now an arbitrary Unruh flow with respect to $\mathcal{R}$,
\begin{equation}
    \xi_{\Psi} = \tilde{c}(\tilde{u})\, \partial_{\tilde{u}} + \tilde{b}(\tilde{v})\, \partial_{\tilde{v}}\,,
    \label{eq:rhapsody}
\end{equation}
obeying the purity condition \eqref{eq:purity}. We have written the vector field in tilde coordinates because our goal is to construct a conformal map $\varphi:(u,v) \rightarrow (\tilde{u}(u),\tilde{v}(v))$  such that $\xi_{\Psi}$ is the pushforward of $\xi_{\Omega}$ under $\varphi$, \ie
\begin{equation}
    \tilde{c}(\tilde{u}(u))=c(u)\,\tilde{u}'(u) \qquad  \tilde{b}(\tilde{v}(v))=b(v)\,\tilde{v}'(v)\,.
    \label{eq:pushforward}
\end{equation}
These equations have a scaling symmetry, made manifest by rewriting them for $\varphi^{-1}$:
\begin{equation}
    \tilde{c}(\tilde{u})\,u'(\tilde{u})=c(u(\tilde{u}))\qquad  \tilde{b}(\tilde{v})\,v'(\tilde{v})=b(v(\tilde{v}))\,.
    \label{eq:diffeq}
\end{equation}
If $u=u(\tilde{u})$ is a solution, so is $\lambda u$, because $c(u)$ is linear in $u$, and
similarly for $v(\tilde{v})$; see eq.~\eqref{eq:vacflow}.
This symmetry expresses the invariance of the boost vector field \eqref{eq:vacflow} under independent scaling of $u$ and $v$. That is, $\xi_{\Omega} = - u \partial_u + v\partial_v$ commutes with $u \partial_u$ and $v\partial_v$. 

We focus on the $u$ equation. The most general solution to eq.~\eqref{eq:diffeq} with $\varphi(\mathcal{R}) = \mathcal{R}$ is
\begin{equation}
    u(\tilde{u}) = 
    \begin{cases} 
    \exp[{f_{\alpha}(\tilde{u})}]\,, &\tilde{u}>0 \\ 
    - \exp[{g_{\alpha'}(\tilde{u})}]\,, &\tilde{u}<0\,,
    \end{cases}
    \label{eq:uvsolutionRindler}
\end{equation}
where 
\begin{equation}
    f_{\alpha}(\tilde{u}):=\alpha- \int_1^{\tilde{u}}\frac{\rd\tilde{u}^*}{\tilde{c}(\tilde{u}^*)}\qquad{\rm and} \qquad
    g_{\alpha'}(\tilde{u}):=\alpha'- \int_{-1}^{\tilde{u}}\frac{\rd\tilde{u}^*}{\tilde{c}(\tilde{u}^*)}\,.
\label{eq:fg}
\end{equation}
Due to scaling symmetry, the integration constants $\alpha$ and $\alpha'$ correspond to a single degree of freedom. We now summarize various properties of the conformal map $u(\tilde{u})$. In particular, we discuss how to constrain $\alpha'(\alpha)$ to ensure that the map $u(\tilde{u})$ is invertible and smooth.

\paragraph{Behaviour at the origin.}
Continuity of $u(\tilde{u})$ at $\tilde{u} = 0$ is guaranteed by the boost-like property that $\tilde{c}(\tilde{u}) \rightarrow 0^{\mp}$ as $\tilde{u}\rightarrow 0^{\pm}$.
This implies $f_{\alpha}(\tilde{u})$ and $g_{\alpha'}(\tilde{u})$ diverge to $-\infty$, giving $u(0) = 0$.
As shown in appendix \ref{app:areabalance}, thrice-differentiability of $u(\tilde{u})$ at $\tilde{u} = 0$ is satisfied if and only if $\alpha$ and $\alpha'$ are related as follows
\begin{equation}
    \alpha' + A' = \alpha + A\,,
    \label{eq:alphapfromalpha}
\end{equation}
where we have defined finite constants
\begin{equation}
    A := -\int_{1}^{0}\rd\tilde{u}\left( \frac{1}{\tilde{c}(\tilde{u})}+\frac{1}{\tilde{u}} \right)\,,
     \qquad
    A' := -\int_{-1}^{0}\rd\tilde{u}\left( \frac{1}{\tilde{c}(\tilde{u})}+\frac{1}{\tilde{u}} \right)\,.
    \label{eq:uprime0}
\end{equation}
Further, $u'(0) = e^{\alpha + A}$ is a positive constant. 

\paragraph{Behavior on $\mathbf{(-\infty,0) \cup (0, \infty)}$.}
Smoothness of $u(\tilde u)$ on $(-\infty,0) \cup (0, \infty)$ follows from $\tilde{c}(\tilde{u})$ being smooth and nonvanishing away from $\tilde{u} = 0$.
Taking a derivative of eq.~\eqref{eq:uvsolutionRindler} and using from the Unruh flow properties\footnote{The Unruh flow being timelike in $\mathcal{R}$ implies $\tilde{c}(\tilde{u})\tilde{b}(\tilde{v}) > 0$, while future-directedness gives $\tilde{c}(\tilde{u})>0$.} 
that $\tilde{c}(\tilde{u})>0$ in $\mathcal{R}$ while $\tilde{c}(\tilde{u})<0$ in the left Rindler wedge, one finds that $u(\tilde u)$ is strictly increasing on $(-\infty,0) \cup (0, \infty)$.

\paragraph{Invertibility.}
Since $u'(0)>0$, $u(\tilde{u})$ is strictly increasing on $(-\infty, \infty)$. 
To ensure that the inverse function is supported on the entire real line, we should check that $u(\tilde{u})\to \pm \infty$ as $\tilde{u} \to \pm \infty$, rather than saturating at finite values.
Indeed, the vacuum-sector property that $\tilde{c}(\tilde{u}) \rightarrow - \tilde{u}$ as $\tilde{u}\rightarrow \pm \infty$ implies that
\begin{equation}
    u'(\infty) = e^{\alpha+A_{\infty}} \,,
    \qquad 
u'(-\infty) = e^{\alpha'+A_{-\infty}}
\label{eq:inftyslope}
\end{equation}
where we have again introduced finite constants
\begin{equation}
    A_{\infty} := - \int_1^{\infty}\rd\tilde{u}\left(\frac{1}{\tilde{c}(\tilde{u})}+\frac{1}{\tilde{u}}\right)
    \qquad
    A_{-\infty} := - \int_{-1}^{-\infty}\rd\tilde{u}\left(\frac{1}{\tilde{c}(\tilde{u})}+\frac{1}{\tilde{u}}\right)\,.
    \label{eq:Ainfty}
\end{equation}
Here, finiteness of $A_{\pm\infty}$ relies on the vacuum sector falloffs \eqref{eq:vacFalloffsRindler}.
See appendix \ref{app:areabalance} for details.

\paragraph{Behavior at infinity.}
Smoothness at infinity is necessary to check because the vacuum sector has special conformal symmetry.\footnote{The vacuum is the unique algebraic state invariant under the global conformal group. Consequently, this group is unitarily implemented in the vacuum GNS Hilbert space (see theorem 23 in \cite{Fewster:2019ixc} and \cite{Haag:1996hvx}).
This theorem does not apply to the thermal sector; e.g.,~the thermal state is not boost invariant.\label{foot:sctFootnote}}
Special conformal transformations map infinity to points at finite distance. Therefore, they can transform states with bad behavior at infinity to states with bad behavior at finite distance.
In appendix \ref{app:areabalance}, we demonstrate that 
\begin{equation}
    \alpha + A_{\infty}=\alpha'+A_{-\infty}
    \label{eq:alphafrom2}
\end{equation} 
is the condition which ensures thrice-differentiability at infinity, as one might expect from eq.~\eqref{eq:inftyslope}.
For compatibility with eq.~\eqref{eq:alphapfromalpha}, the flow must obey the purity condition \eqref{eq:purity}.

\paragraph{Connectedness to identity.}
To show $u(\tilde{u})$ is connected to the identity, we interpolate between $u(\tilde{u})$ and the identity map via the following family of transformations:
\begin{equation}
    u_{\lambda}(\tilde{u}) = 
    \begin{cases} 
    \exp[{f_{(\lambda)}(\tilde{u})}]\,, &\tilde{u}>0 \\ 
    - \exp[{g_{(\lambda)}(\tilde{u})}]\,, &\tilde{u}<0\,,
    \end{cases}
\end{equation}
with
\begin{equation}\label{eq:connectedToId}
f_{(\lambda)}(\tilde{u}):=\alpha_\lambda- \int_1^{\tilde{u}}\frac{\rd\tilde{u}^*}{\tilde{c}_\lambda(\tilde{u}^*)} \qquad{\rm and}\qquad
g_{(\lambda)}(\tilde{u}):=\alpha'_\lambda- \int_{-1}^{\tilde{u}}\frac{\rd\tilde{u}^*}{\tilde{c}_\lambda(\tilde{u}^*)}\,,
\end{equation}
and
\begin{equation}
    \tilde{c}_{\lambda}(\tilde{u}) := -(1-\lambda) \tilde{u} + \lambda \tilde{c}(\tilde{u})\,.
\end{equation}
For every choice of $\lambda$, $\tilde{c}_{\lambda}(\tilde{u})\partial_{\tilde{u}} + \tilde{b}_{\lambda}(\tilde{v}) \partial_{\tilde{v}}$ is an Unruh flow, which is furthermore $C^3(\mathbb{R})$ and invertible upon  fixing $\alpha'_{\lambda}(\alpha_{\lambda})$ appropriately.
This demonstrates that the mapping we have constructed is connected to the identity.

\bigskip

This concludes our construction of a well-defined conformal mapping from $|\Omega\rangle$ to $|\Psi\rangle$.
It follows that, given any Unruh flow with respect to $\mathcal{R}$ satisfying the purity condition, there exists a unitary $U_{\varphi}$ in the CFT which maps the vacuum state to a state with modular flow implementing the Unruh flow.
This demonstrates, as desired, that every such flow coincides with the geometric modular flow for some state in the vacuum sector of the CFT.

We emphasize the importance of eq.~\eqref{eq:purity} to ensure smoothness; see the end of section \ref{sec:uniqueness} below.
The purity property arises because eq.~\eqref{eq:diffeq} is a first order equation, and yet we are imposing two independent constraints \eqref{eq:alphapfromalpha} and \eqref{eq:alphafrom2} on the relationship $\alpha'=\alpha'(\alpha)$.
The purity condition ensures that these two constraints are actually equivalent.

Let us close this subsection by explaining why a generic vacuum-sector geometric modular flow approaches $-\tilde{u}\, \partial_{\tilde{u}} + \tilde{v}\,\partial_{\tilde{v}}$ at large $|\tilde{u}|$ and $|\tilde{v}|$.
The special conformal transformation
\begin{equation}
\begin{aligned}
    \tilde{u} \to \tilde{u}_{\cD} =-R\,\frac{\tilde{u}+2R}{\tilde{u}-2R}
    \qquad
    \tilde{v} \to \tilde{v}_{\cD} = R\,\frac{\tilde{v}-2R}{\tilde{v}+2R}
    \label{eq:specialconf}
\end{aligned}
\end{equation}
 maps spatial infinity to the point $(\tilde{t}_\cD,\tilde{x}_\cD) = (0,R)$, such that $\mathcal{R}$ is mapped to $\mathcal{D}:=D(A)$, where $A$ is the interval $\tilde{x}_\cD\in(-R,R)$ at $\tilde{t}_\cD=0$.
Since this map preserves the vacuum sector (see footnote \ref{foot:sctFootnote}), the pushforward of a vacuum-sector modular flow $\tilde{c}\,\partial_{\tilde{u}}+\tilde{b}\,\partial_{\tilde{v}}$ under this map is a vacuum-sector modular flow associated with the diamond $\mathcal{D}$.
Since a generic state in the CFT is expected to have boost-like modular flow at $\partial A$, we expect
\begin{equation}
    \tilde{c}_\cD(\tilde{u}_\cD) = (\tilde{u}_\cD + R) + \cO\left((\tilde{u}_\cD + R)^2\right)\,,
\end{equation}
and similarly for $\tilde{b}_\cD(\tilde{v}_\cD)$.
Mapping this condition back to the Rindler wedge reveals
\begin{equation}
    \lim_{\tilde{u} \to \pm \infty}\tilde{c}(\tilde{u}) = -\tilde{u} + \cO(1).
\end{equation}
Similar reasoning motivates our assumption that $\tilde{c}(\tilde{u})$ is thrice-differentiable at infinity; if this was not the case, then the modular flow obtained after the special conformal transformation acts in a singular way on the stress tensor at the entangling surface.\footnote{See further discussion in section \ref{sec:exDiamond}.}

\subsection{Excited states in the thermal sector}
\label{sec:thermalsector}

In the previous section, we observed that the CFT Hilbert space splits into sectors based on a choice of GNS state.
We constructed vacuum-sector states, which we argued should look like the vacuum at infinity.
Hence, the corresponding geometric modular flows asymptote to the boost flow for large $x$ and finite $t$.
This allowed us to demonstrate that for every vacuum-sector Unruh flow, there is a CFT state with the corresponding modular flow. 

In this section, we explain how to extend this analysis to the thermal state $|\beta\rangle$ at inverse temperature $\beta$ in a Minkowski CFT.
The state $|\beta\rangle$ is the thermofield double state \cite{Brunetti:2015vmh,Witten:2021jzq} (see also \cite{Martin:1959jp,Schwinger:1960qe,Mahanthappa:1962ex,Bakshi:1963bn,Keldysh:1964ud,Takahashi:1975jc}), which can be thought of as a large-volume limit of the Gibbs state $\rho=e^{-\beta H}/Z$.
While the latter describes a mixed state, $|\beta\rangle$ is a vector in its associated GNS representation, which is the thermofield double Hilbert space ${\cal H}_{\beta}$.
In ${\cal H}_{\beta}$, the algebra of observables of the entire Minskowski spacetime  is a Type III$_1$ von Neumann algebra $\mathcal A_1$ which acts reducibly on $\mathcal H_\beta$, because it is accompanied by the nontrivial action of its type III$_1$ commutant $\mathcal A_{1}'=\mathcal A_2$, related to $\mathcal A_1$ by $\mathsf{CPT}$ conjugation.
In this sense, the Hilbert space ${\cal H}_{\beta}$ describes two copies of the original theory.

\begin{figure}
    \centering
    \includegraphics[width=0.95\linewidth]{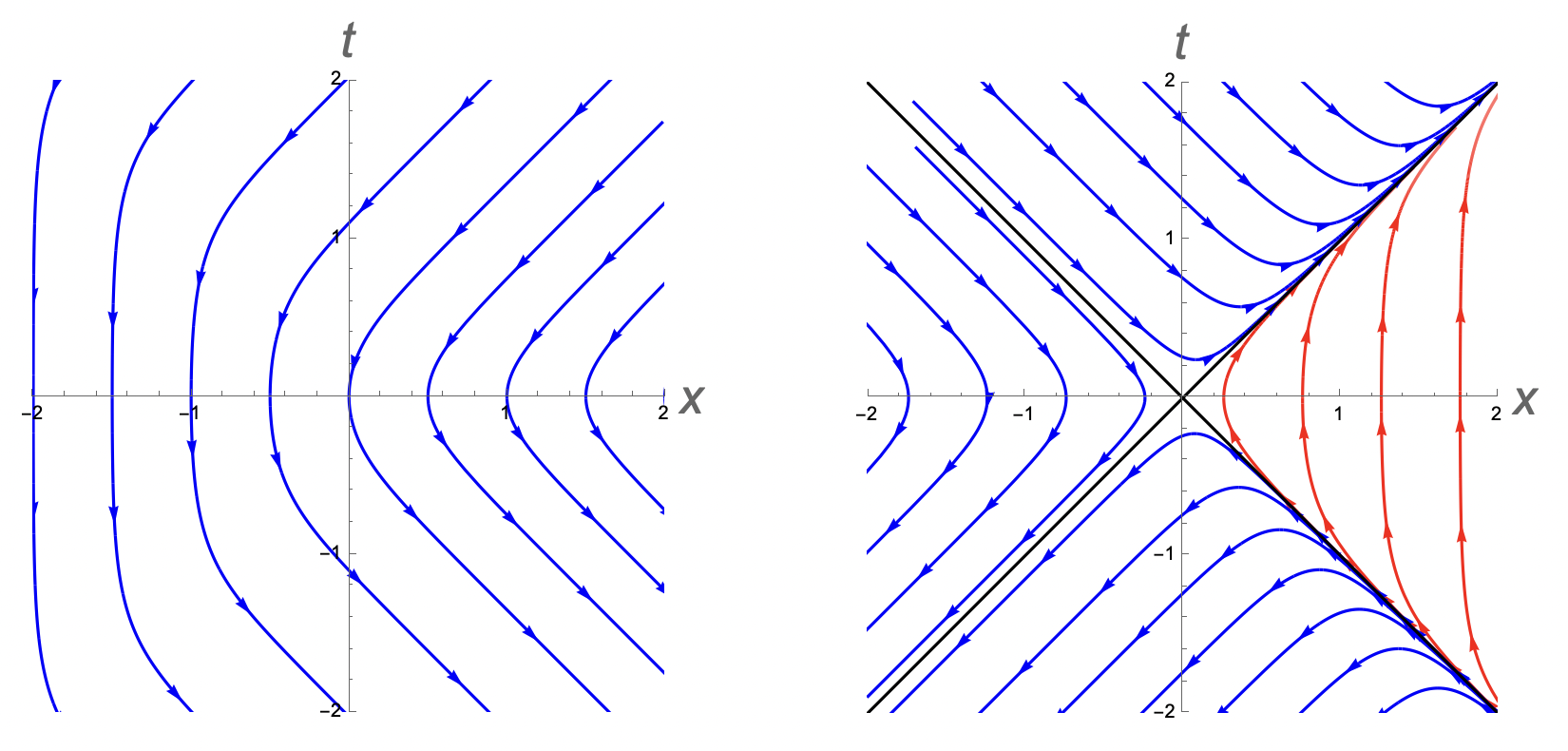}
    \caption{Left: The thermal-state geometric modular flow with respect to $\mathcal{R}$, as it acts in the auxiliary copy of the theory in the thermofield double system.
    This flow is provided in eq.~\eqref{eq:copyflow}.
    Right: For reference, we show the thermal-state geometric modular flow \eqref{eq:borchers1} with respect to $\mathcal{R}$, as it acts in the original copy of the theory, which was shown also in figure \ref{fig:flows}.}
    \label{fig:copy}
\end{figure}

With respect to the Rindler wedge, $|\beta\rangle$ has modular flow \cite{Borchers:1998ye}\footnote{The modular flow for a finite interval in a thermal state of a two-dimensional CFT was explored in \cite{Mintchev:2022fcp}. The Rindler wedge flow \eqref{eq:borchers1} is obtained there via the limit $a=0$ and $b\to\infty$.}
\begin{equation}
    c(u) = 
        \frac{\beta}{2\pi}\left(1-e^{\frac{2\pi u}{\beta}}\right),
\qquad
b(v)=\frac{\beta}{2\pi}\left(1-e^{-\frac{2\pi v}{\beta}}\right)\,,
\label{eq:borchers1}
\end{equation}
The flow is boost-like at the origin.
As $x\to \infty$, it asymptotes to time translations $\frac{\beta}{2\pi} \partial_t$, reflecting the thermality of the underlying state, while at $x\to-\infty$, it asymptotes to $-\frac{\beta}{2\pi}e^{\frac{2\pi |x|}{\beta}}\partial_t$, which is exponentially growing.
The flow is not symmetric in the left and right wedges because the thermofield double state $|\beta\rangle$ is defined on two copies of the original Minkowski theory. The complement of the right Rindler wedge is comprised of both the left Rindler wedge and the entire auxiliary theory in the thermofield double system. The flow is derived in appendix \ref{app:annulus} using the methods of \cite{Cardy:2016fqc}, and shown in figure \ref{fig:copy}.
Note that in the auxiliary theory, the flow asymptotes to $-\frac{\beta}{2\pi} \partial_t$ as $x\to -\infty$ and $-\frac{\beta}{2\pi}e^{\frac{2\pi x}{\beta}}\partial_t$ as $x\to\infty$.

The purity condition \eqref{eq:purity} does not apply to $|\beta\rangle$.
Given an IR regulator $L$, one finds
\begin{equation}
\begin{aligned}
      \lim_{L\rightarrow \infty}\fint_{-L}^{L}\frac{\dd{u}}{c(u)}=\frac{2\pi}{\beta}L\,,
\end{aligned}
\label{eq:notpure}
\end{equation}
which is not vanishing and diverges in $L$.
Nonetheless, the same integral over the auxiliary theory, that is, over the second copy of Minkowski spacetime in the thermofield double system, gives 
\begin{equation}
\begin{aligned}
    \lim_{L\rightarrow \infty}  \fint_{-L}^{L}\frac{\dd{u}}{c_{\mathrm{aux}}(u)}=-\frac{2\pi}{\beta}L,
\end{aligned}
\label{eq:notpure2}
\end{equation}
which exactly cancels eq.~\eqref{eq:notpure}.
This implies that there is the same amount of entropy in the two copies of the system, and thus the state is pure, as one can derive from eq.~\eqref{eq:TheEntropy2d} below. As we derive later, in the thermal sector, the purity condition eq.~\eqref{eq:purity} becomes
\begin{equation}
    \lim_{L\to\infty}\fint_{-L}^L \frac{\rd\tilde{u}}{\tilde{c}(\tilde{u}) }
    + 
    \lim_{L\to\infty}\int_{-L}^L \frac{\rd\tilde{u}}{\tilde{c}_{\mathrm{aux}}(\tilde{u}) } = 0\,.
    \label{eq:purityofpurification}
\end{equation}
This constraint reflects the fact that $|\beta\rangle$ and our excited states in this section are pure states in the full thermofield double system (see eqs.~\eqref{eq:pure1} and \eqref{eq:pure2} below). 
As before, we emphasize  that eq.~\eqref{eq:purityofpurification} constrains the excitations in the left-moving sector alone, and there is an analogous constraint for the right-movers.
Therefore, the constraint we are imposing on the excited thermofield double states is strictly stronger than purity.

We now turn to the construction of Unruh flows.
Any geometric modular flow with respect to $\mathcal{R}$ associated with a state in the thermal sector approaches $\frac{\beta}{2\pi} \partial_t$ at $x\to\infty$.
As $\tilde{u}\to -\infty$ and $\tilde{v}\to \infty$, we thus assume the thermal-sector falloffs
\begin{equation}
    \tilde{c}(\tilde{u})\to \frac{\beta}{2\pi} + \cO\left(\frac{1}{\tilde{u}^2}\right)\qquad
    \tilde{b}(\tilde{v})\to \frac{\beta}{2\pi} + \cO\left(\frac{1}{\tilde{v}^2}\right)\,.
    \label{eq:thermalFalloffs}
\end{equation}
These falloffs are more stringent than their analogue in eq.~\eqref{eq:vacFalloffsRindler}.\footnote{Note that the arguments used in eq.~\eqref{eq:specialconf} to fix the asymptotic behavior do not apply here, due to the absence of special conformal symmetry (see footnote \ref{foot:sctFootnote}).} 
We excluded $\cO(1)$ contributions, since the leading $\frac{\beta}{2\pi}$ term ensures the state reduces to the
thermal state with inverse temperature $\beta$ at spatial infinity.
We moreover excluded $\cO(\frac{1}{\tilde{u}})$ contributions because otherwise, the resulting states would have infinitely more entanglement entropy in the Rindler wedge compared to $|\beta\rangle$, see section \ref{sec:thermalProperties} for 
details.
Note, however, that a similar analysis may apply for more general falloffs, as we discuss further in sections \ref{sec:thermalProperties} and \ref{sec:discuss}.

For the left asymptote at $\tilde{u}\to \infty$, given eq.~\eqref{eq:borchers1}, we assume that
\begin{equation}
    \tilde{c}(\tilde{u})\to -\nu e^{\frac{2\pi \tilde{u}}{\beta}} + {\cal O}(1)
    \label{eq:thermalFalloffs2}
\end{equation}
where $\nu>0$ is an arbitrary constant; it does not affect the energy density eq.~\eqref{eq:manifest} below.

We now show that every Unruh flow with respect to $\mathcal{R}$ corresponds to the modular flow of a state in the thermal sector.
We do so by constructing a conformal map $\varphi:(u,v) \to (\tilde{u}(u),\tilde{v}(v))$ relating the thermal state's modular flow to the Unruh flow of interest, since then $U_{\varphi}|\beta\rangle$ furnishes the desired state. The equation to solve is the same as in eq.~\eqref{eq:diffeq},
\begin{equation}
    \tilde{c}(\tilde{u})u'(\tilde{u})=c(u(\tilde{u}))\,,
    \label{eq:diffeqThermal}
\end{equation}
which in the thermal case becomes
\begin{equation}
    \frac{2\pi \rd u}{\beta\left(1-e^{\frac{2\pi u}{\beta}}\right)}= \frac{\rd\tilde{u}}{\tilde{c}(\tilde{u})}\,.
    \label{cases}
\end{equation}
The solution is
\begin{equation}
    u(\tilde{u}) = -\frac{\beta}{2\pi}
    \begin{cases}
        \log\left(1-e^{f_{\alpha}(\tilde{u})}\right),
        &\tilde{u}>0\\
        \log\left(1+ e^{g_{\alpha'}(\tilde{u})}\right),
        &\tilde{u}<0\,,
    \end{cases}
    \label{eq:uvsolutionThermal}
\end{equation}
with the functions $f_{\alpha}$ and $g_{\alpha'}$ as in eqs.~\eqref{eq:fg}.

We fix the relationship between $\alpha$ and $\alpha'$ by demanding smoothness of $u(\tilde{u})$.
As shown in apppendix \ref{app:thermalApp}, smoothness at the origin requires
\begin{equation}
    \alpha +A + \frac{2\pi}{\beta} = A'+\alpha'\,.
    \label{eq:alpha2pibeta}
\end{equation}
Moreover, to preserve the Rindler wedges, we must set (see appendix \ref{app:thermalApp})
\begin{equation}
    \alpha + A_{\infty} = \log(1-e^{-\frac{2\pi}{\beta}})\,.
    \label{eq:alphafromgen}
\end{equation}

We now discuss how to properly handle the behavior of the map at infinity. In the vacuum sector analysis, we imposed smoothness at infinity by leveraging special conformal symmetry. Since special conformal transformations map left and right spatial infinity to a single finite-distance point, we implicitly identified right and left spatial infinity. 

Since special conformal transformations are not symmetries of the thermal sector, there is no reason here to identify right and left spatial infinity. To properly treat asymptotics, we use the Euclidean path integral framework.
There, the flat-space thermal partition function $Z(\beta)=\Tr e^{-\beta H}$ is associated with the path integral over an infinite cylinder with periodic Euclidean time $\beta$.
The thermofield double state is associated with half of the cylinder, where the range of the Euclidean time is $\frac{\beta}{2}$.
Indeed, in its finite-dimensional form, $|\beta\rangle = \sum_n e^{-\frac{\beta E_n}{2}}\ket{n}\otimes\ket{n}_{\mathrm{aux}}$, where $\ket{n}$ is an energy eigenstate, and $Z(\beta)=\langle \beta|\beta\rangle$.
The two boundaries of the half-cylinder are the $t=0$ slices of the original and auxiliary systems \cite{Hartman:2015}. Conformally mapping the cylinder to $S^2$, the two asymptotic regions of the cylinder become two points, connected by the $t=0$ slices. Thus, the original and auxiliary systems are conformally connected.
Using that the two systems are related by  $\mathsf{CPT}$, the right spatial infinity of one system is then identified with left spatial infinity of the other, and vice versa. We use this crucial identification to address the behavior of the map at infinity in appendix \ref{app:thermalApp}, finding that $u(\tilde{u})$ can be made smooth everywhere if the condition \eqref{eq:purityofpurification} holds.

We thus have shown that every thermal-sector Unruh flow obeying the condition \eqref{eq:purityofpurification} is the modular flow for some state in the thermal sector.

\subsection*{Purity condition and singular behavior}
\label{sec:uniqueness}

The main result of the previous analysis is the identification of a unique\footnote{These states are unique because the global modular flow uniquely determines expectation values in $\mathcal{V}$ and its causal complement \cite{Summers:2003tf}, which can be thought of as spanning a Cauchy surface for the spacetime (assuming suitable treatment of $\partial A$).} state $U_{\varphi}|\Omega\rangle$ or $U_{\varphi}|\beta\rangle$, by the explicit construction of the conformal map $\varphi$, for any vacuum-sector or thermal-sector Unruh flow satisfying the appropriate purity condition.

What happens when an Unruh flow $\xi$ does not satisfy our purity condition?
In that case, the vacuum modular flow can be transformed to $\xi$ using a conformal map $\varphi$ which fails to be differentiable at the origin.\footnote{One can instead use a conformal map $\varphi$ which is not differentiable at infinity.
Similar statements hold for flows not satisfying the boost-like property; then, $\varphi$ is non-invertible at the origin.
See section \ref{sec:exRindler}.}
Due to its Schwarzian transformation rule (eq.~\eqref{eq:schwarzian} below), the resulting stress tensor is divergent
along $\tilde{u}=0$ and $\tilde{v} = 0$. These divergences propagate to other fields in the theory.\footnote{Similarly, the Rindler vacuum stress tensor acquires divergences on the horizon due to the breaking of entanglement between left and right wedges \cite{Hartman:2015, Chen:2022}.}
The singular behavior resulting from violations of our purity condition reflects the idea that a pure state cannot be mapped to a mixed state by a unitary operator.\footnote{For the vacuum sector, eq.~\eqref{eq:purity} requires the entanglement entropy to be the same individually for both the left-moving and right-moving contributions.
This indicates that our conformal unitaries cannot be used to introduce entanglement between right- and left-movers.}
Note that this is different from a  shockwave state produced by the insertion of a primary operator $O_1$ at the origin; in this case, only primary operators with the same scaling dimension receive divergences at null separation.

Nevertheless, due to the Stone-Weierstrass theorem, a continuous function $u(\tilde{u})$ can be obtained as a limit of smooth maps $u_i(\tilde{u})$, even if $u(\tilde{u})$ is not differentiable at $\tilde{u}=0$.
This allows us to understand the behavior of states which violate the purity condition by studying states $U_{\varphi_i}|\Omega\rangle$ for $\varphi_i$ approaching the non-differentiable map of interest.
We use this feature in section \ref{sec:disentangling} to understand disentangling and super-entangling conformal maps.

We conclude this section with an important remark. If one is interested in studying the subregion $\mathcal{R}$ without specifying the behavior of the modular flow outside $\mathcal{R}$,
the results of the previous sections guarantee the existence of a state for every vacuum or thermal-sector Unruh flow specified only in $\mathcal{R}$. 
This comes about because such a flow can be always smoothly extended to an Unruh flow on the entire spacetime which obeys the appropriate purity and boundary conditions.
As an example, consider in the vacuum sector any smooth extension of $\tilde{b}(\tilde{v})$ in $\mathcal{R}$ to the left wedge $\tilde{v}<0$.
This function generically does not satisfy the purity condition, but it can be deformed into a new function which does.
First, multiply the function by a smooth step function which equals 1 for $\tilde{v}\ge -1$ and equals 0 for $\tilde{v} \le -2$.\footnote{This can be obtained from $f(x) = (1+e^{\frac{1}{x-1}+\frac{1}{x}})^{-1}$, satisfying $\lim_{x\to 0^+}f(x)=0$ and $\lim_{x\to 1^-}f(x)=1$.}
Then, add a perturbation which smoothly vanishes outside of the interval $\tilde{v}\in (-3,-2)$. Finally, add a smooth function which is nonzero only for $\tilde{v}<-3$ and approaches $\tilde{b}(\tilde{v})=\tilde{v}$ at infinity.
The intensity of the perturbation can be tuned to ensure the purity condition.

\section{Properties of the excited states}
\label{sec:stateProperties}

In this section, we examine the energy and entanglement properties of the excited states constructed in the previous section, providing a family of examples where the various concepts introduced before can be explicitly constructed and analyzed.
We conclude with a discussion of disentangled states.

\subsection{Vacuum sector}
\label{sec:vacuumProperties}

We begin with the vacuum sector, constructing the energy and entanglement entropy of states with geometric modular flows.
We then analyze the relative entropy.

\subsubsection*{Energy}

Let us evaluate the energy-momentum tensor in the vacuum-sector excited states $|\Psi\rangle$, expressing all quantities in terms of the corresponding geometric modular flow.
While the energy density can be negative, the total energy is always positive and the ANEC is satisfied.

In a $2d$ CFT on Minkowski spactime, the energy-momentum tensor satisfies $\partial_{\mu} T^{\mu \nu} = 0 = T^{\mu}_{\;\;\mu}$. Hence the only non-vanishing components are $T_{uu}=T_{uu}(u)$ and $T_{vv}=T_{vv}(v)$.
Furthermore, in the vacuum state,
\begin{equation}\label{vevT}
   \langle T_{uu}(u) \rangle_{\Omega} := \langle \Omega| T_{uu}(u) |\Omega\rangle
    =0\qquad   \langle T_{vv}(v) \rangle_{\Omega} :=\langle \Omega| T_{vv}(v) |\Omega\rangle = 0\,,
\end{equation}
whereas the state $|\Psi\rangle = U_{\varphi}|\Omega\rangle$ can have non-vanishing energy $T_{tt}= T_{uu}+ T_{vv}$ and momentum density $T_{tx} = - T_{uu}+ T_{vv}$,
due to the Weyl anomaly \cite{Davies:1976hi}.

In order to compute $\langle T_{\tilde{u}\tilde{u}}(\tilde{u}) \rangle_{\Psi}:=\langle \Psi | T_{\tilde{u}\tilde{u}}(\tilde{u}) |\Psi \rangle = \langle \Omega | U_{\varphi}^{\dag}\,  T_{\tilde{u}\tilde{u}}(\tilde{u})\, U_{\varphi} |\Omega \rangle$, we use the Schwarzian transformation rule for the stress tensor, using the conventions in \cite{DiFrancesco:1997nk} for our conformal map $\varphi:(u,v) \rightarrow (\tilde{u}(u),\tilde{v}(v))$,  
\begin{equation}
   U_{\varphi}^{\dag}\,  T_{\tilde{u}\tilde{u}}(\tilde{u})\,U_{\varphi}=\tilde{T}_{\tilde u\tilde u}(\tilde u) = \left(\tilde{u}'\right)^{-2}\left(T_{uu}(u) +\frac{c}{24\pi}\{\tilde{u},u\}\right) 
    \,,
    \label{eq:schwarzian}
\end{equation}
where $\{f(x),x\}$ denotes the Schwarzian derivative
\begin{equation}\label{sch}
    \{f(x) ,x\}=\left(\frac{f''}{f'}\right)'  -\frac{1}{2} \left(\frac{f''}{f'}\right)^2.
\end{equation}

Now using eq.~\eqref{vevT} together with the identity $\{\tilde{u},u\}=-\left(\tilde{u}'\right)^{2}\{u,\tilde{u}\}$, we have
\begin{equation}
    \langle T_{\tilde{u}\tilde{u}}(\tilde{u}) \rangle_{\Psi}
    = -\frac{c}{24\pi}\{u(\tilde{u}),\tilde{u}\} \,,
\label{eq:schwarzianNullEnergyDensity}
\end{equation}
and similarly for $T_{\tilde{v}\tilde{v}}$. From eq.~\eqref{sch}, the energy density at $\tilde{t}=0$ ($\tilde{u}=-\tilde{x}$ and $\tilde{v}=\tilde{x}$) is
\begin{equation}
    \langle T_{\tilde{t}\tilde{t}}(\tilde{x})  \rangle_\Psi
    = \frac{c}{24\pi}\left[\frac{1}{2} \left(\frac{u''}{u'}\right)^2 - \left(\frac{u''}{u'}\right)'  
    + 
    \frac{1}{2} \left(\frac{v''}{v'}\right)^2 -\left(\frac{v''}{v'}\right)' \right]\,,
    \label{eq:energydensity}
\end{equation}
where $u$, $v$, etc.~are functions of $\tilde{x}$; for example, $u' = u'(\tilde{u})\big|_{\tilde{u}=-\tilde{x}}$.

While the energy density $\langle T_{\tilde{t}\tilde{t}}(\tilde{x})  \rangle_\Psi$ can be locally negative, the total energy,
\begin{equation}
    \langle E \rangle_{\Psi}
    =  \int_{-\infty}^{\infty} \rd\tilde{x}\, \langle  T_{\tilde{t}\tilde{t}}(\tilde{x}) \rangle_\Psi \,,
\end{equation}
which after integrating by parts simplifies to\footnote{The boundary term is $(\frac{u''}{u'}+\frac{v''}{v'})\big|^{\infty}_{-\infty}\,$. Rewritten in terms of the flow via $u'(\tilde{u}) = - \frac{u(\tilde{u})}{\tilde{c}(\tilde{u})}$ and $v'(\tilde{v}) =  \frac{v(\tilde{v})}{\tilde{b}(\tilde{v})}$, it becomes  $\left(-\frac{1+\tilde{c}'}{\tilde{c}}+ \frac{1-\tilde{b}'}{\tilde{b}}\right)\big|^{\infty}_{-\infty}$, which, using $\tilde{c}(\tilde{u}) = - \tilde{u} + \cO(1)$ and $\tilde{b}(\tilde{v}) =  \tilde{v} + \cO(1)$, identically vanishes.\label{ft19}}
\begin{equation}
    \langle E \rangle_{\Psi}
    = \frac{c}{48 \pi} \int_{-\infty}^{\infty} \rd\tilde{x}\,\left[ \left(\frac{u''}{u'}\right)^2 
     + \left(\frac{v''}{v'}\right)^2\right] \,,
     \label{eq:positiveenergy}
\end{equation}
is manifestly positive.
In the following subsection, we show that while $\langle E \rangle_{\Psi}$ is bounded from below, it is unbounded from above.

From these results, it is also straightforward to show that $|\Psi\rangle$ satisfies the averaged null energy condition (ANEC).
In two dimensions, the ANEC reads, for any fixed $\tilde{v}$,
\begin{equation}
    \int_{-\infty}^{\infty}\rd \tilde{u}\, \langle T_{\tilde{u}\tilde{u}}(\tilde{u})\rangle_{\Psi} \ge 0\,,
    \label{eq:anec}
\end{equation}
and similarly with the roles of $\tilde{u}$ and $\tilde{v}$ swapped.
Satisfying the ANEC is to be expected, as various proofs \cite{Faulkner:2016mzt, Hartman:2016lgu, Casini:2017roe, Casini:2017vbe} suggest the ANEC holds for generic states in any Minkowski QFT.\footnote{See also \cite{Kontou:2020bta} for a review on energy conditions.}
Thanks to the boost-like property at the origin, the half-sided integrals are also positive, 
\begin{equation}
    \int_{0}^{\infty}\rd \tilde{u}\, \langle T_{\tilde{u}\tilde{u}}(\tilde{u})\rangle_{\Psi} \ge 0\, , \qquad     \int_{-\infty}^{0}\rd \tilde{u} \,\langle T_{\tilde{u}\tilde{u}}(\tilde{u})\rangle_{\Psi} \ge 0\,,
\end{equation}
but we cannot guarantee that the averaged null energy on other semi-infinite intervals is positive; see related discussions in \cite{Davies:1976hi}.

These  properties of the energy of states in the vacuum conformal module are well-known, e.g., see \cite{Birrell:1982ix}. 
Here, we are interested in exploring connections to $|\Psi\rangle$'s geometric modular flow.
Since $u'(\tilde{u}) = -\frac{u(\tilde{u})}{\tilde{c}(\tilde{u})}$ and $v'(\tilde{v}) =  \frac{v(\tilde{v})}{\tilde{b}(\tilde{v})}$, we have
\begin{equation}
    \langle E \rangle_{\Psi}
    = \frac{c}{48 \pi} \int_{-\infty}^{\infty} \rd\tilde{x}\,\left[ \left(\frac{1 + \tilde{c}'}{\tilde{c}}\right)^2 
     + \left(\frac{1 - \tilde{b}'}{\tilde{b}}\right)^2\right]\,,
     \label{eq:ENERGY}
\end{equation}
which vanishes when $\tilde{b}'=1=-\tilde{c}'$, as is the case for the vacuum modular flow. 

Moreover, we remark that the null energy densities read\footnote{Note that, due to the $u$ and $v$ dependence of the energy components, we can compute energy densities everywhere --- also, say, in the future and past lightcones of the origin --- using the following equation. \label{garmond}}
\begin{equation}
    \langle T_{\tilde{u}\tilde{u}}(\tilde{u}) \rangle_\Psi 
    = \frac{c}{48\pi}\left[\frac{1}{\tilde{c}^2}
    +
    4\frac{(\sqrt{\tilde{c}})''}{\sqrt{\tilde{c}}}
    \right]\,,
    \qquad
    \langle T_{\tilde{v}\tilde{v}}(\tilde{v}) \rangle_\Psi 
    = \frac{c}{48\pi}\left[\frac{1}{\tilde{b}^2}
    + 
    4\frac{(\sqrt{\tilde{b}})''}{\sqrt{\tilde{b}}}
    \right]\,.
    \label{eq:manifest}
\end{equation}
These expressions 
make the transformation properties of the $2d$ stress tensor manifest.
Under a conformal map, the first term transforms as a spin-$2$ primary of conformal weight $\Delta = 2$, while the second term encodes the anomalous transformation rule.\footnote{Explicitly, the relation $\tilde{c}(\tilde{u}(u)) = \tilde{u}'(u)c(u)$ implies
\begin{equation}
    (\tilde{u}')^2 \frac{(\sqrt{\tilde{c}})''}{\sqrt{\tilde{c}}}-\frac{1}{2}\{\tilde{u},u\}
   =\frac{(\sqrt{{c}})''}{\sqrt{c}}\,.
\end{equation}
Comparing to eq.~\eqref{eq:schwarzian}, we see the expression $(\sqrt{\tilde{c}})''/\sqrt{\tilde{c}}$ encodes the anomalous transformation.
Note, this decomposition is not unique; for example, we could add $1/\tilde{c}^2$ to $(\sqrt{\tilde{c}})''/\sqrt{\tilde{c}}$ and $1/c^2$ to $(\sqrt{c})''/\sqrt{c}$.
\label{foot:manifest}} We discuss discuss the applicability of eq.~\eqref{eq:manifest} to more general contexts in sections \ref{sec:thermalProperties}, \ref{sec:exDiamond}, and \ref{sec:energytomatch}.

\subsubsection*{Entropy}

We now perform a similar analysis for the entanglement entropy.

The entanglement entropy of a state $\ket{\Phi}$ is given by the von Neumann entropy of $\rho_{\Phi}$, 
\begin{equation}\label{vNe}
S(\rho_{\Phi})=\Tr(\rho_{\Phi} h_{\Phi}) 
= \langle h_{\Phi} \rangle_{\Phi}
\,,\qquad\rm{with}\quad h_{\Phi} = -\log \rho_{\Phi},
\end{equation}
where $h_{\Phi}$ is the one-sided modular Hamiltonian.
When we use the language of reduced density matrices $\rho_{\Phi}^{\mathcal{V}} := \Tr_{\bar{A}}|\Phi\rangle \langle \Phi |$
rather than states,
the subscript labels the global state and the superscript labels the subregion of interest. 
In most of this section, we are interested in the Rindler wedge $\mathcal{V}=\mathcal{R}$; in this case, we drop the superscript.
As we comment further below, the use of density matrices requires the use of a UV regulator.

For $h_\Phi^\mathcal{V}$ of the form
\begin{equation}
    h_\Phi^\mathcal{V} = \int_A \dd x\, \beta(x)\, T_{tt}(x)\,,
    \label{eq: EE Wongham}
\end{equation}
the following formula for the entanglement entropy has been proposed in \cite{Wong:2013gua}:
\begin{equation}
    S(\rho_\Phi^\mathcal{V}) =\frac{c \pi}{3} \int_A \frac{\dd x}{\beta(x)}\,.
    \label{eq: EE Wong}
\end{equation}
The function $\beta(x)$ in eq.~\eqref{eq: EE Wongham} plays the role of a spatially varying inverse temperature \cite{Susskind:2005js}, and so the thermal equilibrium formula for the entropy density,\footnote{This can be derived using standard thermodynamics for uniform thermal states.
Applying the first law $\rd E = T \rd S$ at the level of entropy and energy densities, we have $\rd \langle T_{tt}\rangle_{\beta} = T \rd s$.
From eq.~\eqref{eq:cftThermalEnergyDensity}, we have $\langle T_{tt}(x)\rangle_{\beta}= \frac{c\pi}{6\beta^2}$, and thus to ensure $T \rd s =\rd\langle T_{tt}\rangle_{\beta}= \frac{c\pi}{3\beta} \rd T $, we must have $s = \frac{c\pi}{3\beta}$.} 
$s = \frac{c\pi}{3\beta}$, is being generalized to the case of a local inverse temperature $\beta=\beta(x)$ and then integrated over the region of interest.
In \cite{Wong:2013gua, Cardy:2016fqc}, it was found that eq.~\eqref{eq: EE Wong} correctly reproduces the entanglement entropy of an interval in the Minkowski vacuum, in the Minkowski thermal state, and in the vacuum state on the cylinder, among other cases.

Our states $|\Psi \rangle = U_{\varphi}|\Omega\rangle$ satisfy the condition \eqref{eq: EE Wongham} with respect to $\mathcal{V} = \mathcal{R}$ if $\tilde{b}(\tilde{x})= \tilde{c}(-\tilde{x})$.
Then, the modular flow $\xi$ at $\tilde{t}=0$ points only in the $t$ direction, and eq.~\eqref{eq:localHam} yields
\begin{equation}
    h_{\Psi}^\mathcal{R}
    = 2\pi \int_0^{\infty} \dd{\tilde x}\,  T_{\tilde{t}\tilde{t}}(\tilde x)\,\xi^{\tilde{t}}(\tilde{x})\,,
    \label{eq:betaIsPsit}
\end{equation}
leading to the identification $\beta=2\pi \xi^{\tilde{t}}$.
Physically, this says that the more the modular flow vector extends upwards into the time direction, the lower the temperature associated with an observer following integral curves of $\xi$, and hence a corresponding decrease results in the entanglement entropy \eqref{eq: EE Wong}.
We prove in appendix \ref{app:annulus} that eq.~\eqref{eq: EE Wong} provides the entanglement entropy of our states $|\Psi\rangle$ obeying the condition $\tilde{b}(\tilde{x}) = \tilde{c}(-\tilde{x})$, both in the vacuum and thermal sector.

Since most of our states do not satisfy the condition $\tilde{b}(\tilde{x}) = \tilde{c}(-\tilde{x})$, let us propose a generalization of eq.~\eqref{eq: EE Wong} using similar reasoning about local temperatures. Following \cite{SEWELL198023, Martinetti:2002sz,Hartman:2015}, note that for an observer traveling tangent to an Unruh modular flow $\xi^{\mu}$, the instantaneous relationship between the proper time $\tau$ and modular time $s$ is $\tau = \norm{\xi} s$.\footnote{Integral curves $x^{\mu}(s)$ of $\xi^{\mu}$ obey, by definition, $\frac{\rd x^{\mu}}{\rd s} = \xi^{\mu}$.
Then, we have $\rd \tau  =\sqrt{-\frac{\rd x}{ds}\cdot\frac{\rd x}{\rd s}}\,\rd s = \norm{\xi} \rd s$.\label{foot:dtauds}}
This implies, locally and instantaneously,
\begin{equation}
    Q[\partial_s] = \norm{\xi}\,Q[\partial_\tau]\,,
\end{equation}
where $Q[\zeta]$ denotes the generator of the diffeomorphism $\zeta$.
We observe that, from eq.~\eqref{eq:localHam}, the modular Hamiltonian $h_{\Psi}$ is $2\pi$ times the generator of $\xi = \partial_s$. Moreover, since $\partial_{\tau}$ coincides with time-translations for the observers, we have $Q[\partial_\tau]=H_{\mathrm{obs}}$. Then, we have
\begin{equation}
    h_{\Psi} = 2\pi \norm{\xi} H_{\mathrm{obs}}\,.
\end{equation}
Given that $\rho_{\Psi} = e^{- h_{\Psi}}/Z = e^{-2\pi\norm{\xi} H_{\mathrm{obs}}}/Z$, it appears that we can assign to each observer a local temperature $\beta = 2\pi \norm{\xi}$.
One has to be cautious with this interpretation of this ``modular temperature'' --- see the discussion in section \ref{sec:discuss}.
For now, we note that this reasoning generalizes the identification $\beta = 2\pi \xi^{\tilde{t}}$ in \cite{Wong:2013gua}, and it precisely retrieves $\beta = \frac{2\pi}{a}$ experienced by observers with constant acceleration $a$ along the vacuum modular flow in the Rindler wedge.
Notice that Unruh temperatures for accelerated observers in de Sitter space and anti-de Sitter space can also be obtained via $\beta = 2\pi \norm{\xi}$, \eg see \cite{Narnhofer:1996zk,Deser:1997ri,Jacobson:1997ux}.

In each observer's rest frame, we have the entropy density
\begin{equation}
    s_{\mathrm{obs}}= 
    \frac{c\pi}{3\beta}
    =\frac{c}{6\norm{\xi}}\,. 
    \label{eq:entropydensity}
\end{equation}
To covariantize this expression, recall that an electric charge density $\rho$ at rest with respect to a unit normal vector $t^{\mu}$ has current four-vector $j^{\mu}= \rho \,t^{\mu}$, and conversely $\rho = - j^{\mu}\,t_{\mu}$.
Analogously, setting $t^{\mu} = \frac{\xi^{\mu}}{\norm{\xi}}$, the entropy density \eqref{eq:entropydensity} corresponds to the current
\begin{equation}
    j_S^{\mu}
    = \frac{c}{6}\cdot \frac{\xi^{\mu}}{\norm{\xi}^2}\,,
    \label{eq:news}
\end{equation}
such that $s_{\mathrm{obs}}= - j_S^{\mu}\,t_{\mu}= - \frac{j_S^{\mu}\,\xi_{\mu}}{ \norm{\xi}}$, as desired.

We accordingly propose the following entanglement entropy formula in $d=2$ for a general state $\Psi$ with geometric modular flow $\xi$ in a subregion $\mathcal{V}=D(A)$:
\begin{equation}
    S(\rho^{\mathcal{V}}_{\Psi})
    =- \int_A  \rd \Sigma_{\mu}\, j_{S}^{\mu}= - \frac{c}{6}\int_A  \rd \Sigma_{\mu}\, \frac{\xi^{\mu}}{\norm{\xi}^2}\,,
    \label{eq:covariantEntropy}
\end{equation}
where $\rd\Sigma_{\mu} =\sqrt{h} \, \rd x\,  n_{\mu} $ is the appropriate volume form on $A$, with $n^{\mu}$ the future-pointing unit normal.
As a consistency check, when the flow is geometric throughout $\mathcal{V}$, $S$ is independent of the choice of Cauchy slice $A$ for $\mathcal{V}$,
\begin{equation}
\nabla_{\mu}j_S^{\mu} \sim
\nabla_{\mu}\left(\frac{\xi^{\mu}}{\norm{\xi}^2}\right)=0\,,
\end{equation}
thanks to the conformal Killing property \eqref{eq:ckv} of $\xi$ in $d=2$.
Typically, $S$ will be divergent, and UV and/or IR cutoffs must be imposed on the domain of integration.
In this case, for $S$ to still be conserved, the cutoff surfaces must run tangent to the modular flow $\xi$.
The entropy formula \eqref{eq:covariantEntropy} will be discussed further in section \ref{sec:entanglementcurrents} and proved in section \ref{sec:discuss}.

Applying eq.~\eqref{eq:covariantEntropy} to our excited states with geometric modular flow, we have\footnote{We used $\rd\Sigma_{u} = -\frac{1}{2} \rd\Sigma^v = -\frac{1}{2}\sqrt{h}\,n^v \rd v$. 
Studying the line $v=-\frac{u}{\epsilon}$ as $\epsilon\to0$, one finds $\sqrt{h}\,n^v\to 1$.} 
\begin{equation}
    S(\rho^{\mathcal{V}}_{\Psi}) = -\frac{c}{6}\left( \int_A \rd \Sigma_{\tilde{u}} \frac{1}{\tilde{b}(\tilde{v})} + \int_A \rd \Sigma_{\tilde{v}} \frac{1}{\tilde{c}(\tilde{u})} \right)
    = \frac{c}{12}\left( \int_A \frac{\rd \tilde{v}}{\tilde{b}(\tilde{v})} + \int_A \frac{\rd \tilde{u}}{\tilde{c}(\tilde{u})} \right)\,.
    \label{eq:TheEntropy2d}
\end{equation}
From this expression, we confirm that the vanishing of eqs.~\eqref{eq:purity} and \eqref{eq:purityofpurification} is indeed related to the purity of the underlying state, as discussed in sections \ref{sec:vacuumsector} and \ref{sec:thermalsector}. Indeed, when eqs.~\eqref{eq:purity} or \eqref{eq:purityofpurification} hold, the entropy \eqref{eq:TheEntropy2d} in the Rindler wedges coincide.

Comparing this result to the modular Hamiltonian\footnote{We can derive this result either from eq.~\eqref{eq:localHam} or by using $h_{\Omega}=2\pi\int_{0}^\infty \rd x \, x \, T_{tt}(x)= 2\pi\int_{0}^\infty \rd u \,u\, T_{uu}(u)+2\pi\int_{0}^\infty \rd v \,v\, T_{vv}(v)$, and dropping the operator-free part coming from the Schwarzian transformation rule, which is relevant only for normalization considerations.}
\begin{equation}
\begin{aligned}
    h_{\Psi} &= U_{\varphi}\, h_{\Omega}\, U_{\varphi}^{\dag}
    = 
    2\pi\int_{0}^\infty \rd \tilde{x} \left[ \tilde{b}(\tilde{x})\,T_{\tilde{v}\tilde{v}}(\tilde{x})
    + \tilde{c}(-\tilde{x})\,T_{\tilde{u}\tilde{u}}(\tilde{x})\right]\,,
    \label{eq:modhamsplits}
\end{aligned}
\end{equation}
we see $\tilde{c}$ and $\tilde{b}$ play the role of inverse temperatures for the right- and left-moving sectors.

As the simplest example, consider the vacuum state.
In this case, $\frac{1}{c(-\tilde{x})} = \frac{1}{b(\tilde{x})} = \frac{1}{\tilde{x}} = (\log \tilde{x})'$, so eq.~\eqref{eq:TheEntropy2d} yields 
\begin{equation}
    S(\rho_{\Omega})
    = \frac{c}{6}\int_0^{\infty} \rd \tilde{x} \, (\log \tilde{x})'
    =\frac{c}{6}\log\frac{L}{\epsilon}
    \,,
    \label{eq:2dentropy}
\end{equation}
where we have introduced a UV cutoff $\epsilon$ and an IR cutoff $L$ to regulate the divergences at small and large $x$, respectively.
The UV divergence, $\frac{c}{6}\log\frac{1}{\epsilon}$,
comes from the modular flow vanishing in a boost-like manner at the entangling surface.
This is can be compared with the entanglement entropy of a finite interval of length $\ell$,
\begin{equation}
    S=\frac{c}{3}\log\frac{\ell}{\epsilon}\,.
    \label{eq:loft}
\end{equation}
In this case, the entangling surface is comprised of the two endpoints and the modular flow vanishes at both. Hence, each of these contributes $\frac{c}{6}\log\frac{1}{\epsilon}$ to the UV divergence.

Now consider our excited states. Their vacuum-subtracted entanglement entropy is
\begin{equation}
\Delta S=S(\rho_{\Psi})-S(\rho_{\Omega}) = \frac{c}{12}\int_0^{\infty}\rd\tilde{x}\left(\frac{1}{\tilde{b}(\tilde{x})}+\frac{1}{\tilde{c}(-\tilde{x})}-\frac{2}{ \tilde{x}}\right)\,.
\label{eq:FINITE_ENTROPY}
\end{equation}
The vacuum-subtraction ensures that the UV and IR divergences cancel and thus $\Delta S$ is finite. To confirm  this, recall the push-forward equation 
\begin{equation}
    \frac{1}{\tilde{c}(\tilde{u})} =\frac{u'(\tilde{u})}{c(u(\tilde{u}))} = -\frac{u'(\tilde{u})}{u(\tilde{u})}=-\left[\log(u(\tilde{u}))\right]'\,,
\end{equation}
and similarly for $\tilde{b}(\tilde{v})$.
Therefore, the entanglement entropy reduces to a boundary term:
\begin{equation}
    S(\rho_{\Psi}) = \frac{c}{12}\log\frac{u(-L)v(L)}{u(-\epsilon)v(\epsilon)}
    =
    \frac{c}{12}\log\frac{u'(-\infty)v'(\infty) L^2}{u'(0)v'(0) \epsilon^2}
    = 
    \frac{c}{12}\log\frac{u'(-\infty)v'(\infty)}{u'(0)v'(0)} + S(\rho_{\Omega})
    \,.
    \label{eq:EEExcitedVacState}
\end{equation}
From section \ref{sec:AStateForEveryUnruhFlow}, we have that $u'(0)=e^{\alpha + A}$, $u'(\infty)=e^{\alpha' + A_{-\infty}}$ are finite nonzero constants,  and similarly for $v$. Hence, $\Delta S$ is finite and explicitly given by
\begin{equation}
    \Delta S=S(\rho_{\Psi})-S(\rho_{\Omega}) = \frac{c}{12}(A_{-\infty}-A'+B_{\infty}-B).
    \label{eq:ABentropy}
\end{equation}

Going from eq.~\eqref{eq:2dentropy} to eq.~\eqref{eq:EEExcitedVacState} amounts to taking $\epsilon\to \epsilon\sqrt{u'(0)v'(0)}$ and $L\to L\sqrt{u'(-\infty)v'(\infty)}$.
This is reminiscent of the well-known result \cite{Callebaut:2023fnf, Holzhey:1994we} that the UV cutoff $\epsilon:=\sqrt{\epsilon_1 \epsilon_2}$ for the entanglement entropy of a finite interval from $z_1$ to $z_2$ in an Euclidean $2d$ CFT transforms under $z\to f(z), \bar{z}\to\bar{f}(\bar{z})$ as
\begin{equation}
    \epsilon_{f_1}= \sqrt{f'(z_1)}\epsilon_{z_1}
    \qquad
    \epsilon_{\bar{f}_1}= \sqrt{\bar{f}'(\bar{z}_1)}\epsilon_{\bar{z}_1}
    \qquad
    |\epsilon_{f_1}|= \sqrt{f'(z_1)\bar{f}'(\bar{z}_1)} |\epsilon_{z_1}|\,.
    \label{eq:cutoffTransformation}
\end{equation}
The entropy formula \eqref{eq:covariantEntropy} is thus consistent with the Lorentzian analog of $\sqrt{f'(z_1)
\bar{f}'(\bar{z}_1)}$, \ie $\sqrt{u'(\tilde{u}_1)v'(\tilde{v}_1)}$. Moreover, in the vacuum sector the IR cutoffs transform in the same way as the UV cutoffs, in contrast to the thermal case discussed below.

We close this subsection with a consistency check of the  entropy formula \eqref{eq:covariantEntropy}: We test that it yields a relative entropy with the expected properties.\footnote{For an analysis of the relative entropy of CFT$_2$ states excited by a conformal transformation, see \cite{deBoer:2023lrd}.}
The relative entropy of $\rho_{\Psi}$ with respect to $\rho_{\Omega}$ is defined as
\begin{equation}
S_{\mathrm{rel}}(\rho_{\Psi}||\rho_{\Omega})=\text{Tr}(\rho_{\Psi} \log\rho_{\Psi})-\text{Tr}(\rho_{\Psi} \log\rho_{\Omega})
=\langle h_{\Omega}\rangle_{\Psi}-S(\rho_{\Psi})\,,
\end{equation} and it measures the distinguishability of $|\Psi\rangle$ relative to $|\Omega\rangle$, given access to only $\mathcal{V} = \mathcal{R}$.

Adding and subtracting the von Neumann entropy of $\rho_{\Omega}$ and using $S(\rho_{\Omega}) = \langle h_{\Omega}\rangle_{\Omega}$,
\begin{equation}   S_{\mathrm{rel}}(\rho_{\Psi}||\rho_{\Omega})
=\left(\langle h_\Omega\rangle_{\Psi}-\langle h_\Omega\rangle_{\Omega}\right)
-\left(S(\rho_{\Psi})-S(\rho_{\Omega})\right).
\label{eq:relEntropyDef}
\end{equation}
While the entropy difference $S(\rho_{\Psi})-S(\rho_{\Omega})$ can have either sign, $S_{\mathrm{rel}}$ is by definition non-negative, as it  measures the distinguishability of states, and it vanishes only when $\rho_\Psi = \rho_\Omega$.

Substituting the results from the previous subsection,
we have\footnote{For computational simplicity, we have assumed $\tilde{c}$ is an odd function.}
\begin{equation}
    S_{\mathrm{rel}}(\rho_{\Psi}||\rho_{\Omega})= \frac{ c}{12}\int_0^{\infty}\rd\tilde{x}\left[ \frac{1}{\tilde{c}(\tilde{x})}+\frac{1}{\tilde{x}}
    + 
    \frac{\tilde{x}}{2 \tilde{c}(\tilde{x})^2} \left(1 - \tilde{c}'(\tilde{x})^2\right)
    + \tilde{x}\frac{\tilde{c}''(\tilde{x}) }{\tilde{c}(\tilde{x})}\right] + b\textrm{ terms}\,.
    \label{eq: Srel}
\end{equation}
As expected, this quantity is extremized in the vacuum state, since, focusing on the $u$-sector terms only, a small perturbation $\tilde{c}(\tilde{u}) = -\tilde{u}-\delta c(\tilde{u})$ yields
\begin{equation}
    \delta S_{\mathrm{rel}} = \frac{ c}{6}\int_0^{\infty}\frac{\rd\tilde{x}}{\tilde{x}^2}\left(\delta c(\tilde{x}) - \tilde{x}\delta c'(\tilde{x}) 
    + \tilde{x}^2\delta c''(\tilde{x}) \right) + \cO(\delta c^2)\,,
\end{equation}
which vanishes after integrating by parts.
Since $0=\delta S_{\mathrm{rel}} = \delta \langle h_{\Omega} \rangle - \delta S$, this gives the well-known first law $\delta S = \delta \langle h_{\Omega}\rangle$ for infinitesimal perturbations of the vacuum.
Together with the positivity of the relative entropy in various examples in section \ref{sec:exRindler} (see figure \ref{fig:entropy}), this provides a consistency check of our proposed entropy formula eq.~\eqref{eq:covariantEntropy}.

\subsection{Thermal sector}
\label{sec:thermalProperties}

We now turn to the thermal sector, constructing the energy and entanglement entropy of states with geometric modular flows. We then discuss generalizations of this sector's falloffs.

\subsubsection*{Energy}

Let us compute the energy of our excited states $|\Psi\rangle := U_{\varphi}|\beta\rangle$ in the thermal sector.

We begin by calculating the one-point function $\expval{T_{\mu \nu}}_\beta$ in the thermal state on the plane. 
The state at $t=0$ is prepared by a Euclidean path integral on a cylinder with periodic Euclidean time $\tau\sim\tau + \beta$.
Noting that the Euclidean cylinder can be mapped to the Euclidean plane by $w=-\frac{\beta}{2\pi}\log z$, the Schwarzian transformation rule \eqref{eq:schwarzian} gives \cite{DiFrancesco:1997nk}
\begin{equation}
    \expval{T_{ww}(w)}_{\beta} = \expval{\left(\dv{w}{z}\right)^{-2}\left[ T_{zz}(z)-\frac{c}{24\pi}\{w, z\} \right]}_{\mathrm{Plane}}
    = -\frac{\pi c}{12\beta^2}\,.
\end{equation}
By construction, this Euclidean derivation yields the one-point function at the $t=0$ hypersurface of the Lorentzian theory.
That is, identifying $w=iu$, we have
\begin{equation}
    \langle T_{uu}(u)\rangle_{\beta} =\frac{\pi c}{12\beta^2}\,,
    \label{eq:nullenergyThermal}
\end{equation}
so that, using the time-translation symmetry of the thermal state, 
\begin{equation}
    \expval{T_{tt}(t,x)}_\beta=\expval{T_{xx}(t,x)}_\beta= \frac{\pi c}{6\beta^2}
    \qquad \expval{T_{tx}(t,x)}_\beta=0\,.
    \label{eq:cftThermalEnergyDensity}
\end{equation}
These are the desired $|\beta\rangle$-expectation values on the entire Lorentzian spacetime.

We then compute the expectation values in the excited state $|\Psi\rangle$
\begin{equation}
\begin{aligned}
     \expval{T_{\tilde{u}\tilde{u}}(\tilde{u})}_\Psi &= \expval{U_{\varphi^{-1}}T_{\tilde{u}\tilde{u}}(\tilde{u})U_{\varphi^{-1}}^{\dag}}_\beta \\
     &= \expval{\left(\tilde{u}'\right)^{-2}\left(T_{uu}(u) +\frac{c}{24\pi}\{\tilde{u},u\}\right) }_\beta \\
     &= \left(u'\right)^{2} \frac{\pi c}{12\beta^2} - \frac{c}{24\pi}\{u,\tilde{u}\}\,,
\end{aligned}
\end{equation}
and similarly for $T_{\tilde{v}\tilde{v}}$. With this, we  evaluate
\begin{equation}
\begin{aligned}
     \expval{T_{\tilde{t}\tilde{t}}(\tilde{t}, \tilde{x})}_\Psi &= \left[\left(u'\right)^{2}+\left(v'\right)^{2}\right]\frac{\pi c}{12\beta^2}-\frac{c}{24\pi} \left( \{u, \tilde{u}\} + \{ v,\tilde{v}\} \right)\,,
\end{aligned}
\end{equation}
and similarly for $T_{\tilde{t}\tilde{x}}$ (one simply flips the sign on the $u$ terms).
Note that the second term takes the same form as eq.~\eqref{eq:energydensity}, the energy density in the analogous vacuum sector state.

Rewriting these quantities in terms of the modular flows $\tilde{c}$ and $\tilde{b}$, we obtain
\begin{equation}
\begin{aligned}
     \expval{T_{\tilde{v}\tilde{v}}(\tilde{v})}_\Psi &= \frac{c}{24\pi}\left[\frac{2\pi^2}{\beta^2}\left(\frac{b}{\tilde{b}}\right)^{2}+\frac{b'^2-\tilde{b}'^2 }{2\tilde{b}^2}+\frac{\tilde{b}\tilde{b}''-b b''}{\tilde{b}^2} \right]\,,
     \label{eq:Thermalenergydensity}
\end{aligned}
\end{equation} 
and similarly for $T_{\tilde{u}\tilde{u}}$, where one replaces $(b,\tilde{b})$ with $(c,\tilde{c})$.
Here, $b'$ denotes $b'(v)\big|_{v=v(\tilde{v})}$.

Upon substituting eq.~\eqref{eq:borchers1} for $b$, eq.~\eqref{eq:Thermalenergydensity} simplifies to 
\begin{equation}
\begin{aligned}
    \langle T_{\tilde{v}\tilde{v}}(\tilde{v}) \rangle_\Psi 
    = \frac{c}{48\pi}\left[\frac{1}{\tilde{b}^2}
    + 
    4\frac{(\sqrt{\tilde{b}})''}{\sqrt{\tilde{b}}}
    \right]\,.
     \label{eq:Thermalenergydensity2}
\end{aligned}
\end{equation}
which precisely matches the vacuum sector result \eqref{eq:manifest}.
This simplification was expected, since as discussed in section \ref{sec:intro}, the modular flow is a generalization of the density matrix, and thus we expect two states whose modular flows match in an open subregion to share the same local properties, such as the energy density.
Stated differently, eq.~\eqref{eq:Thermalenergydensity2} applies for all states with geometric modular flow in $d=2$. We confirm this in section \ref{sec:exDiamond}, when studying a finite diamond $\mathcal{D}$, instead of $\mathcal{R}$.

As a consequence of eq.~\eqref{eq:Thermalenergydensity}, if  the modular flow vector field vanishes somewhere, then the null energy density is finite if and only if the flow vanishes there in a boost-like manner: $\tilde{b}(\tilde{v}) = \tilde{v} + O(\tilde{v}^2)$.
Similar observations are discussed in section 4.3 of \cite{Jensen:2023yxy}. It would be interesting in future work to seek a general relationship between energy and modular flow in $d>2$ CFTs, using the $d=2$ result \eqref{eq:Thermalenergydensity2} as a starting point.

\subsubsection*{Entropy}

We now apply the entropy formula eq.~\eqref{eq:covariantEntropy} to excited states in the thermal sector.

Let us start with the thermal state $|\beta\rangle$. Using eq.~\eqref{eq:borchers1}, we get
\begin{equation}
    \frac{1}{c(-\tilde{x})} = \frac{1}{b(\tilde{x})}
    =
    \frac{2\pi e^{\frac{2\pi \tilde{x}}{\beta}}}{\beta\big(e^{\frac{2\pi \tilde{x}}{\beta}}-1\big)}
    =
    \left( \log\left(-1+e^{\frac{2\pi \tilde{x}}{\beta}}\right)\right)'\,.
    \label{finishline}
\end{equation}
Therefore, eq.~\eqref{eq:TheEntropy2d} yields
\begin{equation}
    S(\rho_\beta) 
    = \frac{c}{6}\log\frac{-1+e^{\frac{2\pi L}{\beta}}}{-1+e^{\frac{2\pi \epsilon}{\beta}}}
    =\frac{c}{6}\left[ \frac{2\pi L}{\beta} + \log\frac{\beta}{2\pi \epsilon}\right]\,,
    \label{eq:Lepsilon}
\end{equation}
where we only displayed the leading order contributions in the UV and IR cutoffs, $\epsilon$ and $L$.

We can compare eq.~\eqref{eq:Lepsilon} to the standard formula \cite{Calabrese:2004eu,Calabrese:2005zw} for the entanglement entropy of an interval $I$ of width $\ell$ in a thermal bath,
\begin{equation}
    S_I = \frac{c}{3}\log\left(\frac{\beta}{\pi\epsilon}\sinh\frac{\pi\ell}{\beta}\right)\,,
\end{equation}
by taking $\ell = L \to\infty$.
We obtain
\begin{equation}
    S_I  \to \frac{c}{3}\log\left(\frac{\beta}{2\pi\epsilon}\exp\frac{\pi L}{\beta}\right)
    =  \frac{c}{6}\left[\frac{2\pi L}{\beta}+2\log\frac{\beta}{2\pi \epsilon}\right]
    \,.
\end{equation}
While the IR (or ``thermal'') contributions match, the UV contributions differ by a factor of two.
This is because the IR contribution is extensive, while, as discussed below eq.~\eqref{eq:2dentropy}, the UV contribution is tied to the number of points in the entangling surface.

We now compare eq.~\eqref{eq:Lepsilon} with the entropy of the left Rindler wedge. Using eq.~\eqref{eq:borchers1},
\begin{equation}
    S(\rho_\beta^{\mathrm{left}}) = \frac{c}{6}\log \frac{\beta}{2\pi \epsilon}\,,
    \label{eq:pure1}
\end{equation}
which matches the UV divergence in eq.~\eqref{eq:Lepsilon} but is not IR divergent. Indeed the inverse modular temperature $\beta=2\pi\norm{\xi}$ and the corresponding entropy density \eqref{eq:entropydensity} vanish as $\tilde{x}\to -\infty$. 
The entropy contribution from the auxiliary system in the thermofield double state, calculated using eq.~\eqref{eq:copyflow}, reads
\begin{equation}
    S(\rho_\beta^{\mathrm{aux}}) = \frac{c}{6}\left(\frac{2\pi L}{\beta}\right)\,,
    \label{eq:pure2}
\end{equation}
and matches the IR divergence in eq.~\eqref{eq:Lepsilon}  but is not UV divergent.
Note that in the auxiliary theory, the entropy density becomes constant as $x\to-\infty$, leading to the linear divergence in the IR cutoff shown, but dies off rapidly as $x\to \infty$.
Hence we see that $S(\rho_\beta)=S(\rho_\beta^{\mathrm{left}})+S(\rho_\beta^{\mathrm{aux}})$, as expected from the fact that $|\beta\rangle$ is a pure state when considering both copies in the thermofield double system.
By subtracting $S(\rho_\beta^{\mathrm{left}})$ from both sides, one recovers the balance between eqs.~\eqref{eq:notpure} and \eqref{eq:notpure2} observed in section \ref{sec:thermalsector}.

For an excited state $|\Psi\rangle$ in the thermal sector, we obtain from eq.~\eqref{eq:TheEntropy2d}
\begin{equation}
    \Delta S =S(\rho_\Psi)
    -
    S(\rho_\beta)
    =
    \frac{c}{12}\int_{0}^{\infty}\rd\tilde{x}
    \left[-
    \frac{1}{c(-\tilde{x})}+\frac{1}{\tilde{c}(-\tilde{x})} 
    - \frac{1}{b(\tilde{x})}+ 
    \frac{1}{\tilde{b}(\tilde{x})}
    \right]\,,
    \label{eq:thermalEntropyDiff}
\end{equation}
where $c$ and $b$ describe the Rindler modular flow in the thermal state, given by eq.~\eqref{eq:borchers1}.
This expression is both UV and IR finite, with the IR-finiteness crucially relying on $\tilde{c}$ and $\tilde{b}$ obeying the falloffs \eqref{eq:thermalFalloffs}.

As in the vacuum sector, the finiteness of eq.~\eqref{eq:thermalEntropyDiff} comes from the fact that
\begin{equation}
    \frac{1}{\tilde{c}(\tilde{u})} = \frac{u'(\tilde{u})}{c(u(\tilde{u}))} 
    = \frac{2\pi u'(\tilde{u})}{\beta\big(1-e^{\frac{2\pi u(\tilde{u})}{\beta}}\big)}
    =-\left(\log\left(-1+e^{-\frac{2\pi u(\tilde{u})}{\beta}}
    \right)\right)'
\end{equation}
is a total derivative, and similarly for $\tilde{b}(\tilde{v})$. Hence, for a general $|\Psi\rangle = U_{\varphi}|\beta\rangle$,
\begin{equation}
\begin{aligned}
    S(\rho_\Psi)
    &= \frac{c}{12}\log\left( \frac{\big(-1+e^{-\frac{2\pi u(-L)}{\beta}}\big)\big(-1+e^{\frac{2\pi v(L)}{\beta}}\big)}{\big(-1+e^{-\frac{2\pi u(-\epsilon)}{\beta}}\big)\big(-1+e^{\frac{2\pi v(\epsilon)}{\beta}}\big)}\right).
\end{aligned}
\end{equation}
From appendix \ref{app:thermalApp}, at leading orders in $\frac{\epsilon}{\beta}\to 0^+$ and $\frac{L}{\beta}\to \infty$, we have
\begin{equation}
\begin{aligned}
u(-\epsilon) &= -u'(0)\epsilon = -e^{\alpha' + A' + f(\beta)}\epsilon
\qquad
u(-L) = -L- \frac{\beta}{2\pi}\left(\alpha'+A_{-\infty}+f(\beta)\right)
\end{aligned}
\label{eq:linearize}
\end{equation}
and similarly for $v$, where all these constants $A'$, $A_{-\infty}$, $f(\beta)$, \dots, are finite. This leads to
\begin{equation}
    \Delta S=S(\rho_\Psi)-S(\rho_{\beta})=
    \frac{c}{12}\left[A_{-\infty}-A'+(v\textrm{-sector contribution})\right]\,,
    \label{eq:thermalEEAB}
\end{equation}
which is then manifestly finite.

We therefore obtained that the entanglement entropy difference $\Delta S$ is a boundary term depending on the conformal transformation $(u(\tilde{u}),v(\tilde{v}))$.
While in the vacuum sector both UV and IR cutoffs transformed with the first derivatives of $u(\tilde{u})$ and $v(\tilde{v})$ (see eq.~\eqref{eq:cutoffTransformation}),
in the thermal sector the UV cutoff  still transforms as $\epsilon \to \sqrt{u'(0)v'(0)}\epsilon$, but the IR cutoff transforms differently as
$L\to L+\cO(1)$.

We conclude this subsection by noting that the relative entropy in the thermal sector has a similar form to that of the vacuum sector.
Indeed, the stress tensor one-point functions \eqref{eq:manifest} and \eqref{eq:Thermalenergydensity2} have the same form, and the entropy formulas \eqref{eq:FINITE_ENTROPY} and  \eqref{eq:thermalEntropyDiff} differ only by the choice of reference state ($|\beta\rangle$ or $|\Omega\rangle$) which determine the functions $b(\tilde{v})$ and $c(\tilde{u})$.
This fact will be used hereafter to discuss thermal sector falloffs.

\subsubsection*{Generalizing the thermal sector falloffs}

We here explore the consequences of relaxing the falloffs discussed in section \ref{sec:thermalsector}.
That is, we weaken the thermal sector falloffs required in eq.~\eqref{eq:thermalFalloffs}, 
\begin{equation}
    \tilde{c}(\tilde{u})\to  \frac{\beta}{2\pi} + \cO\left(\frac{1}{\tilde{u}^2}\right)\qquad
    \tilde{b}(\tilde{v})\to  \frac{\beta}{2\pi} + \cO\left(\frac{1}{\tilde{v}^2}\right)
    \label{eq:thermalFalloffsWeak}
\end{equation}
at the right asymptotic boundary, $\tilde{u}\to -\infty$ and $\tilde{v}\to \infty$.
We first do so by including $\cO\left(\frac{1}{\tilde{u}}\right)$ and $\cO\left(\frac{1}{\tilde{v}}\right)$ terms. We show this leads to the energy and entropy with respect to the original thermal state being infinite, while the relative entropy remains finite. 
We then explore the possibility of relating different thermal sectors, by  changing the leading inverse temperature as $\tilde{x}\to\infty$.\footnote{This does not change the temperature near the left asymptotic boundary.}
In this case, relative to the original thermal state, the energy, entropy, and relative entropy all diverge, consistent with the fact that these states are in different GNS sectors, and different sectors are not unitarily equivalent.\footnote{For the unitary inequivalence of the vacuum and finite-temperature sector, see \cite{Brunetti:2015vmh}, p.~231 or \cite{Fewster:2019ixc}, p.~23. 
For the unitary inequivalence of two sectors at finite, nonzero temperature, see theorem 5.3.35 in \cite{Bratteli:1996xq}.\label{foot:inequiv}} 

We start with two states which appear to belong to the same thermal sector, but have different ${\cal O}\left(\frac{1}{\tilde{u}}\right)$ and ${\cal O}\left(\frac{1}{\tilde{v}}\right)$ terms in the expansion \eqref{eq:thermalFalloffsWeak}.
Let $|\Psi\rangle$ be a state with an ${\cal O}\left(\frac{1}{\tilde{v}}\right)$ term in its $\tilde{b}(\tilde{v})$ falloff, \ie $\delta\tilde{b}\sim\frac{\eta_1}{\tilde{v}}$, and similarly let $\delta\tilde{c}\sim\frac{\eta_2}{\tilde{u}}$.
Then, eq.~\eqref{eq:thermalEntropyDiff} gives
\begin{eqnarray}
    \Delta S
    \simeq
    -\frac{c \pi^2 (\eta_1-\eta_2)}{3\beta^2}\int^{L}\rd\tilde{x}
    \left(\frac{1}{\tilde{x}} + \cO\left(\frac{1}{\tilde{x}^2}\right) \right)
    = -\frac{c \pi^2(\eta_1-\eta_2)}{3\beta^2}\log \frac{L}{\beta}\,,
\end{eqnarray}
where again $L$ is the IR cutoff. Avoiding the IR divergence in the case $\eta_2 \ne \eta_1$\footnote{The case $\eta_2 = \eta_1$ is an asymptotic perturbation of only the spatial component of the vector field.} provided the rationale for our restrictive choice of falloffs in eq.~\eqref{eq:thermalFalloffs}. 
Similarly, the energy difference between $|\Psi\rangle$ and $|\beta\rangle$ generically diverges:
\beqn
    \langle E_{\beta}\rangle_{\Psi}- \langle E_{\beta}\rangle_{\beta}
    &=& \int_0^{\infty} \rd \tilde{x}\,  \left(\langle T_{tt}(\tilde{x})\rangle_{\Psi}- \langle T_{tt}(\tilde{x})\rangle_{\beta}\right)\nonumber\\
    &=& \int_0^{\infty} \rd \tilde{x}\, 
    \left(\frac{c}{48\pi}\left[\frac{1}{\tilde{b}^2}
    + 
    \frac{2\tilde{b}\tilde{b}''-\tilde{b}'^2}{\tilde{b}^2}  +( \tilde{b}\leftrightarrow \tilde{c})\right]
     - \frac{\pi c}{6\beta^2}\right)\nonumber\\
    &\simeq&-\frac{c \pi^2(\eta_1-\eta_2)}{3\beta^3} \int^{L} \rd \tilde{x} 
    \left(\frac{1}{\tilde{x}} + \cO\left(\frac{1}{\tilde{x}^2}\right)
    \right)\,.
    \label{eq:energyDifferenceDiverges}
\eeqn

Nevertheless, the difference of modular Hamiltonians
\beqn
    \langle h_{\beta}\rangle_{\Psi}- \langle h_{\beta}\rangle_{\beta}
    &=& 2\pi \int_0^{\infty} \rd \tilde{x}\, \xi^t_{\beta} \left(\langle T_{tt}(\tilde{x})\rangle_{\Psi}- \langle T_{tt}(\tilde{x})\rangle_{\beta}\right)\nonumber\\
    &\simeq&-\frac{c \pi^2(\eta_1-\eta_2)}{3\beta^2} \int^{L} \rd \tilde{x} 
    \left(\frac{1}{\tilde{x}} + \cO\left(\frac{1}{\tilde{x}^2}\right)
    \right)\,,
    \label{eq:modHamDifferenceDiverges}
\eeqn
precisely cancels the $-\frac{c  \pi^2(\eta_1-\eta_2)}{3\beta^2}\log \frac{L}{\beta}$ divergence in the entanglement entropy difference.
Therefore, the relative entropy \eqref{eq:relEntropyDef} remains finite.
We discuss this result in section \ref{sec:discuss}. 

On the other hand, two states with $\cO(1)$ contributions in eq.~\eqref{eq:thermalFalloffsWeak}, and thus in manifestly different thermal sectors, have diverging relative entropy.
Choosing for simplicity ordinary thermal states $|\beta_1\rangle$ and $|\beta_2\rangle$, and keeping only leading order divergences, we have 
\begin{equation}
    \langle h_{\beta_1}\rangle_{\beta_2}-\langle h_{\beta_1}\rangle_{\beta_1}
    \simeq  \beta_1  \int^{L}\rd \tilde{x} \left(\langle T_{tt}(\tilde{x})\rangle_{\beta_2}-\langle T_{tt}(\tilde{x})\rangle_{\beta_1}\right)\simeq \frac{\pi c}{6}L \beta_1 \left(\frac{1}{\beta_2^2} - \frac{1}{\beta_1^2}\right)
\end{equation}
for the difference in modular energy, via \eqref{eq:Thermalenergydensity}, and
\begin{equation}
    S(\rho_{\beta_2})-S(\rho_{\beta_1})
    \simeq 
    \frac{c \pi}{3}\int^{L}\rd \tilde{x} \left(\frac{1}{\beta_2}-\frac{1}{\beta_1}\right)
    \simeq \frac{\pi c}{3}L\left(\frac{1}{\beta_2}-\frac{1}{\beta_1}\right)
\end{equation}
for the difference in entanglement entropy, via 
\eqref{eq:thermalEntropyDiff}, and finally
\begin{equation}
    S_{rel}(\rho_{\beta_2}||\rho_{\beta_1}) \simeq  \frac{\pi c}{6} \beta_1 L\left(\frac{1}{\beta_2} + \frac{1}{\beta_1}\right)^2\,\label{eq:divrel}
\end{equation}
for the relative entropy, via \eqref{eq:relEntropyDef}.
From the quantum information point of view, the divergence in the relative entropy \eqref{eq:divrel} indicates that two states at different temperatures are easily distinguished, as expected.
This linear divergence in the IR cutoff also indicates that these two states belong to different GNS sectors.

\subsection{An example}
\label{sec:exRindler}

In this section, we construct an explicit family of vacuum-sector Unruh flows in the Rindler wedge and discuss the energetic and entropic properties of the associated excited states.

From section \ref{sec:vacuumsector}, given any Unruh flow characterized by $\tilde{c}(\tilde{u})$ and $\tilde{b}(\tilde{v})$ in the Rindler wedge, there exists a conformal mapping $(u(\tilde{u}),v(\tilde{v}))$ such that
\begin{equation}
    \tilde{c}(\tilde{u})=-\frac{u(\tilde{u})}{u'(\tilde{u})} \qquad
    \tilde{b}(\tilde{v})=\frac{v(\tilde{v})}{v'(\tilde{v})} \,.
    \label{eq:pushforwardAgain}
\end{equation}
Consider the following ansatz for the conformal map:\footnote{Our discussion easily extends to a more general ansatz: $u(\tilde{u})=\tilde u + \delta u(\tilde{u}),$ $v(\tilde{v})=\tilde v + \delta v(\tilde{v})$ for a wide class of functions $\delta u(\tilde{u})$ and $\delta v(\tilde{v})$ satisfying $\delta u(\tilde{u}=0)=0=\delta v(\tilde{v}=0)$
and $\delta u(\tilde{u}\to\pm\infty)=0=\delta v(\tilde{v}\to\pm\infty)$. Of course, this generalized ansatz introduces right- and left-moving disturbances determined with $\delta u(\tilde{u})$ and $\delta v(\tilde{v})$, respectively.}
\begin{equation}
u(\tilde{u})=\tilde u + \delta u(\tilde{u}) = \tilde{u}+ z\,\tilde{u}\, e^{-\tilde{u}^2}\,,
\qquad
v(\tilde{v}) = \tilde{v}\,,
\label{eq:exAnsatz}
\end{equation}
where for simplicity we are only perturbing the coordinate $u\ (=t-x)$. As we will see below this introduces a right-moving excitation over the original vacuum state. The exponential factor in $\delta u$ is designed to suppress the perturbation at large distances, and thus preserve the boost-like property at infinity. Similarly, the factor of $\tilde{u}$
produces boost-like behaviour at the origin and ensures that the Rindler wedges are preserved. 

\begin{figure}
    \centering
    \includegraphics[width=1\linewidth]{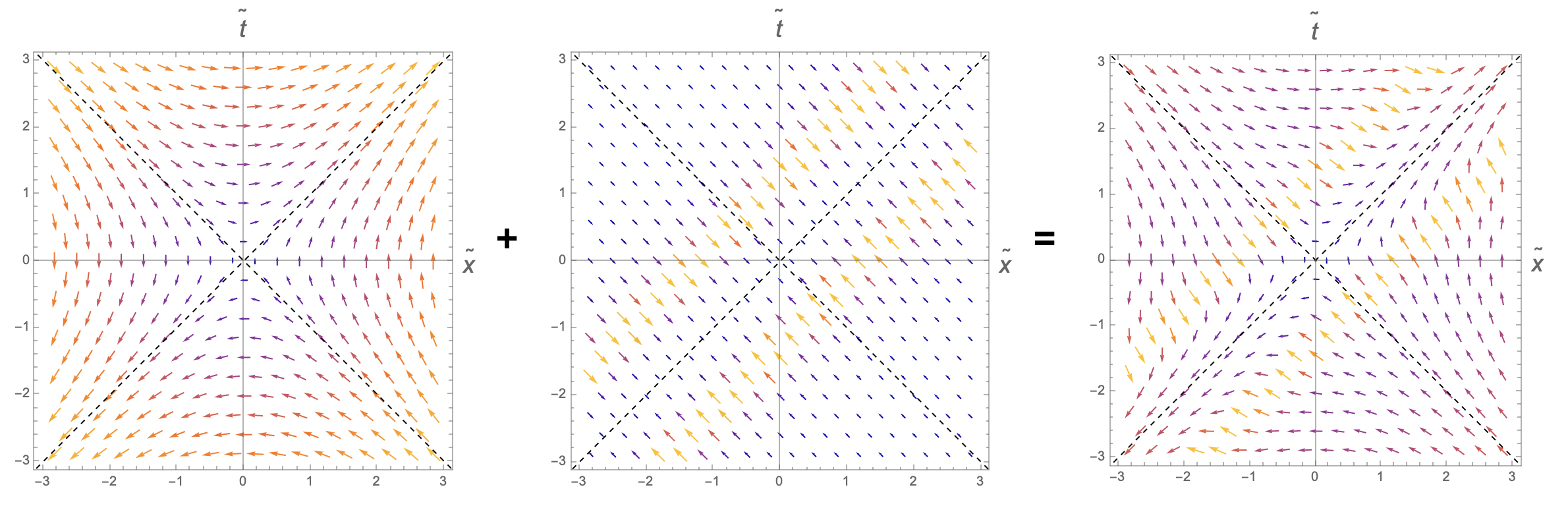}
    \caption{Left: vacuum modular flow for $\mathcal{R}$.
    Right: an Unruh flow described by eqs.~\eqref{eq:pushforwardAgain}-\eqref{eq:exAnsatz} with $z=2$. Middle: the difference between the two flows, which is future-directed in $\mathcal{R}$.
    By contrast, when $z < 0$, the perturbation is past-directed. Future/past directedness of the perturbation is related to the sign of the entanglement entropy difference as seen from eq.~\eqref{eq:covariantEntropy} -- see also figure \ref{fig:entropy}. 
    The color of the arrows indicates the intensity of the vector field. 
    Note that along the $\tilde{x}$-axis, the new modular flow becomes nearly null near $\tilde x\simeq \pm 1.22$. 
    This behavior arises due to the critical point which emerges at $z=z_{\max}$; we are close to the maximum at $z_{\max}$ with $z=2$ in this plot.
    Compare figure \ref{fig:energies} and caption.}
    \label{fig:unruhflow}
\end{figure}

\begin{figure}
    \centering
    \includegraphics[width=.5\linewidth]{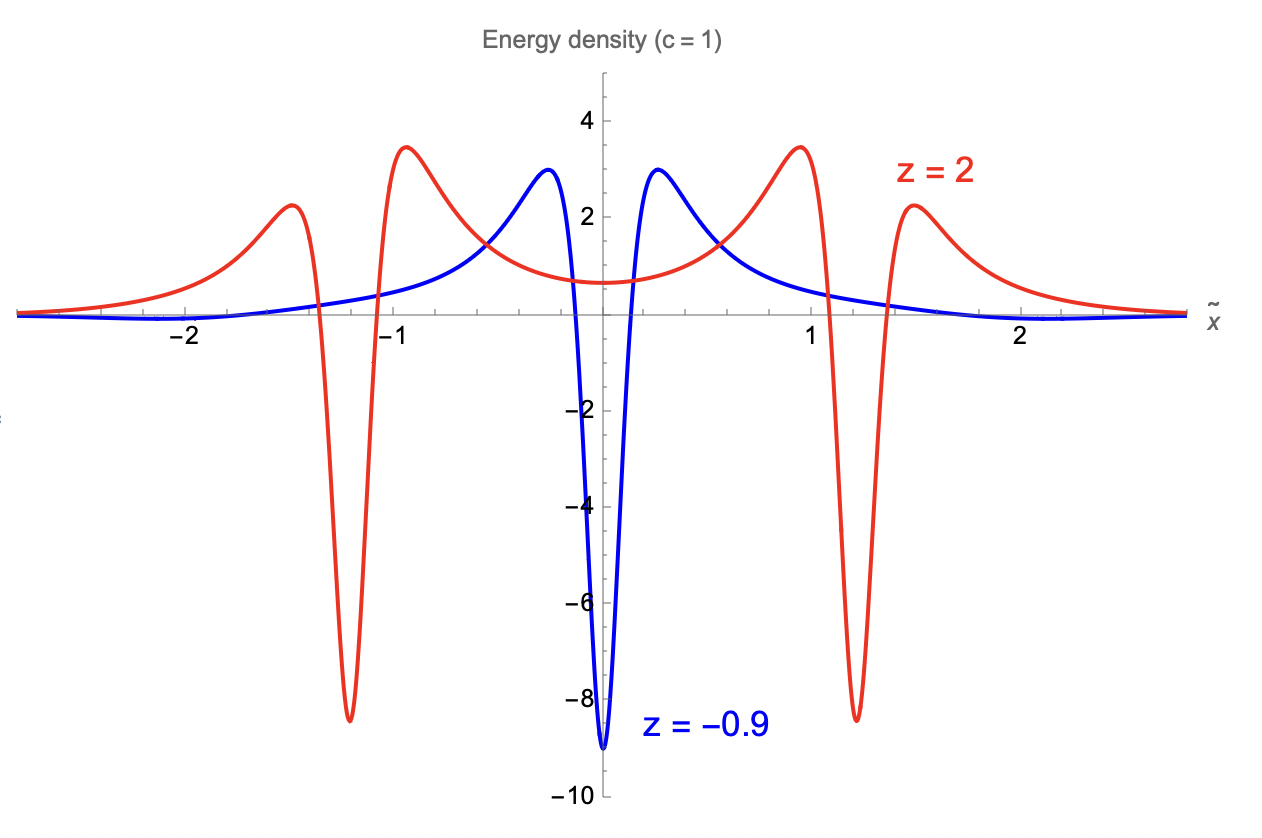}
    \hfill
    \includegraphics[width=0.48\linewidth]{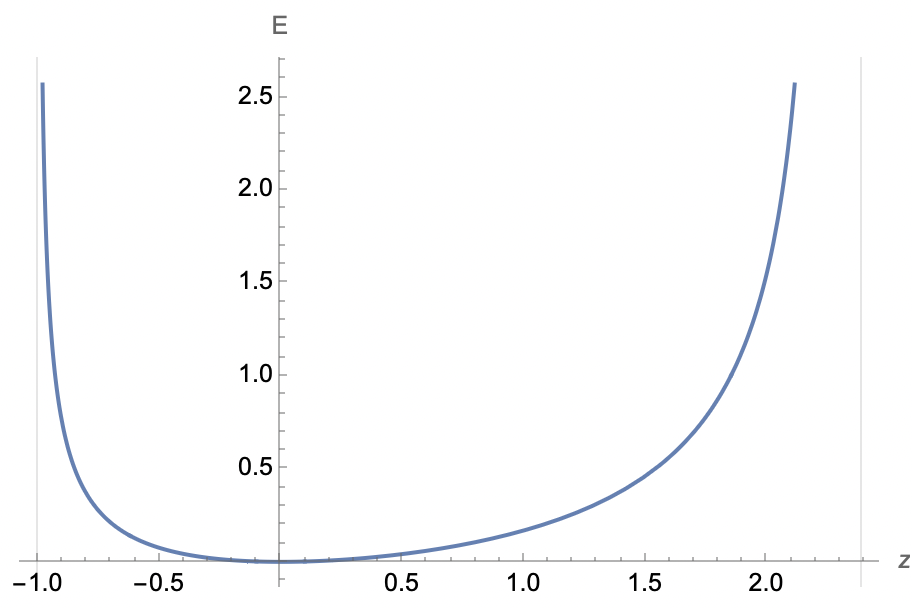}
    \caption{Energy density for $z=-0.9$ and $z=2$ along the $\tilde{t}=0$ slice (left), and total energy as a function of $z$ (right) for states associated with eqs.~\eqref{eq:pushforwardAgain}--\eqref{eq:exAnsatz}. Note that at the critical value $z=z_{\max}$ the energy density is diverging at $\tilde x\simeq \pm 1.22$ .  
    }
    \label{fig:energies}
\end{figure}

Further, $z$ is a constant controlling the amplitude of the perturbation. We must restrict this amplitude to realize an invertible conformal map with the Unruh flow described by eq.~\eqref{eq:pushforwardAgain}. As $z$ approaches $-1$, $u'(\tilde{u}=0)$ approaches zero. In fact, with $z=-1$ , the map $u(\tilde{u})$ has vanishing first and second derivatives at  $\tilde{u}=0$ and so we must require $z>-1$. Increasing $z$, we find  also that $u(\tilde{u})$ has vanishing first derivative at $\tilde{u}\simeq\pm 1.2248$ when $z \simeq2.2408$. Hence by restricting our attention to $z\in (z_{\min}=-1,z_{\max}\simeq2.2408)$, we ensure that $u(\tilde{u})$ is an everywhere monotonically increasing function, and thus invertible. The corresponding Unruh flow \eqref{eq:pushforwardAgain}, with $z=2$, is compared to the vacuum modular flow in figure \ref{fig:unruhflow}. The difference between the two flows shown in the middle panel reveals that our perturbation $\delta u$ creates a right-moving excitation symmetrically arranged about the Rindler horizon at $\tilde{u}=0$. 
The latter can also be seen by examining the energy density \eqref{eq:energydensity} along the $\tilde{x}$-axis, which is plotted in the left panel of figure \ref{fig:energies} for $z=-0.9$ and $z=2$.

By design, for $z=0$ we recover the vacuum modular flow which yields vanishing  energy density everywhere. 
The right panel shows the total energy \eqref{eq:ENERGY} of the corresponding states across the full allowed range of $z$.
As anticipated in section \ref{sec:stateProperties}, 
the energy density can be negative in some places, but the total energy is always positive.
For $z=-0.9\approx z_{\min}$ (blue curve), the total energy becomes large due to the large ridges of positive energy hugging $\tilde{x}=0$, where the conformal map \eqref{eq:exAnsatz} is nearly degenerate.
Similarly for $z=2\approx z_{\max}$ (red curve), the total energy is large because there are pairs of large positive-energy ridges now centered about $\tilde{x}\approx \pm 1.22$, where the conformal map is nearly degenerate. 
More generally, the sign of $z$ controls whether the energy density has a single set of ridges near $\tilde{x}=0$ ($z<0$ case) or two pairs near $\tilde{x}\approx \pm 1.22$ ($z>0$ case).  In principle, these profiles for the energy density have contributions coming from both the left-movers and from the right-movers, \ie $T_{\tilde{t}\tilde{t}}=T_{\tilde{v}\tilde{v}} +T_{\tilde{u}\tilde{u}}$. However, in the present case with  eq.~\eqref{eq:exAnsatz}, we have $\langle T_{\tilde{v}\tilde{v}}(\tilde{v})\rangle_{\Psi}=0$
and $\langle T_{\tilde{u}\tilde{u}}(\tilde{u})\rangle_{\Psi}\ne 0$.
Hence the profiles of the energy density shown in figure \ref{fig:energies} correspond to pulses traveling to the right, as one can also see directly from the Unruh flow in figure \ref{fig:unruhflow}. 

\begin{figure}
    \centering
    \includegraphics[width=1\linewidth]{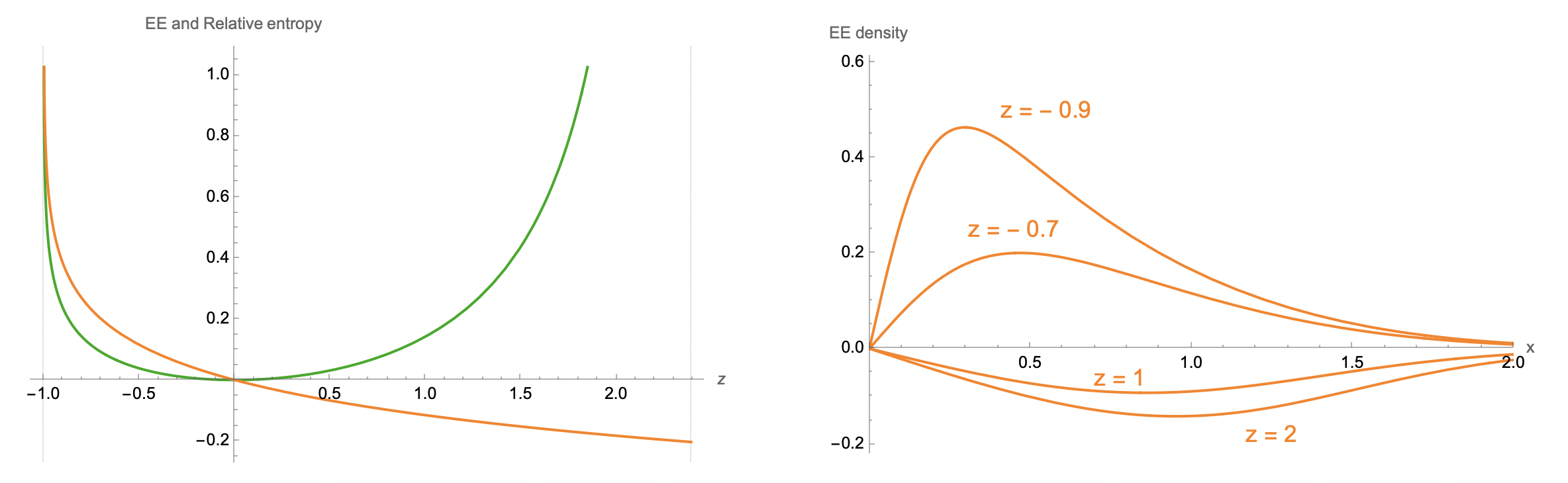}
    \caption{Left: The orange curve indicates the relative-to-vacuum change in the total entanglement entropy  of excited states associated with eqs.~\eqref{eq:pushforwardAgain}-\eqref{eq:exAnsatz}, while the green curve indicates their relative entropy. Right: Change in the entropy density, \ie the integrand in eq.~\eqref{eq:FINITE_ENTROPY}, for various choices of $z$, including the values $z=-0.9$ and $z=2$ considered also in figure \ref{fig:energies}.}
    \label{fig:entropy}
\end{figure}

Let us now consider the difference in the entanglement entropy \eqref{eq:FINITE_ENTROPY} between the excited states and the vacuum state,
plotted in orange in the left panel of figure \ref{fig:entropy} as a function of $z$.
According to eq.~\eqref{eq: EE Wong}, a future-directed perturbation ($z>0$) corresponds to $S_{\Psi}<S_{\Omega}$, and a past-directed perturbation ($z<0$) corresponds to $S_{\Psi}>S_{\Omega}$. 
Hence, the latter corresponds to a family of states more entangled than the vacuum.
Further, we find a divergence in $\Delta S$ as $z\to z_{\min}$, corresponding to a perturbation of the flow in the proximity of the entangling surface and indicating an enhancement of the UV entanglement between the two Rindler wedges.
That is, at $z= z_{\min}$, the boost-like property at the origin is violated, with $\tilde{c}(\tilde{u})$ scaling as $-\frac{\tilde{u}}{3}$ instead of $-\tilde{u}$ near $\tilde{u}=0$.\footnote{ 
The factor of $3$ comes from the cubic critical point in $u(\tilde{u})$ when $z=1$, see eq.~\eqref{eq:stillBoost} in appendix \ref{app:pushforward}. One can construct examples with critical points of degree $>3$, leading to an even larger entanglement entropy.}
By contrast, near $z\simeq z_{\max}$, the perturbations of the modular flow are far from the entangling surface and the entanglement entropy remains finite.
 
The integrand in eq.~\eqref{eq:FINITE_ENTROPY} gives the vacuum-subtracted entropy density for the excited states. This is plotted in the right panel of figure \ref{fig:entropy} for $z=-0.9$, $-0.7$, $1$, and $2 $. 
For $z=0$, the corresponding state is the vacuum and hence the change in the entropy density vanishes everywhere.
Compared with the energy density, the entropy density has a simpler profile which exhibits a single lump.
For $z<0$, the lump is everywhere positive, yielding the increased entanglement entropy indicated by the orange curve in the left panel. Further, the peak rises and move to the left to become a divergence at $\tilde{x}=0$ as $z\to z_{\min}$. Similarly, $z>0$ yields a negative lump to produce a decrease in the overall entanglement entropy. As shown in the figure, the profile is much broader in this case and the minimum is only loosely in the vicinity of the critical point $\tilde{x}\simeq 1.22$, even as $z \to z_{\max}$.  

The relative entropy as a function of $z$ is shown in green in the left panel of figure \ref{fig:entropy}. 
Since the relative entropy is a measure of distinguishability, it vanishes at $z=0$, where $\ket{\Psi}=\ket{\Omega}$, and grows monotonically as move away from $z=0$, ultimately diverging as $z$ approaches either $z_{\max}$ or $z_{\min}$. This indicates that the states are becoming more and more distinguishable as the absolute value of $z$ increases. 
While the relative entropy is always positive, the relative entropy density, \ie the integrand of eq.~\eqref{eq: Srel}, can be negative.

\subsection{Disentangled states}
\label{sec:disentangling}

In section \ref{sec:exRindler}, 
we found a ``super-entangling'' map which perturbs the modular flow near the entangling surface (from $\tilde{c}\sim -\tilde{u}$ to $-\frac{\tilde{u}}{3}$) such that the entanglement entropy increases by a divergent amount.
Can a conformal mapping be used to do the opposite, i.e.~to disentangle the two Rindler wedges? 

A disentangling conformal map cannot be regular at the entangling surface, since regular conformal maps preserve the boost-like property, and from eq.~\eqref{eq:covariantEntropy}, boost-like flows at the entangling surface contribute a prescribed UV divergence to the entanglement entropy.
Indeed, in the context of free theory, states which are not entangled between the two wedges are not Hadamard states,
leading to shockwaves at the boundaries of both the wedges \cite{Wald:1995yp, Fewster:2013lqa, Unruh:2017uaw}.\footnote{This can also be seen from the CFT$_2$ energy density formula eq.~\eqref{eq:Thermalenergydensity2}, which diverges when the modular flow is not boost-like at the entangling surface.}
Nevertheless, one can still study disentangling maps that can be approximated using a series of regular maps, as discussed in section \ref{sec:uniqueness}. We have encountered a similar scenario for a super-entangling map in the example above, in the $z\to z_{\mathrm{min}}$ limit.

From the discussion around eq.~\eqref{eq:covariantEntropy}, rescaling $\xi^{\mu}$ to $\lambda \xi^{\mu}$ with $\lambda \to \infty$ corresponds to sending the temperature of each individual observer to zero, and also the entanglement entropy current \eqref{eq:news} to zero.
Thus, to disentangle the vacuum state, we can seek a conformal map implementing 
\begin{equation}
    (c(u)=-u\,,\; b(v)=v)
    \quad \to \quad 
    (\tilde{c}(\tilde{u})=-\lambda\tilde{u}\,, \;
    \tilde{b}(\tilde{v})
    =\lambda\tilde{v})\,.
    \label{eq:rescaleMe}
\end{equation}
Solving eq.~\eqref{eq:pushforward} with these inputs yields the conformal map
\begin{equation}
u(\tilde{u})=\tilde{u}^{\frac{1}{\lambda}}\,, \qquad v(\tilde{v})=\tilde{v}^{\frac{1}{\lambda}}\,,
    \label{eq:chart}
\end{equation}
with appropriate sign flips when the arguments are negative.
As anticipated, it is singular at the entangling surface, but it can be obtained by taking a limit of regular maps.\footnote{For example, consider a map similar to eq.~\eqref{eq:exAnsatz}, with $u(\tilde{u}) = \tilde{u}+ e^{-\tilde{u}^2-\frac{z}{\tilde{u}^2}} \tilde{u}^{\frac{1}{\lambda}}$.
The perturbation is highly suppressed near the origin for finite $z$, thus preserving the boost-like property, but for $z=0$ ensures a leading $\tilde{u}^{\frac{1}{\lambda}}$ behavior of $u(\tilde{u})$ at the origin. Furthermore, unlike the map discussed below eq.~\eqref{eq:rescaleMe}, the ansatz \eqref{eq:exAnsatz} ensures the perturbation is suppressed at infinity, thus preserving the vacuum sector property.}

From eqs.~\eqref{eq:manifest} and \eqref{eq:Thermalenergydensity2}, the null energy density of a general state $\ket{\Psi}$ with geometric modular flow $\tilde{c}$, $\tilde{b}$ is
\begin{equation}
    \langle T_{\tilde{u}\tilde{u}} \rangle_\Psi 
    =\frac{c}{48\pi}\left[\frac{1}{\tilde{c}^2}
    +
    4\frac{(\sqrt{\tilde{c}})''}{\sqrt{\tilde{c}}}
    \right]\,.
    \label{eq:stateEnergy1}
\end{equation}
If $\xi$ can be rescaled using a conformal map such as eq.~\eqref{eq:chart}, then after $\lambda\to\infty$, we have 
\begin{equation}
    \langle T_{\tilde{u}\tilde{u}} \rangle_{\Psi_{\mathrm{dis}}}
    = \frac{c}{12\pi}\frac{(\sqrt{\tilde{c}})''}{\sqrt{\tilde{c}}}\,,
    \label{eq:disEnergy1}
\end{equation}
as the null energy density of  $\ket{\Psi_{\mathrm{dis}}}$, the new state built from $\ket{\Psi}$ after disentangling.

Applying this procedure to the vacuum, using the disentangling map \eqref{eq:chart}, we gather
\begin{equation}
    \langle T_{\tilde{u}\tilde{u}} \rangle_{\Omega_{\mathrm{dis}}} 
    = -\frac{c}{48\pi\, \tilde{u}^2}
    \qquad \langle T_{\tilde{v}\tilde{v}} \rangle_{\Omega_{\mathrm{dis}}} = -\frac{c}{48\pi\, \tilde{v}^2}\,.
    \label{eq:2drindlerTuu}
\end{equation}
For a central charge of $c=1$, this matches the expectation value of the stress-energy tensor in the Rindler vacuum state of a massless free scalar field theory \cite{Takagi:1986kn}, as introduced in footnote \ref{foot:rindlervac}. 
This suggests that our disentangling map
\eqref{eq:chart} acting on the vacuum produces the Rindler vacuum state in a general two-dimensional CFT.

Note, eq.~\eqref{eq:2drindlerTuu} implies the energy density $\langle T_{\tilde{t}\tilde{t}} \rangle_{\Omega_{\mathrm{dis}}}$ is negative for all nonzero $\tilde{u}$ and $\tilde{v}$.
On the other hand, we expect the total energy of $\ket{\Omega_{\mathrm{dis}}}$ is positive, because we can approximate $\ket{\Omega_{\mathrm{dis}}}$ using a sequence of smooth conformal maps, each of which yields a positive total energy according to eq.~\eqref{eq:positiveenergy}.
In practice, the positive total energy arises due to a singular, positive lump of energy density localized at the entangling surface, as explained in \cite{Parentani:1993yz} for the Rindler vacuum.

We now repeat the disentangling procedure for the Borchers flow \eqref{eq:borchers1}:
\begin{equation}
    c(\tilde{u}) = 
        \frac{\beta}{2\pi}\left(1-e^{\frac{2\pi \tilde{u}}{\beta}}\right)\,,
        \qquad
    b(\tilde{v}) = \frac{\beta}{2\pi}\left(1-e^{-\frac{2\pi \tilde{v}}{\beta}}\right)\,.
\end{equation}
A conformal map which rescales the corresponding flow by $\lambda$ is given by
\begin{equation}
    u(\tilde{u})
    =
    -\frac{\beta}{2\pi}
    \log\left(1-(e^{-\frac{2\pi \tilde{u}}{\beta}}-1)^{\frac{1}{\lambda}}\right)\,,
\end{equation}
and similarly for $v$.
Applying eq.~\eqref{eq:disEnergy1}, we obtain that the null energy density of the thermal state after the disentangling procedure reads
\begin{equation}
    \langle T_{\tilde{u}\tilde{u}} \rangle_{\beta_{\mathrm{dis}}}
    = \frac{c \pi}{48\beta^2}\left(3-\coth\frac{\pi \tilde{u}}{\beta}\right)
    \left(1+\coth\frac{\pi \tilde{u}}{\beta}\right)
\end{equation}
At small $\tilde{u}$, this agrees with the disentangled vacuum state result \eqref{eq:2drindlerTuu}. At large $\tilde{u}$, we obtain $\frac{c\pi}{16 \beta^2}$, which is $\frac34$ of $\frac{c\pi}{12 \beta^2}$, the constant thermal state null energy density in eq.~\eqref{eq:nullenergyThermal}.
As with the Rindler vacuum, we expect the total energy here is positive.

We have applied our procedure to disentangle the Rindler wedges both in the vacuum and thermal states, such that observers along the respective modular flows experience zero temperature. 
Since, according to eq.~\eqref{eq:disEnergy1}, we obtain different energy densities for every choice of $\tilde{c}$ and $\tilde{b}$, there are many different states in which the Rindler wedges are disentangled.
This is expected: if we act on the Rindler vacuum with local operators restricted to one wedge, the resulting excited state is still completely disentangled. 

We revisit the topic of disentangled states in section \ref{sec:energytomatch}, where we find that the quantity
\begin{equation}
    \langle T_{\tilde{u}\tilde{u}} \rangle_{\Psi} -\langle T_{\tilde{u}\tilde{u}}\rangle_{\Psi_{\mathrm{dis}}} 
    = \frac{c}{48\pi\,\tilde{c}(\tilde{u})^2}\,,
    \label{eq:psiMinusPsidis2d}
\end{equation}
is the energy-momentum Unruh observers attribute to the state based on their experience of a thermal bath. That is, Unruh observers naturally measure excitations relative to the state in which they experience zero temperature, which by definition is $\ket{\Psi_{\mathrm{dis}}}$.

\section{To Infinity \texorpdfstring{$\ldots$}{...} And Beyond!}
\label{sec:beyondRindler}

In the previous section, we constructed various states with geometric modular flows, examined their energetic and entropic properties, and analyzed them for an explicit family of excited states. 
Up to this point, we have focused on the Rindler wedge in two-dimensional Minkowski space, restricting our analysis to conformal maps $\varphi$ that preserve this region.

In this section, we extend our findings to more general backgrounds. 
In section \ref{sec:warmup}, we begin with the simple case of the vacuum state reduced to a finite causal diamond, to explain the general procedure and its associated subtleties. 
Then, we analyze generic Unruh flows in a finite causal diamond on the cylinder in section \ref{sec:exDiamond}. 
This geometry is useful for understanding more general configurations, since various spacetimes, \eg de Sitter, anti-de Sitter, flat space, and FLRW, can be conformally embedded in the cylinder. 
In section \ref{sec:exCylinder}, we discuss how our results on Unruh flows extend to these geometries.

\subsection{Vacuum flows}
\label{sec:warmup}

Following \cite{Casini:2011kv, Hislop:1981uh}, we review how to use a conformal transformation to relate the vacuum modular flow in the Rindler wedge to the vacuum flow in a finite causal diamond.

\begin{figure}
    \centering
    \includegraphics[width=0.9\linewidth]{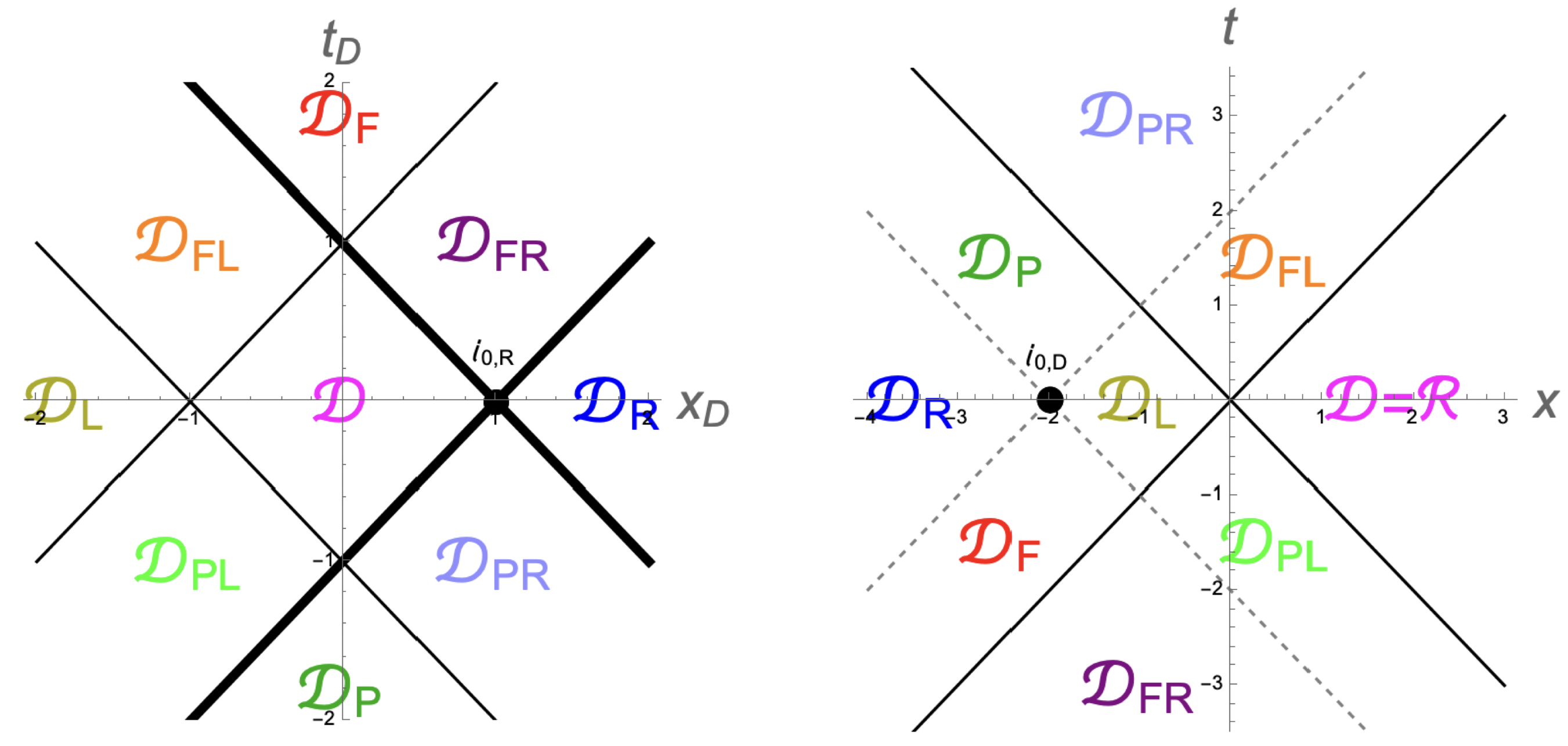}
    \caption{A visualization of the conformal map \eqref{eq:DtoRmap}, with $R=1$. Here, $P$ and $F$ stand for ``past'' and ``future'', while $R$ and $L$ stand for ``right'' and ``left''.
   }
    \label{fig:DtoRmap}
\end{figure}

The key tool is the conformal transformation
\begin{equation}
\begin{aligned}
    u_{\mathcal{R}}(u_\cD) = - 2 R\,\frac{R-u_\cD}{R+u_\cD}\qquad
    v_{\mathcal{R}}(v_\cD) = 2 R\,\frac{R+v_\cD}{R-v_\cD}
    \,,
\end{aligned}
\label{eq:DtoRmap}
\end{equation}
which maps spatial infinity in the $(u_\cD,v_\cD)$ frame to finite distance in the $(u_{\mathcal{R}},v_{\mathcal{R}})$ frame, and the diamond, \ie double-cone, region $\mathcal{D}$ to the Rindler wedge $\mathcal{R}$ in figure \ref{fig:DtoRmap}.

Compare eq.~\eqref{eq:DtoRmap} to the general form of a special conformal transformation (SCT)
\begin{equation}
    x_T^{\mu}(x)=\frac{x^{\mu}-b^{\mu}x^2}{1-2 b\cdot x+b^2 x^2}\,,
\end{equation}
which in chiral coordinates reads
\begin{equation}
    u_T(u)=\frac{u}{1+u b^v}
    \qquad
    v_T(v) = \frac{v}{1+v b^u}\,.
\end{equation}
Setting $b^v = \frac{1}{2R}$ and $b^u = -\frac{1}{2R}$, i.e.~$(b^t, b^x)=(0,\frac{1}{2R})$, we obtain
\begin{equation}
    u_T(u)=2R\frac{u}{2R+u}
    \qquad
    v_T(u) = 2R\frac{v}{2R-v}\,,
    \label{eq:baseSCT}
\end{equation}
so we conclude that eq.~\eqref{eq:DtoRmap} is nothing but a special conformal transformation $u_T = u_{\mathcal{R}}$, $v_T = v_{\mathcal{R}}$, combined with a shift $u= u_\cD-R$, $v= v_\cD+R$, i.e.~$x=x_\cD+R$. Since both special conformal transformations and translations preserve the vacuum state \cite{Haag:1996hvx}, we can use eq.~\eqref{eq:DtoRmap} to relate the vacuum modular flows with respect to $\mathcal{R}$ and $\mathcal{D}$.

As shown in appendix \ref{app:pushforward}, the vacuum modular flow with respect to $\mathcal{D}$ is simply the push-forward of the boost flow \eqref{eq:vacflow} under the map \eqref{eq:DtoRmap}, yielding
\begin{equation}
    c_\cD(u_\cD):= \frac{1}{2R}(R^2-u_\cD^2)
    \qquad
    b_\cD(v_\cD):= \frac{1}{2R}(R^2-v_\cD^2)\,,
    \label{eq:diamondflowVacuum}
\end{equation}
which recovers the well-known results of \cite{Hislop:1981uh,Casini:2011kv} for the infinitesimal modular flow.

An interesting subtlety arises when studying the finite modular flow: while $\mathcal{D}$ is preserved, points at finite distance outside $\mathcal{D}$ can reach infinity in finite modular time!
Explicitly, the exponentiation of the vacuum flow \eqref{eq:diamondflowVacuum} for $\mathcal{D}$ is
\begin{equation}
u_s(u_0)=R\,\frac{u_0+R\,\tanh\frac{s}{2}}{R+ u_0\,\tanh\frac{s}{2}}\,,
\label{eq:exponentiated}
\end{equation}
and identically for $v_s(v_0)$. 
For $(u_0,v_0)$ placed near right spatial infinity, eq.~\eqref{eq:exponentiated} gives that as $s$ increases towards a critical value $s_0$ where the denominator of eq.~\eqref{eq:exponentiated} vanishes, the point $(u_s,v_s)$ reaches $u_s=-\infty$ at finite $v_s$, crossing $\cal{I}^-$,\footnote{We use standard notation for the asymptotic conformal structure of Minkowski spacetime where $i^\pm$ denotes future and past timelike infinity, $\mathcal{I}^{\pm}$ denote future and past null infinity, and $i^0$ is spacelike infinity.}  and then suddenly reappears at $u_s=\infty$, as if $\cal{I}^-$ was antipodally identified with $\cal{I}^+$.
This effect arises because in the Rindler frame, the boost flow can evolve a point from $\mathcal{D}_R$ to $\mathcal{D}_F$ in finite modular time $s$.
In the diamond frame, the corresponding point evolves to infinity \dots and beyond!

Moreover, since $(u_0,v_0)$ is spacelike separated from $\mathcal{D}$, while  $(u_s, v_s)$ is timelike separated from $\mathcal{D}$ for $s>s_0$, we conclude eq.~\eqref{eq:exponentiated} can send spacelike separated points to timelike separated points.
This is a special feature of Minkowski CFTs and does not arise when considering CFTs on the cylinder.
As discussed in appendix \ref{app:nonlocal}, making this effect consistent with microcausality requires a modification of the usual, local transformation rule eq.~\eqref{eq:primary} for primary operators.\footnote{For further explorations of causality and modular flows in 2$d$ CFT, see \cite{Jovanovic:2025mwe}.}
While operators transform in the usual local way at finite distances in the spacetime, they transform nonlocally upon crossing $\mathcal{I}^{\pm}$.

The nonlocal behavior discussed in appendix \ref{app:nonlocal} plays an important role in any geometric modular flow which sends operators beyond $\mathcal{I}^{\pm}$ in finite modular time. Indeed, this phenomenon also occurs for the thermal modular flow with respect to $\mathcal{R}$, given in eqs.~\eqref{eq:borchers1} and \eqref{eq:copyflow}.
Having pointed out this subtlety, we are now prepared to construct more general modular flows with respect to the diamond.

\subsection{A state for every flow in the diamond and in the cylinder}

In this section, we use conformal mappings such as eq.~\eqref{eq:DtoRmap} above to study Unruh flows in causally complete regions other than the Rindler wedge of Minkowski spacetime.

\subsection*{A state for every flow in the diamond}\label{sec:exDiamond}
\noindent
In this section, we demonstrate that every Unruh flow $\xi_\cD = \tilde{c}_\cD(\tilde{u}_\cD) \partial_{\tilde{u}_\cD} + \tilde{b}_\cD(\tilde{v}_\cD) \partial_{\tilde{v}_\cD}$ with respect to the causal diamond for an interval of length $2R$, namely
\begin{equation}
    \mathcal{D}:=\{(u_\cD,v_\cD): |u_\cD|< R, |v_\cD|<R\}\,,
\end{equation} 
obeying the purity condition 
\begin{equation}
    \lim_{L\rightarrow \infty,\epsilon\to0}  
    \left(
    \int_{-L}^{-R-\epsilon}\frac{\dd{\tilde{u}_{\cD}}}{\tilde{c}_{\cD}(\tilde{u}_{\cD})}
    +
    \int_{-R+\epsilon}^{R-\epsilon}\frac{\dd{\tilde{u}_{\cD}}}{\tilde{c}_{\cD}(\tilde{u}_{\cD})}
    +
    \int_{R+\epsilon}^{L}\frac{\dd{\tilde{u}_{\cD}}}{\tilde{c}_{\cD}(\tilde{u}_{\cD})}
    \right)=0\,,
    \label{eq:purityD}
\end{equation}
coincides with the modular flow for some state in a Minkowski CFT$_2$.
Using boosts, translations, and other conformal mappings, this statement can be generalized to all possible causal diamonds in $\mathbb{R}^{1,1}$. This is achieved applying the map \eqref{eq:DtoRmap} and its inverse \eqref{eq:specialconf}. 

\begin{figure}
    \centering
    \includegraphics[width=0.9\linewidth]{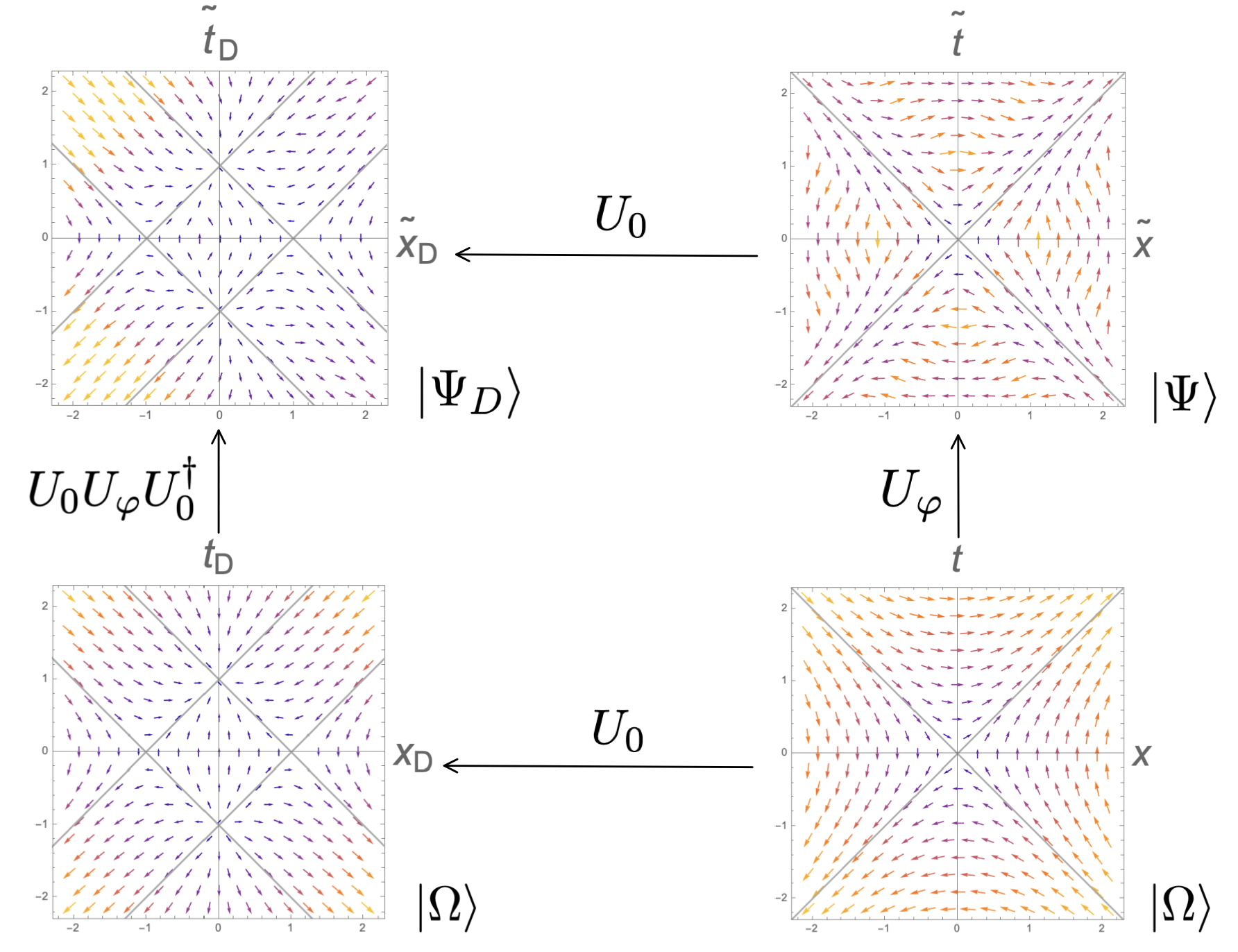}
    \caption{Diagram of the conformal  maps employed to transform the vacuum $|\Omega\rangle$ in ${\cal R}$ to an excited state $|\Psi_\cD\rangle$ with modular flow $\xi_{\cD}$ on the diamond.}
    \label{fig:box}
\end{figure}

Consider any Unruh flow $\xi_\cD$ in the diamond $\mathcal{D}$ and take the pushforward of $\xi_\cD$ under the conformal transformation \eqref{eq:DtoRmap}, which takes $\mathcal{\cD}$ to $\mathcal{R}$,
with associated unitary $U_0^{\dag}$.
Because conformal mappings preserve the Unruh flow properties, we are left with a vacuum-sector Unruh flow $\tilde{c}(\tilde{u}) \partial_{\tilde{u}} + \tilde{b}(\tilde{v}) \partial_{\tilde{v}}$ on $\mathcal{R}$.
As usual, this flow is required to obey the purity property eq.~\eqref{eq:purity}, which we write here as
\begin{equation}
    \lim_{L\rightarrow \infty,\epsilon\to0^+}  
    \left(\int_{-L}^{-\epsilon}\frac{\dd{\tilde{u}}}{\tilde{c}(\tilde{u})}
    +
    \int_{\epsilon}^{2R-\epsilon}\frac{\dd{\tilde{u}}}{\tilde{c}(\tilde{u})}
    +
    \int_{2R+\epsilon}^{L}\frac{\dd{\tilde{u}}}{\tilde{c}(\tilde{u})}\right)=0\,,
    \label{eq:purityR}
\end{equation}
to make explicit that it translates to eq.~\eqref{eq:purityD} using the change of integration variables \eqref{eq:DtoRmap}.
With this constraint, the flow on $\cal R$ is the modular flow of some excited state of the form $|\Psi \rangle = U_\varphi |\Omega \rangle$.
Mapping back to $\mathcal{D}$ using eq.~\eqref{eq:specialconf}, we conclude $|\Psi_\cD\rangle := U_0 |\Psi \rangle =U_0 U_\varphi |\Omega \rangle$ has modular flow $\xi_\cD$ in $\mathcal{D}$, demonstrating the existence of a state for the Unruh flow in $\mathcal{D}$.
See figure \ref{fig:box} for an illustration of this reasoning.

In this argument, the conformal mapping from the vacuum modular flow in $\mathcal{R}$ to the Unruh flow $\xi_\cD$ in $\mathcal{D}$ is a composition of two maps.
The first is $\varphi:(u,v)\to(\tilde{u},\tilde{v})$ as determined by eq.~\eqref{eq:uvsolutionRindler}, with the functions $\tilde{c}$ and $\tilde{b}$ there implicitly determined by $\tilde{c}_\cD$ and $\tilde{b}_\cD$.\footnote{Interestingly, $\varphi$ preserves $\mathcal{R}$ but need not preserve, for example, $\mathcal{D}_L$ and $\mathcal{D}_R$, or the position of $i_{0,\cD}$;
thus, the conformal map $U_0 U_\varphi U^\dag_0$ that constructs a generic state with Unruh flow on the diamond from $|\Omega\rangle$ 
will preserve the diamond itself, but otherwise mix the regions of figure \ref{fig:DtoRmap}.}
This map generally yields a nonvanishing Schwarzian in eq.~\eqref{eq:schwarzian}.
The second is the map $\tilde{u}_\cD(\tilde{u}), \tilde{v}_\cD(\tilde{v})$ of eq.~\eqref{eq:specialconf}, for which the Schwarzian vanishes.
From this, one confirms that the energy density formula \eqref{eq:Thermalenergydensity2} directly applies to Unruh flows in the diamond.

As an application of the procedure described above, we can start from the Minkowski vacuum state and create a local excitation such that the resulting state is thermal in the diamond ${\cal D}$.
Given a global thermal state, its modular flow with respect to $\mathcal{D}$ \cite{Borchers:1998ye,Wong:2013gua} yields
\begin{equation}
\tilde{c}_\cD(\tilde{u}_\cD) = \frac{\beta}{2\pi}\,\frac{\cosh\frac{2\pi R}{\beta}-\cosh\frac{2\pi\tilde{u}_\cD}{\beta}}{\sinh\frac{2\pi R }{\beta}}
\qquad 
\tilde{b}_\cD(\tilde{v}_\cD) = \frac{\beta}{2\pi}\,\frac{\cosh\frac{2\pi R}{\beta}-\cosh\frac{2\pi\tilde{v}_\cD}{\beta}}{\sinh\frac{2\pi R }{\beta}}\,.
\label{eq:diamondflowThermal}
\end{equation} 
Hence, we seek to excite the vacuum such that the resulting state's modular flow $\xi_{\cal D}$ in $\mathcal{D}$ matches eq.~\eqref{eq:diamondflowThermal}.
Thanks to theorem 2.2 in \cite{Summers:2003tf}, the modular flow in a subregion $\mathcal{V}$ uniquely specifies expectation values in $\mathcal{V}$. Therefore, we are demonstrating how an excitation of the Minkowski vacuum can spoof the physics of the thermal state inside $\mathcal{D}$.\footnote{The ability to spoof various states in a compact subregion via local excitations is closely tied to the mathematical notion of quasiequivalence; see e.g.~\cite{Verch:1992eg} and upcoming work revisiting this topic in \cite{Caminiti:JonPaper}.}

First, we map eq.~\eqref{eq:diamondflowThermal} in $\mathcal{D}$ to the Rindler wedge, yielding a vector field $(\tilde{c}, \tilde{b})$.
To ensure that the purity condition \eqref{eq:purityR} is satisfied,  we extend the modular flow to the left Rindler wedge by demanding $\tilde{c}$ and $\tilde{b}$ are odd functions.
We observe that $\tilde{c}$ contains nonzero ${\cal O}(\tilde{u}^2)$ terms near $\tilde u =0$, and similarly for $\tilde{b}$, so the resulting flow is not smooth at the origin.
Mapping everything back to $\cD$, we obtain eq.~\eqref{eq:diamondflowThermal} inside $\mathcal{D}$, and
\begin{equation}
\tilde{c}_\cD(\tilde{u}_\cD) = -\frac{\beta u_{\cD}^2}{2\pi R^2}\,\frac{\cosh\frac{2\pi R}{\beta}-\cosh\frac{2\pi R^2}{\beta \tilde{u}_\cD}}{\sinh\frac{2\pi R }{\beta}}
\qquad 
\tilde{b}_\cD(\tilde{v}_\cD) = -\frac{\beta v_{\cD}^2}{2\pi R^2}\,\frac{\cosh\frac{2\pi R}{\beta}-\cosh\frac{2\pi R^2}{\beta \tilde{v}_\cD}}{\sinh\frac{2\pi R }{\beta}}\,,
\label{eq:diamondflowThermal2}
\end{equation}
outside $\mathcal{D}$.
This flow $\xi_{\cD}$ satisfies, by construction, the purity condition \eqref{eq:purityD}; however, it is only once-differentiable at $\tilde{u}_{\cD}=\pm R$, $\tilde{v}_{\cD}=\pm R$.
Thus, while we can still apply the above procedure to obtain a state with modular flow $\xi_{\cal D}$ from the vacuum state, the map $\varphi$ is not smooth everywhere.
As in section \ref{sec:disentangling}, we expect this is not a problem, as it is possible to find a family of smooth maps which limit to the map of interest.

\begin{figure}
    \centering
    \includegraphics[width=0.9\linewidth]{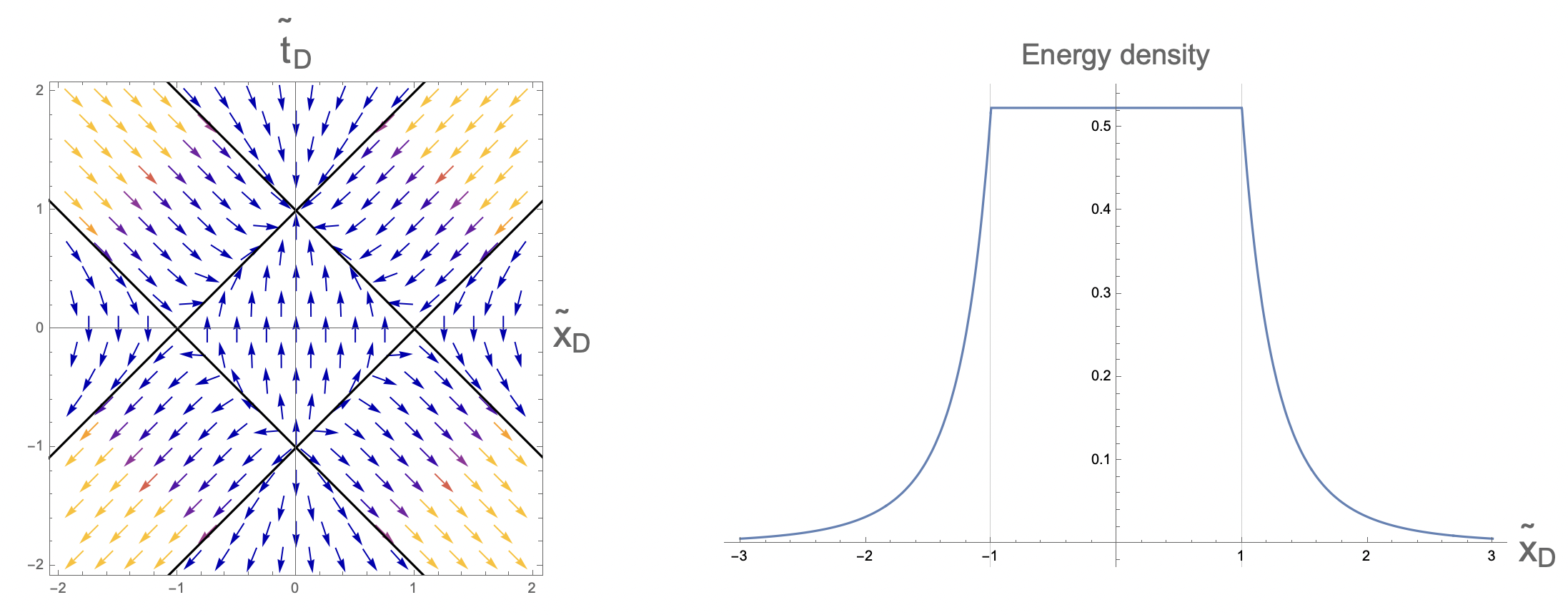}
    \caption{Left: The modular flow in eqs.~\eqref{eq:diamondflowThermal}-\eqref{eq:diamondflowThermal2}. 
    This is the thermal state flow inside $\mathcal{D}$, while in its complement it behaves differently.
    Right: The energy density \eqref{eq:sharpenergy} of the corresponding state on the $\tilde{t}=0$ slice.
    Inside the diamond, the constant energy density is given by eq.~\eqref{eq:cftThermalEnergyDensity}. Outside the diamond, the energy density decays to zero as $\frac{R^4}{\tilde{x}_{\cD}^4}$.}
    \label{fig:heatedup}
\end{figure}

The flow $\xi_{\cD}$ is shown in the left panel of figure \ref{fig:heatedup}.
The energy density, calculated using eq.~\eqref{eq:manifest}, matches the thermal-state energy density \eqref{eq:cftThermalEnergyDensity} in $\mathcal{D}$ but rapidly decays to zero outside $\mathcal{D}$.
That is,
\begin{equation}
    \langle T_{\tilde{t}_{\cD}\tilde{t}_{\cD}}(\tilde{x}_{\cD})\rangle = \frac{\pi}{6\beta^2}
    \begin{cases}
        1, & \tilde{x}_{\cD} \in[-R,R]\\
        \frac{R^4}{\tilde{x}_{\cD}^4}, & |\tilde{x}_{\cD}|>R\,.
    \end{cases}
    \label{eq:sharpenergy}
\end{equation}
This profile is shown on the right-hand side side of figure \ref{fig:heatedup}.

In general, our procedure also reveals subtleties in the asymptotic behavior of modular flows associated with the diamond.
Since the vacuum state has modular flow eq.~\eqref{eq:diamondflowVacuum} with respect to $\mathcal{D}$ \cite{Hislop:1981uh, Casini:2011kv},
which is defined both in $\mathcal{D}$ and in the causal complement of $\mathcal{D}$,
one might expect all of the Unruh flows $\xi_\cD$ constructed above to obey
\begin{equation}
    \tilde{c}_\cD(\tilde{u}_\cD)= -\frac{1}{2R}\tilde{u}_\cD^2 + \cO(\tilde{u}_\cD)
    \qquad
    \tilde{b}_\cD(\tilde{v}_\cD)= -\frac{1}{2R}\tilde{v}_\cD^2 + \cO(\tilde{v}_\cD)\,
    \label{eq:vacAsymptoticsDiamond}
\end{equation}
at large distances $|\tilde{x}_\cD|$.
This is not the case, however. Recall from figure \ref{fig:DtoRmap} that spatial infinity in the diamond frame corresponds to the single contact point of $\mathcal{D}_L$ and $\mathcal{D}_R$ in the left Rindler wedge under the map \eqref{eq:DtoRmap}. Perturbing the modular flow in the left Rindler wedge turns out to modify the flow's asymptotic behavior in the diamond.

Explicitly, consider the point $(\tilde{t},\tilde{x})=(0, -2R)$ in the Rindler frame, which corresponds to spatial infinity in the diamond frame.
Since we assume past-directedness of the modular flow in the left Rindler wedge, the coefficients $\tilde{c}(2R)$ and $\tilde{b}(-2R)$ of the Unruh flow in the Rindler frame at that point are guaranteed to be negative. Using the map \eqref{eq:DtoRmap}, the corresponding Unruh flow with respect to $\mathcal{D}$ is given by
\begin{equation}
\begin{aligned}
    \tilde{c}_\cD(\tilde{u}_\cD) = \frac{\tilde{c}(\tilde{u}(\tilde{u}_\cD))}{\tilde{u}'(\tilde{u}_\cD)}\qquad 
    \tilde{b}_\cD(\tilde{v}_\cD) = \frac{\tilde{b}(\tilde{v}(\tilde{v}_\cD))}{\tilde{v}'(\tilde{v}_\cD)} \,,
\end{aligned}
\end{equation}
which as $\tilde{x}_\cD\to\infty$ for finite $\tilde{t}_\cD$ becomes
\begin{equation}
\begin{aligned}
    \tilde{c}_\cD(\tilde{u}_\cD) = \frac{\tilde{u}_\cD^2}{4 R^2}\, \tilde{c}\left(2R\right) + \cO(\tilde{u}_\cD)\qquad
    \tilde{b}_\cD(\tilde{v}_\cD) = \frac{\tilde{v}_\cD^2}{4 R^2} \, \tilde{b}\left(-2R\right)  + \cO(\tilde{v}_\cD)\,.
    \label{eq:psiDFalloffs}
\end{aligned}
\end{equation}
We thus see that $\xi_\cD$ has the expected asymptotic behavior \eqref{eq:vacAsymptoticsDiamond}, but only up to rescaling the $\tilde{u}$ and $\tilde{v}$ parts of the flow by the positive constants $-\frac{\tilde{c}(2R)}{2R}$ and $-\frac{\tilde{b}(-2R)}{2R}$.

In footnote \ref{foot:sctFootnote}, we remarked that SCTs are symmetries of the vacuum sector. 
Section \ref{sec:AStateForEveryUnruhFlow} further argued that states within the same sector share similar asymptotic properties, as GNS Hilbert spaces are generated by locally smeared quantum fields. This appears to be in tension with the multiplicative ambiguity in the modular flow falloffs \eqref{eq:psiDFalloffs}, arising from the use of SCTs to bring infinity to a finite point. The resolution lies in the energy density formula \eqref{eq:Thermalenergydensity2}. In the Rindler frame, rescaling modular flow falloffs \eqref{eq:vacFalloffsRindler} alters the energy density falloffs from 
$\langle T_{tt}\rangle\sim \frac{1}{x^4}$ to $\langle T_{tt}\rangle\sim \frac{1}{x^2}$. 
Contrastingly, in the diamond frame, the modular flow rescalings in eq.~\eqref{eq:vacAsymptoticsDiamond} maintain the decay 
$\langle T_{tt}\rangle\sim \frac{1}{x^4}$. 
Therefore, if vacuum-sector states are identified by the asymptotic behavior of the energy density, there is no tension; we should allow for the multiplicative ambiguity in the diamond frame because it leaves the energy decay invariant.
We implicitly employed this reasoning in eq.~\eqref{eq:thermalFalloffs2}.

\subsection*{A state for every flow in the cylinder}
\label{sec:exCylinderAside}

In this section, we briefly summarize how the above analysis can be repeated for the case of a diamond on the cylinder.
We will be interested in $\mathcal{D}_c$, the causal development of the interval $t_c=0$, $x_c\in (-\frac{L}{4},\frac{L}{4})$ on the cylinder of circumference $L$.
The case of all other diamonds is obtained straightforwardly by conformal transformations.

Consider an Unruh flow $\xi_c$ with respect to $\mathcal{D}_c$ obeying the purity condition:
\begin{equation}
    \fint_{-\frac{L}{2}}^{\frac{L}{2}} \frac{\rd \tilde{u}_c}{\tilde{c}_c(\tilde{u}_c)} = 0 = \fint_{-\frac{L}{2}}^{\frac{L}{2}} \frac{\rd \tilde{v}_c}{\tilde{b}_c(\tilde{v}_c)}\,,
\end{equation}
where these integrals are over the entire circle.
Take the pushforward of $\xi_c$ under the conformal transformation which takes $\mathcal{D}_c$ and its complement to $\mathcal{R}$ and the left Rindler wedge, respectively:
\begin{equation}
    \tilde{u}(\tilde{u}_{c}) = \tan\frac{2\pi \tilde{u}_{c}}{L}
    \qquad\qquad
    \tilde{v}(\tilde{v}_{c}) = \tan\frac{2\pi \tilde{v}_{c}}{L}\,.
    \label{eq:DctoRmap}
\end{equation}
By the Unruh flow property, $\xi_c$ is boost-like at the entangling points. 
Therefore, the corresponding flow in the Rindler frame is boost-like both at the origin and at spacelike infinity, and so it is a vacuum-sector Unruh flow obeying the purity condition \eqref{eq:purity}. 
The unitary operator that takes us from the $\mathcal{R}$ frame to $\mathcal{D}_c$ frame maps the associated excited state $|\Psi\rangle$ to the desired state with modular flow $\xi_c$ in $\mathcal{D}_c$.

Note, various subtleties which pertain to the Minkowski analysis are not present in the cylinder. 
First, since conformal transformations on the cylinder generally do not send spacelike separated to timelike separated points, there are no subtleties of the type discussed in appendix \ref{app:nonlocal}. Moreover, on the cylinder, the boost-like property at the entangling surface plays the role of the boost-like falloffs at infinity of Minkowski spacetime. The former is a property we required of our flows, while the latter is generically satisfied on the cylinder.

\subsection{A unifying picture}
\label{sec:exCylinder}

The modular flows on the cylinder discussed above provide a natural starting point for generalizing our result, since one can perform conformal mappings from various patches of the cylinder to curved spacetimes such as anti-de Sitter, de Sitter and Minkowski spacetimes, as well as various cosmological universes --- see figure 1 of \cite{Candelas79}.
Moreover, many causally complete regions in a given conformally flat spacetime\footnote{We use the expression ``conformally flat spacetime'' to specifically refer to any Lorentzian spacetime which can be mapped to a connected region of the cylinder by a Weyl transformation.} are related to one other by a change of patch on the aforementioned cylinder. As a consequence, we expect the following:
\begin{quote}
    For every Unruh flow in a connected, causally complete region $\mathcal{V}$ of a conformally flat spacetime, there is a state in the CFT$_2$ with the corresponding modular flow.
\end{quote} 
Here, the Unruh flow criteria is applied to the corresponding flow on the cylinder.

Let us demonstrate the utility of the cylinder picture by discussing modular flows in causally complete subregions of Minkowski spacetime.
The examples that we discuss here can be straightforwardly generalized to higher-dimensional CFTs.
For an analogous analysis in de Sitter spacetime, see \cite{Frob:2023hwf}, although note that our formalism additionally allows one to discuss modular flows which extend beyond time-like infinity in this setting.

\begin{figure}
    \centering
    \includegraphics[width=0.242\linewidth]{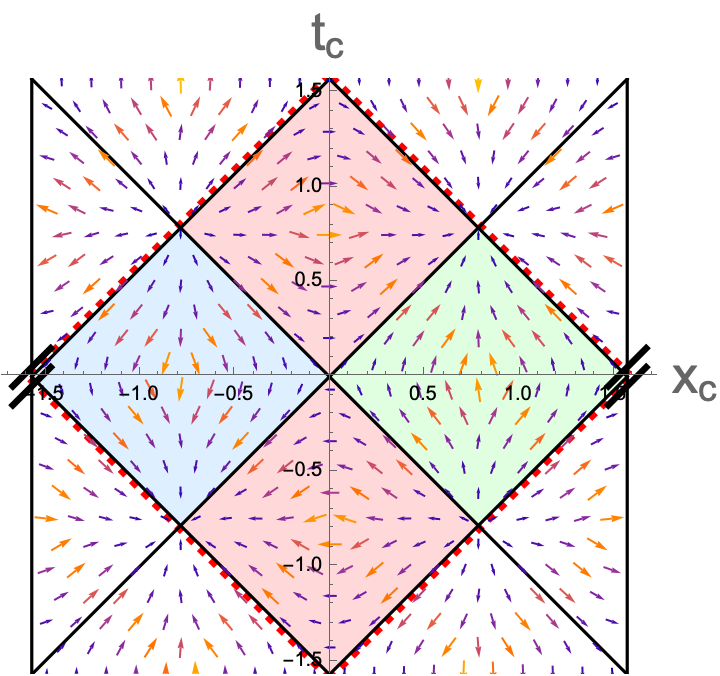}
    \includegraphics[width=0.242\linewidth]{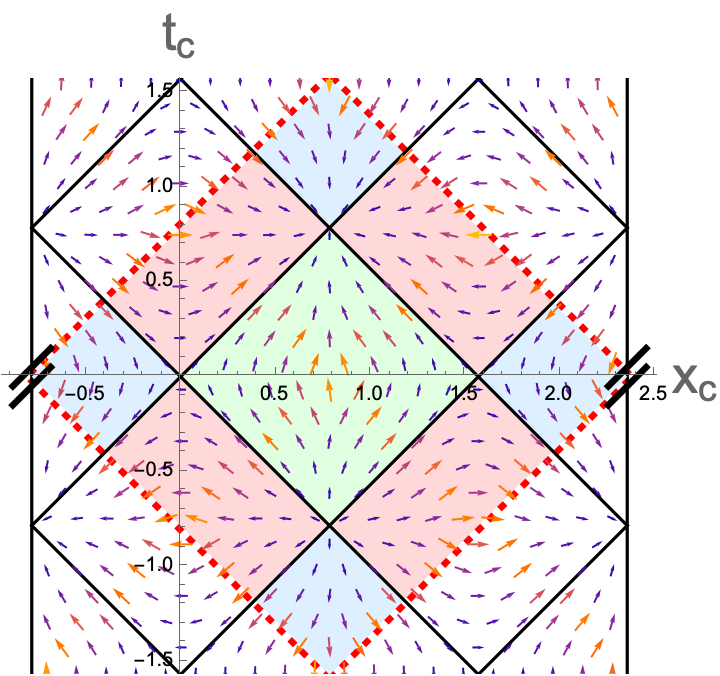}
    \includegraphics[width=0.242\linewidth]{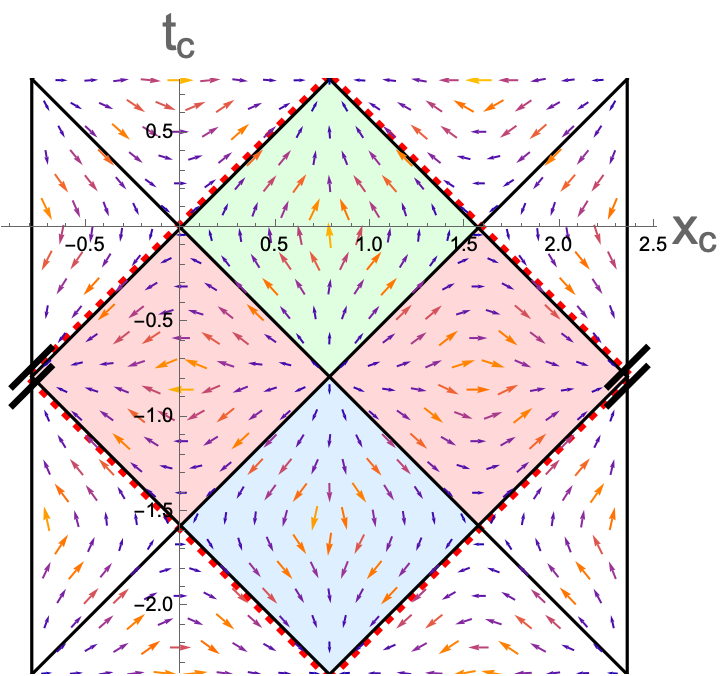}
    
    \includegraphics[width=0.242\linewidth]{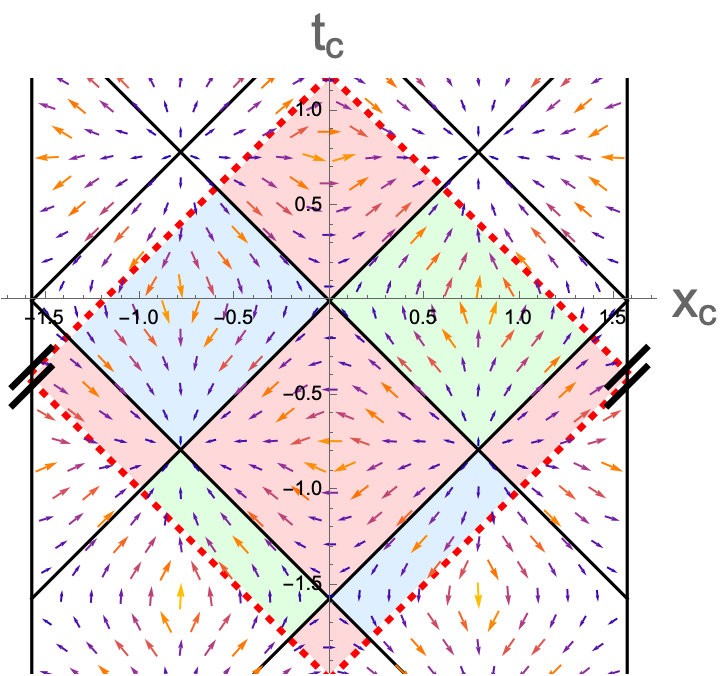}
    \includegraphics[width=0.242\linewidth]{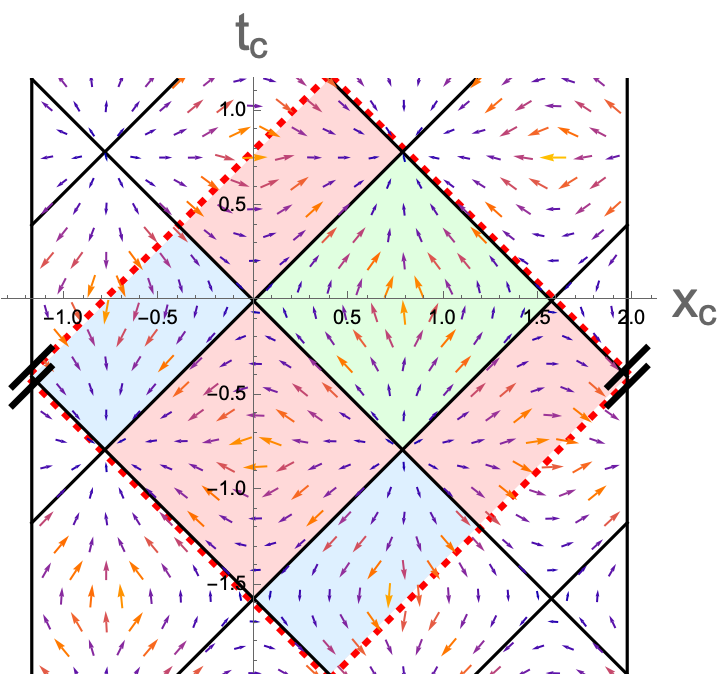}
    \includegraphics[width=0.242\linewidth]{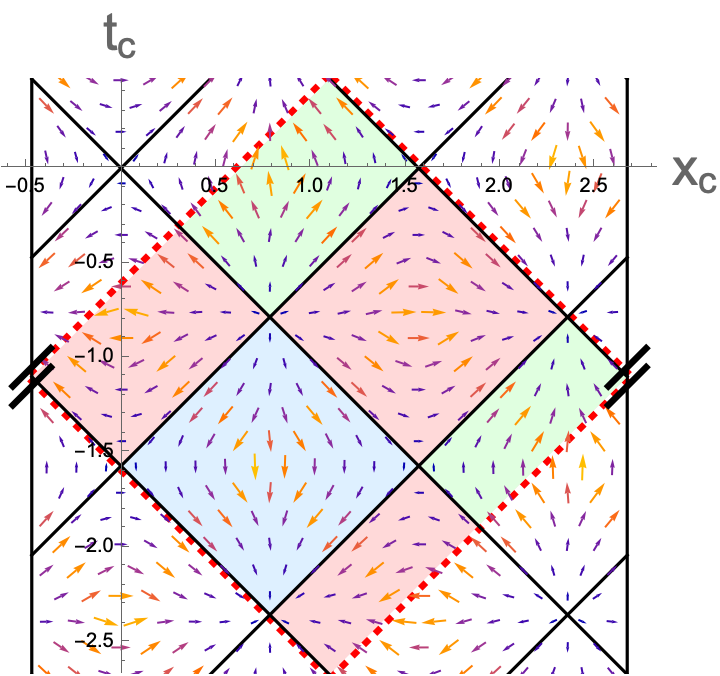}
    \caption{In all six figures, we are plotting the flow depicted in figure \ref{fig:unruhflow}, mapped to the cylinder.
    The figures show six different placements of a dotted red diamond on the cylinder, which correspond to different maps back to Minkowski spacetime using eq.~\eqref{eq:DctoRmap}.
    Each map yields a state on Minkowski with modular flow in the green region $\mathcal{V}$, simply given by pushing forward the vector field shown.
    We expect analogous statements hold in dS$_2$ and AdS$_2$.  
    }
    \label{fig:minkpatches}
\end{figure}

Consider the six choices 
of Minkowksi-patch placement on the cylinder in figure \ref{fig:minkpatches}.
The first choice corresponds to the map \eqref{eq:DctoRmap} discussed in section \ref{sec:exCylinderAside}, reproducing vacuum-sector modular flows associated with Rindler wedge $\mathcal{R}$, such as the example depicted in figure \ref{fig:unruhflow}.
The second choice corresponds to vacuum-sector modular flows associated with the causal diamond $\mathcal{D}$, discussed in section \ref{sec:exDiamond}.

The third case has not yet been discussed.
Here, the region of interest $\mathcal{F}$ is the Milne wedge of Minkowski spacetime: a limit of $\mathcal{D}$ where the future boundaries are pushed to $\cal{I}^{+}$.
It can be related to $\mathcal{R}$ using the following conformal map:
\begin{equation}
    u_{\mathcal{R}} = -\frac{1}{u_{\mathcal{F}}}
    \qquad
    v_{\mathcal{R}} = v_{\mathcal{F}}\,.
    \label{eq:neginv}
\end{equation}
This region is also called the forward lightcone since it is the interior of the future lightcone emanating from a point.
From the perspective of QFT with a UV cutoff, the corresponding density matrix comes from choosing a Cauchy slice which is spacelike in the green region but otherwise hugs $\mathcal{I}^+$, and then tracing out the quantum fields on this asymptotic portion of the Cauchy slice outside the green region.

When the state on the cylinder is the vacuum state of a free, massless scalar field theory, the corresponding vacuum modular flow in the Milne wedge of Minkowski spacetime is a dilatation \cite{Buchholz:1977ze}. 
We expect this statement also generalizes to a general CFT$_d$.
On the other hand, rather than picking the vacuum state, we can consider an arbitrary state with Unruh flow on the cylinder.
This induces a corresponding geometric modular flow on the forward lightcone in Minkowski spacetime, and so our result of finding a state for every vacuum-sector\footnote{Note, not only the vacuum state but also the thermal state $|\beta\rangle$ on Minkowski spacetime has a geometric modular flow in the forward lightcone.
This flow is given in figure 1 of \cite{Borchers:1998ye}.
At future infinity, it approaches time translations, rather than a dilatation, and therefore it cannot be related to the vacuum modular flow by a conformal mapping induced from a regular conformal mapping on the cylinder.} Unruh flow generalizes to the Milne wedge.

We now move to the fourth case, which involves two causally complete but disjoint (green) regions $\mathcal{V}_1\cup \mathcal{V}_2$ in the Minkowski patch.
Let the future-most region be denoted $\mathcal{V}_1$ and the past-most region, $\mathcal{V}_2$. It is puzzling that we obtain a modular flow here, because these regions are timelike separated, and ordinarily reduced density matrices for multi-interval subregions are only considered for spacelike separated intervals.\footnote{Indeed, $\mathcal{V}_1\cup\mathcal{V}_2$ is not a causally complete subregion.
This is because the causal complement of $\mathcal{V}_1\cup\mathcal{V}_2$ is empty, and by definition, a causally complete region is the causal complement of its causal complement.}
To resolve this puzzle, first recall that without a UV cutoff in place, density matrices are ill-defined, and hence we must use the more abstract and general formalism of Tomita-Takesaki theory.
Now, the conformal map relating $\mathcal{V}_1\cup \mathcal{V}_2$ with $\mathcal{R}$,
\begin{equation}
    u_{\mathcal{R}} = 2R \frac{u_{\mathcal{V}}}{2R+u_{\mathcal{V}}}
    \qquad
    v_{\mathcal{R}} =2R \frac{v_{\mathcal{V}}}{2R+v_{\mathcal{V}}}\,,
    \label{eq:toV1V2}
\end{equation}
necessarily sends spacelike separated points to timelike separated points.
As discussed in appendix \ref{app:nonlocal}, this means that operators will transform nonlocally whenever the conformal factor $\Omega(x)$ is negative.
Here, the conformal factor $\Omega(x)$ is negative\footnote{Explicitly, $\Omega(x)$ has the same sign as $(2R-u_{\cal R})(2R-v_{\cal R})$.} in the pre-image of $\mathcal{V}_2$, denoted $\mathcal{R}_2$ and comprised of the points $u_{\cal R}<0$, $v_{\cal R}>2R$.
On the other hand, $\Omega(x)$ is positive in $\mathcal{R}_1$, the pre-image of $\mathcal{V}_1$.
Thus, under the map \eqref{eq:toV1V2}, local operators in $\mathcal{R}_1$ are mapped to local operators in $\mathcal{V}_1$ according to eq.~\eqref{eq:primary}, while local operators in $\mathcal{R}_2$ are mapped to nonlocal operators.
The modular flow obtained using eq.~\eqref{eq:toV1V2} is thus \textit{not} an automorphism of the algebra of local operators in $\mathcal{V}_1\cup \mathcal{V}_2$. It is instead an automorphism of the algebra formed by local operators in $\mathcal{V}_1$, together with certain nonlocal operators.

While $\mathcal{V}_1$ and $\mathcal{V}_2$ seem to play very different roles in the preceding discussion, we can exchange their roles by flipping the sign convention for $\Omega(x)$, as it is defined in appendix \ref{app:nonlocal}. Then, $\Omega(x)$ is positive in the preimage of $\mathcal{V}_2$ and negative in the preimage of $\mathcal{V}_1$.
The difference between these two cases can be understood as the difference between sliding the green Rindler wedge in the first figure of \ref{fig:minkpatches} upwards until it crosses $\mathcal{I}^+$, versus downwards and to the left until it crosses $\mathcal{I}^-$.
The key point is that nonlocal operators appear whenever a conformal map such as eq.~\eqref{eq:toV1V2} is employed to push subregions beyond infinity, and hence the resulting modular flow is associated with an algebra containing nonlocal operators.

The remaining possible configurations include those sketched in cases five and six, 
as well as those obtained by swapping blue and green colorings in each picture.
For an example of the latter, consider swapping the blue and green regions in case 2, so that $\mathcal{V}$ is the union of four causally complete subregions.
For all of these cases, our formalism  allows us to obtain interesting Unruh flows in the chosen subregions.
Again, nonlocal effects occur whenever a conformal map or modular flow pushes an operator past $\mathcal{I}^{\pm}$.

\vspace{.1cm} 

In summary, we have discussed how to construct a state with corresponding geometric modular flow for every Unruh flow in a causally complete subregion of a 2$d$ conformally flat spacetime.
In particular, while there are subtleties involved, we can make sense of regions which reach to infinity and beyond.
An analogous analysis to the one performed above pertains to AdS or dS spacetimes.
This opens the door to the study of modular flows and entanglement entropies of various interesting subregions, such as those anchored at future infinity of dS, generalizing the flat space calculation of \cite{Chen:2023tvj,Chen:2024kuq}.

\section{Lessons for higher \texorpdfstring{$d$}{d}}
\label{sec:entanglementcurrents}

In this section, we generalize two insights of the previous sections to higher-dimensional CFTs. First, we generalize the proposed entanglement entropy formula \eqref{eq:covariantEntropy} to higher dimensions and consider several consistency checks. We then generalize the discussion of the disentangled states and the energy-momentum tensor considered in section \ref{sec:disentangling}.

\subsection{Entanglement currents}
\label{sec:derivation}

In section \ref{sec:stateProperties}, we found that the entanglement entropy of states with geometric modular flows in CFT$_2$ is given by a local integral over the region of interest. We proposed an interpretation of this result in terms of a conserved entropy current associated to Unruh observers following the modular flow. Here we extend this discussion higher dimensions.

In a CFT$_d$, the entropy density of a thermal state at constant inverse temperature $\beta$ is fixed by dimensional analysis to be
\begin{equation}
    s = \alpha \left(\frac{2\pi}{\beta}\right)^{d-1}\,,
    \label{eq:thermalentropydensity}
\end{equation}
where $\alpha$ is a dimensionless constant.
For example, in $d=2$, we have $s=\frac{c\pi}{3\beta}$ and thus $\alpha = \frac{c}{6}$.
In the setting of a geometric modular flow, we use the relation $\norm{\xi}=\frac{\beta}{2\pi}$ from section \ref{sec:vacuumProperties} to determine the local temperature. Then each Unruh observer is assigned a local rest-frame entropy density $s_{obs} = \frac{\alpha}{\norm{\xi}^{d-1}}$,
or, more generally, a covariant entropy current
\begin{equation}
    j_S^{\mu} :=  \alpha \frac{\xi^{\mu}}{\norm{\xi}^{d}}\,.
    \label{eq:current}
\end{equation}
Accordingly, we obtain a generalization of eq.~\eqref{eq:covariantEntropy} for $d>2$,\footnote{This formula is Weyl invariant. 
To see why, write $S = -\alpha \int_A \sqrt{h}\, \rd^{d-1} x \frac{n^{\nu}}{\norm{n}} g_{\mu \nu}  \frac{\xi^{\mu}}{\norm{\xi}^{d}}$ and use that under a Weyl transformation, $g_{\mu \nu}\to \Omega^2 g_{\mu\nu}$, $\sqrt{h}\to \Omega^{d-1}\sqrt{h}$, $\norm{n}\to \Omega \norm{n}$, and $\norm{\xi}\to\Omega \norm{\xi}$.\label{foot:weyl}}
\begin{equation}
    S = -\alpha \int_A \rd \Sigma_{\mu} \frac{\xi^{\mu}}{\norm{\xi}^{d}}\,.
    \label{eq:TheEntropy}
\end{equation}

A consistency check on this result is that the entanglement entropy should not depend on the Cauchy slice $A$ chosen within $\mathcal{V}$.
This is indeed the case, as the current $j_S^{\mu}$ is conserved, i.e.~$\nabla_{\mu}j_S^{\mu}=0$, by the conformal Killing property \eqref{eq:ckv} of $\xi$.
We look for further consistency checks in the two examples that follow, as well as in appendix \ref{App:multiple}, where we consider a subregion comprised of two intervals.

\subsubsection*{Example 1: Rindler wedge}

Consider the vacuum modular flow of the Rindler wedge $\mathcal{R}$ in any QFT \cite{Bisognano:1976za}
\begin{equation}
    \xi\big|_{t=0} = x \partial_t\,.
    \label{eq:lazzy}
\end{equation}
We allow the spacetime to have arbitrary dimension $d$, meaning there are $d-2$ transverse coordinates denoted $y^i$.
The Rindler wedge $\mathcal{R}$ is the region $x>0$, $|t|<x$.

In this case, eq.~\eqref{eq:TheEntropy} yields
\begin{equation}
    S = \alpha \int_{\mathcal{R}} \rd x \, \rd^{d-2}y \frac{\xi^{t}}{\norm{\xi}^{d}}
    = \alpha \int \rd^{d-2}y \int_{0}^{\infty}\frac{\rd x}{x^{d-1}}
    \,.
\end{equation}
Denoting the area of the entangling surface as $R^{d-2} = \int \rd^{d-2}y$, and imposing UV and IR cutoffs $\epsilon$ and $L$, respectively, we recover in $d>2$
\begin{equation}
    S = 
    \frac{\alpha}{d-2} \left(\frac{R}{\epsilon}\right)^{d-2}\,,
    \label{eq:arealaw}
\end{equation}
which is the celebrated area-law divergence of the entanglement entropy \cite{Srednicki:1993im,Amico:2007ag,Eisert_2010,Chandran:2015vuk,Casini:2022rlv}.
The $d=2$ result involves a logarithm, as given in eq.~\eqref{eq:2dentropy}.

From the perspective of eq.~\eqref{eq:TheEntropy}, the divergence arises because the vector field vanishes in a controlled way near the entangling surface.
Since it is widely expected \cite{Fredenhagen:1984dc, Haag:1996hvx, Sorce:2020ama} that generic modular flows approach a boost near the entangling surface, this result then implies that an area-law divergence of the form $(\frac{R}{\epsilon})^{d-2}$ should be present in all states and choices of subregion in QFT.
Conversely, given that the area law eq.~\eqref{eq:arealaw} is quite generally the leading divergence in the entanglement entropy for QFTs, one might change perspective and argue that the modular flow must become boost-like near the boundary, with a universal coefficient $\alpha$.
Note, there are in fact exceptions to the area law in $d>2$, such as the case of free fermions at zero temperature but nonzero chemical potential, \eg see \cite{Chandran:2015vuk}. 
It would be interesting to examine the near-boundary behavior of the modular flow in such examples.

We conclude noticing that our formula \eqref{eq:TheEntropy} exhibits divergences whenever the vector field  becomes null, not only at the subregion boundary but also within the open subregion $\mathcal{V}$ of interest. This echoes the results in \cite{Sorce:2020ama,Arias:2016nip} that generic modular flows are future-directed. We speculate that divergences in the vacuum-subtracted entanglement entropy are in general associated with geometric modular flows that violate future-directedness, thus possibly providing a new rationale for considering these flows unphysical.

\subsubsection*{Example 2: Sphere}
\label{sec:chm}

As a second example, consider the CFT$_d$ vacuum state on Minkowski spacetime, and the double cone region associated to a $(d-1)$-dimensional ball of radius $R$. If we introduce spherical polar coordinates centered on the ball, the metric becomes 
\begin{equation}
  \rd s^2 = -\rd t^2 + \rd r^2 + r^2 \rd \Omega^2_{d-2} \label{flatten}  
\end{equation}
and the double-cone region is described as 
\begin{equation}
\lbrace r<R\,,\ |t|<R-r\rbrace\,. \label{bball}
\end{equation}

The associated entanglement entropy was calculated in \cite{Casini:2011kv} as follows: 
First, with the change of coordinates
\begin{equation}
    \frac{t}{R}= \frac{\sinh(\frac{\tau}{R})}{\cosh{u} + \cosh(\frac{\tau}{R})}
    \qquad{\rm and}\qquad 
    \frac{r}{R}= \frac{\sinh{u}}{\cosh{u}  + \cosh(\frac{\tau}{R})}\,,
    \label{bball2}
\end{equation}
one observes that the flat-space metric \eqref{flatten} becomes\footnote{The precise form of $\Omega^2(\tau,x)$ will not be important here, but the interested reader may find it in \cite{Casini:2011kv}.}
\begin{equation}
    \rd s^2 =  \Omega^2(\tau,u)\left[-\rd \tau^2 + R^2\left( \rd u^2 + \sinh^2 u\, \rd \Omega^2_{d-2}\right)\right]\,.
\end{equation}
Next we note that allowing $\tau$ and $u$ to run over their full ranges, $-\infty\le\tau\le\infty$ and $0\le u\le\infty$, precisely covers the ball region \reef{bball}. 
Now invoking the Weyl invariance of the underlying CFT$_d$, we strip off the conformal factor $\Omega^2(\tau, u)$ and recognize that the ball is conformally mapped into $\mathbb{R}\times H^{d-1}$, where $R$ is the radius of curvature of the hyperbolic geometry. Making use of eqs.~\eqref{eq:definingeq} and \eqref{eq:primary}, the reduced density matrix in the double-cone region \reef{bball} is mapped to a standard Gibbs density matrix on the new space $\mathbb{R}\times H^{d-1}$ with inverse temperature $\beta=2\pi\,R$.\footnote{In the language of eqs.~\eqref{eq:thermalentropydensity}-\eqref{eq:TheEntropy}, $\beta = 2\pi R$ corresponds to the modular flow vector field $\xi = R \partial_{\tau}$.\label{foot:hotxi}} Furthermore, this unitary map leaves the von Neumann entropy $S=-\Tr( \rho \log \rho)$ invariant, meaning the entanglement entropy of the sphere has become a thermal entropy in the hyperbolic spacetime. The latter is given by \cite{Casini:2011kv} 
\begin{equation}
    S  = \alpha_0 \int \sqrt{h}\,  \rd x\, \rd\Omega_{d-2}
    =\alpha_0\,R^{d-1}\,\Omega_{d-2}
    \int^{u_{\max}}_0 
    \rd u\,\sinh^{d-2}\!u
    \,,
    \label{eq:CHM}
\end{equation}
where $h$ is the determinant of the $H^{d-1}$ metric and $\Omega_{d-2}=\frac{2\pi^{\frac{d-1}{2}}}{\Gamma\left(\frac{d-1}{2}\right)}$ is the area of the unit $(d-2)$-sphere. Moreover, $\alpha_0$ is a thermal entropy density; c.p.~eqs.~\eqref{eq:thermalentropydensity}-\eqref{eq:TheEntropy} with $\xi = R \partial_{\tau}$ from footnote \ref{foot:hotxi}.
It can be written as
\begin{equation}
\alpha_0 := \frac{2 \Gamma(\frac{d}{2})}{\pi^{\frac{d-2}{2}}}a_d^*\,,
\label{eq:adstar}
\end{equation}
where the universal coefficient $a_d^*$ characterizes the number of degrees of freedom of the underlying CFT$_d$ \cite{Myers:2010xs,Myers:2010tj}. For even $d$, it corresponds to the coefficient of the A-type trace anomaly $A$, while for odd $d$, $a_d^*$ is another central charge, \eg $a_3^*=\frac{F}{2\pi}$ of the $F$-theorem in $d=3$ \cite{Jafferis:2011zi}. The range of integration must be regulated and we choose a large-distance cutoff $u_{\max}$ as in \cite{Casini:2011kv},
\begin{equation}
    \cosh{u_{\max}}=\frac{R}{\delta}\,,
    \label{stop1}
\end{equation} 
where $\delta$ is a short-distance regulator in the CFT$_d$ in the original Minkowski frame.

The leading and subleading contributions to the entropy in eq.~\eqref{eq:CHM} are then \cite{Casini:2011kv, Hung:2011nu}
\begin{equation}
\label{eq:CHM2}
\begin{aligned}
        S \simeq a_d^\ast \frac{2\Gamma(\frac{d}{2})}{\pi^{\frac{d-2}{2}}(d-2)}\frac{{\cal A}_{d-2}}{\delta^{d-2}}  +\dots 
\end{aligned}
\end{equation}
where ${\cal A}_{d-2}=\Omega_{d-2} R^{d-2}$ is the area of the entangling surface in the original flat-space frame, \ie the area of the ($d$--2)-dimensional sphere at $r=R$. Hence we have recovered the expected area law contribution here.  We also note that the entropy \eqref{eq:CHM2} is an expansion in powers of $\frac{\delta^2}{R^2}$ and the higher order terms can be interpreted in terms of curvatures integrated over the entangling surface divided by the appropriate power of the cutoff, \eg see discussion in \cite{Hung:2011ta}. Furthermore, one finds the following universal contributions \cite{Casini:2011kv}:
\begin{equation}
    S_{univ}=\left\{ \begin{matrix}
(-)^{\frac{d-2}{2}}\, 4\, a^*_d\,\log\frac{2R}{\delta} & \quad & {\rm for\ even}\ d\,,\\
(-)^{\frac{d-1}{2}}\,2\pi\,a^*_d\hfill & \quad& {\rm for\ odd}\ d\,.
\end{matrix}\right.
\label{eq:CHM3}
\end{equation}

Let us compare the previous calculation with our approach to evaluate the entanglement entropy via eq.~\eqref{eq:TheEntropy}.
The vacuum modular flow of the double-cone region is \cite{Longo:2018zib,Casini:2011kv}\footnote{Note that $\xi$ is transformed to $\partial_t$ under the conformal map from ${\cal S}$ to $H$. See eqs.~(2.25), (2.26), and (3.22) in \cite{Casini:2011kv}. This is closely related to the fact that, as mentioned above, the entanglement entropy of the sphere becomes a thermal entropy for the hyperbolic spacetime.}
\begin{equation}
    \xi = \frac{1}{2R}\left((R^2-u^2)\partial_u + (R^2-v^2)\partial_v \right) \,,
    \label{eq:diamondflow}
\end{equation}
where $u=t-r$ and $v=t+r$.
In particular, on the $t=0$ slice, we have
\begin{equation}
    \xi\big|_{t=0} 
    = \frac{1}{2R}(R^2-r^2)\partial_t \,.
\end{equation}
Substituting this into eq.~\eqref{eq:TheEntropy} yields
\begin{equation}
    S  = \alpha\, \Omega_{d-2} \int_0^{R-\epsilon}  \rd r \,  r^{d-2}  \left(\frac{2R}{R^2-r^2}\right)^{d-1}
    = 
    \alpha\,R^{d-1}\, \Omega_{d-2} \int_0^{u'_{max}}  \rd u \,  \sinh^{d-2}\!u\,,
    \label{eq:matchingCHM}
\end{equation}
where we have employed the coordinate transformation\footnote{This is the coordinate transformation in eq.~\eqref{bball2}, restricted to the $t=0$ (or $\tau=0$) slice.}
\begin{equation}
     \sinh{u} = \frac{2R\, r}{R^2-r^2}\,.
     \label{eq:xTor}
\end{equation}
We have also introduced a short-distance cutoff $\epsilon$ in \eqref{eq:matchingCHM} to regulate the divergence which arises as the integration approaches $r=R$. 
Our answer precisely matches eq.~\eqref{eq:CHM} upon setting $\alpha = \alpha_0$ and 
$u'_{\max}=u_{\max}$. 
The first equality fixes the coefficient $\alpha$ in our proposed entropy current \eqref{eq:current} and reinforces the proposal that it should be a universal state- and geometry-independent coefficient. 

However, given the cutoff in eq.~\eqref{stop1} and the coordinate transformation in eq.~\eqref{eq:xTor}, one finds that the cutoffs do not match. Rather in terms of the short-distance cutoffs, 
\begin{equation}
\epsilon \simeq \delta + R\left(1-\frac{\delta}{R}\right)\left( 1-\frac{1}{\sqrt{1-\frac{\delta^2}{R^2}}}\right)=\delta + {\cal O}\left(\frac{\delta^2}{R}\right).
    \label{stop2}
\end{equation}
Hence eq.~\eqref{eq:matchingCHM} reproduces the leading area-law contribution in \eqref{eq:CHM2} but the subleading terms do not agree. That is, some additional input would be needed to fix $\epsilon$ in terms of the short-distance cutoff $\delta$ as in eq.~\eqref{stop2}. Furthermore, there is no reason to expect that the relation eq.~\eqref{stop2} is a universal formula.
Keep in mind that while the detailed choice of the cutoff will not be relevant when studying UV-finite quantities, its importance emerges here because the entanglement entropy is a UV divergent quantity.

Let us add that the naive choice $\epsilon=\delta$ would induce some important discrepancies with respect to eq.~\eqref{eq:CHM}. In particular, expanding the entropy from eq.~\eqref{eq:CHM} in terms of small $\frac{\delta}{R}$ yields a power series where successive terms decrease by $(\frac{\delta}{R})^2$. As noted above, each of the UV-divergent contributions proportional to a power $(\frac{R}{\delta})^{2n}$ has a geometric interpretation. In contrast, expanding the result from eq.~\eqref{eq:matchingCHM} in terms of  small $\frac{\epsilon}{R}$ yields a series where each term decreases by a single power of $\frac{\epsilon}{R}$, and hence the previous geometric interpretation is lost. Not all of the contributions with power-law divergences are universal, and so this discrepancy is not surprising. One confirms that the coefficient of the universal logarithmic divergence for even $d$ in eq.~\eqref{eq:CHM3} is identical for both regulators, as expected \cite{Liu:2012eea,Metlitski:2011pr,Casini:2015woa}. However, the constant contribution for odd $d$ does not agree between the two regulators, reflecting the non-universality of this term, as noted in \cite{Casini:2015woa}. It is important to use a geometric regulator, such as the one introduced using mutual information in \cite{Casini:2015woa}.

\subsection{Stress tensors for modular flows}
\label{sec:energytomatch}

We argued above that the entanglement entropy of states with geometric modular flows can be evaluated as an integral of a local entropy density which can be associated with Unruh observers.
In this section, we explore the possibility that there is a local stress tensor that can be associated with Unruh observers. Here we find an expression for the expectation value of the stress tensor after a vacuum-subtraction suited to the modular flow of interest. 

The entropy formula \eqref{eq:TheEntropy} began with the observation that in a CFT$_d$, a thermal bath has a uniform entropy density $s=\alpha\,(2\pi T)^{d-1}$, for some constant $\alpha$.
Using the first law of thermodynamics (\ie for local densities, $\dd \langle {T}_{tt}\rangle = T \dd{s}$), we can also determine the energy density of the assumed thermal bath
\begin{equation}
    \langle T_{tt}\rangle = \frac{d-1}{d}\,T s = \frac{(d-1)}{2\pi d}\,\alpha\,(2\pi T)^{d}\,.
    \label{eq:actualTtt}
\end{equation}
We then infer the full stress tensor to be
\begin{equation}
    \langle T_{\mu \nu} \rangle = \frac{\alpha}{2\pi d}\,(2\pi T)^{d}\,\mathrm{diag}\left(d-1,1, 1, ..., 1\right)\,,
    \label{eq:hotT}
\end{equation}
where the spatial components are fixed by the tracelessness of $T_{\mu \nu}$ and the symmetry of the thermal state.
To express the right-hand side of this expression in a covariant fashion, we introduce $t^{\mu}$ as the  future-directed unit vector defining the time direction, and then we can rewrite eq.~\eqref{eq:hotT} as
\begin{equation}
   \langle T_{\mu \nu}\rangle =\frac{\alpha}{2\pi d}(2\pi T)^{d} \left(d \,t_{\mu} t_{\nu} + g_{\mu\nu}\right)\,,
\end{equation}
as expected for a CFT$_d$ plasma \cite{Rangamani:2009xk}.

Following the logic in section \ref{sec:derivation}, we can now introduce a spacetime-dependent stress-energy  tensor which describes the local experience of a family of Unruh observers.
Using $t^\mu = \frac{\xi^\mu}{\norm{\xi}}$ and $2\pi T = \frac{1}{\norm{\xi}}$ as above and in section \ref{sec:vacuumProperties}, we have
\begin{equation}
    \langle T_{\mu \nu} \rangle = \frac{\alpha}{2\pi d\norm{\xi}^{d+2}} \left(d \,\xi_{\mu} \xi_{\nu} + \norm{\xi}^2 g_{\mu\nu}\right)\,.
    \label{eq:UnruhEnergyMom}
\end{equation}
We note that this tensor is again conserved, \ie $\nabla^{\mu} \langle T_{\mu\nu}\rangle =0$, because the modular flow vector $\xi^{\mu}$ is a conformal Killing vector \eqref{eq:ckv}.
Moreover, it transforms as a primary with conformal weight $\Delta=d$ and spin $J=2$ under conformal transformations,\footnote{This can be derived as follows: First, under a conformal transformation, $g_{\mu \nu}\to \Omega^2\,g_{\mu \nu}$, while we posit that $\xi^{\mu}$ simply transforms as a vector (with no additional conformal factors). Then recalling that $\norm{\xi}=\sqrt{-g_{\mu \nu}\xi^{\mu}\xi^{\nu}}$, we have $\norm{\xi}\to\Omega \norm{\xi}$.
Similarly, using $\xi_{\mu}=g_{\mu\nu}\xi^\nu$, we see that $\xi_\mu$ transforms with a factor of $\Omega^2$. 
Combining these ingredients, eq.~\eqref{eq:UnruhEnergyMomTrans} follows from the definition \eqref{eq:UnruhEnergyMom}.}
\begin{equation}
    \langle T_{\mu \nu}(x) \rangle \to \langle T'_{\mu\nu}(x')\rangle = \Omega^{-(d-2)}\frac{\partial x^{\rho}}{\partial x'^{\mu}}\frac{\partial x^{\sigma}}{\partial x'^{\nu}}\,\langle T_{\rho \sigma}(x)\rangle
    \,.
    \label{eq:UnruhEnergyMomTrans}
\end{equation}

How do we interpret the expression \eqref{eq:UnruhEnergyMom}? First of all, eq.~\eqref{eq:UnruhEnergyMom} does not describe the usual energy-momentum one-point function in the state of interest.
For example, in the Minkowski vacuum state, we have $\langle T_{\mu \nu}\rangle_{\Omega} = 0$ while eq.~\eqref{eq:UnruhEnergyMom} yields a nonvanishing answer for the vacuum modular flow $\xi$ with respect to the Rindler wedge! We examine this in detail in an example below, but let us anticipate we will find that the energy density diverges (positively) as one approaches the entangling surface with $\langle T_{tt}\rangle \sim \frac{1}{x^d}$ on the $t=0$ slice. Considering our discussion of the entropy current above, we propose that eq.~\eqref{eq:UnruhEnergyMom} describes the excitations that a family of Unruh observers would attribute to the state, relative to the state in which they all experience zero temperature. That is, inertial observers and Unruh observers have different vacuum-subtraction schemes, corresponding to different choices of the zero-point energy-momentum. Hence our proposal is that eq.~\eqref{eq:UnruhEnergyMom} corresponds to
\begin{equation}
    \langle T_{\mu\nu}\rangle = \langle T_{\mu\nu} \rangle_{\Psi} - \langle T_{\mu\nu} \rangle_{\Psi_{\mathrm{dis}}}=: 
    \langle T_{\mu\nu} \rangle_{\Psi-\Psi_{\mathrm{dis}}}\,,
    \label{eq:disdiff}
\end{equation}
where $\ket{\Psi_{\mathrm{dis}}}$ is the ``disentangled'' state in which the Unruh observers all experience zero temperature and the entropy current vanishes everywhere. Such states were explicitly constructed in section \ref{sec:disentangling} for $d=2$. 

Let us make several observations about this proposal. 
First, the fact that a difference between two expectation values appears in eq.~\eqref{eq:disdiff} is consistent with the absence of additional anomalous terms in the transformation rule eq.~\eqref{eq:UnruhEnergyMomTrans}, such as the Schwarzian term in eq.~\eqref{eq:schwarzian} for $d=2$.\footnote{In $d>2$, the stress tensor in flat spacetime transforms as an ordinary conformal primary of spin $J$ and scaling dimension $\Delta$ \cite{Simmons:2020}, so there will be no anomalous terms in any case.
An interesting task for future work would be to constrain $\langle T_{\mu\nu}\rangle_{\Psi}$ as a function of $\xi^{\mu}$ and $g_{\mu \nu}$ from tracelessness, conservation, and symmetry of the indices, with the aim of generalizing the $d=2$ formula \eqref{eq:manifest}.}
Since anomalous contributions depend on the transformation parameters only, they cancel in the difference between the two expectation values.
Finally, our proposal can be explicitly checked in $d=2$; in this case, eq.~\eqref{eq:UnruhEnergyMom} yields
\begin{equation}
    \langle T_{\tilde u\tilde u} \rangle_{\Psi-\Psi_{\rm dis}} = \frac{c}{48\pi\, \tilde c(\tilde u)^{2}}
    \qquad
    \langle T_{\tilde v\tilde v} \rangle_{\Psi-\Psi_{\rm dis}} = \frac{c}{48\pi\, \tilde b(\tilde v)^{2}}
    \qquad 
    \langle T_{\tilde u\tilde v}\rangle_{\Psi-\Psi_{\rm dis}} =0\,.
    \label{eq:match2}
\end{equation}
For $d=2$, the individual expectation values in eq.~\eqref{eq:disdiff} can be evaluated using the methods in section \ref{sec:stateProperties}.  The difference $\langle T_{uu}\rangle_{\Psi}-\langle T_{uu}\rangle_{\Psi_{\mathrm{dis}}}$\footnote{Eq.~\eqref{eq:psiMinusPsidis2d} displays only the $uu$ component but the other components are determined straightforwardly.} was already given in eq.~\eqref{eq:psiMinusPsidis2d} and precisely matches the result above. 
Moreover, as noted in and below eq.~\eqref{eq:manifest}, eq.~\eqref{eq:match2} is the part of $\langle T_{uu}\rangle_{\Psi}$ which transforms non-anomalously under conformal maps.

\subsubsection*{Example: Rindler vacuum}

Let us apply the previous discussion to the Minkowski vacuum restricted to the Rindler wedge in $d$ dimensions. 
Given the Rindler flow
\eqref{eq:lazzy} on the $t=0$ slice, eq.~\eqref{eq:UnruhEnergyMom} yields
\begin{equation}
    \langle T_{\mu \nu}(x)\rangle = \frac{\alpha}{2\pi d} \,\frac1{x^d}\,\mathrm{diag}\left(d-1,1, 1, ..., 1\right)
    \,.
    \label{eq:hotT2}
\end{equation}
From our proposal in eq.~\eqref{eq:disdiff}, this is the difference between the expectation value of the stress tensor in the Minkowski vacuum and in the corresponding disentangled state, \ie the Rindler vacuum $|\Omega_{\mathcal{R}}\rangle$ introduced in footnote \ref{foot:rindlervac}. Since the first of these expectation values vanishes, eq.~\eqref{eq:disdiff} reduces to 
\begin{equation}
 \langle T_{\mu\nu}\rangle =  - \langle T_{\mu\nu} \rangle_{\Omega_{\mathcal{R}}}\,.
    \label{eq:RinComp}
\end{equation}

The Rindler vacuum state is well understood in the context of free field theory \cite{Takagi:1986kn, Birrell:1982ix,Candelas:1977zza,Candelas:1978gg}. 
In this context, one studies the Bogoliubov transformation between creation and annihilation modes $(a^{\dag}_k,a_{k})$ obtained by solving the wave equation in inertial Minkowski coordinates, and modes $(b^{\dag}_k,b_{k})$ obtained by solving the wave equation in Rindler coordinates. 
The Minkowski vacuum $|\Omega\rangle$ is defined as the unique state satisfying
$a_k |\Omega\rangle = 0$ for all momenta $k$, while $|\Omega_{\mathcal{R}}\rangle$ is the unique state satisfying
$b_k|\Omega_{\mathcal{R}}\rangle = 0$ for all $k$.
The latter condition implies a complete absence of Rindler particles, \ie the vanishing of $\langle \Omega_{\mathcal{R}}| b_k^{\dag} b_k|\Omega_{\mathcal{R}}\rangle$ for all $k$, and thus Rindler observers experience zero temperature.

The expectation value of the energy-momentum tensor in the Rindler vacuum of a conformally-coupled massless free scalar field theory in $d$ dimensions was studied in \cite{Takagi:1986kn}. 
The final expression on the $t=0$ slice yields 
\begin{equation}
     \langle T_{\mu\nu} \rangle_{\Omega_{\mathcal{R}}}
     = - \frac{a_d \sigma_d}{d-1}\, \frac1{(2\pi\,x)^d}\, \textrm{diag}(d-1,1,....,1) \label{eq:hotT3}
\end{equation}
where  $\sigma_d$ is the Stefan-Boltzmann constant in $d$ dimensions and the numerical factor $a_d$ encodes the deviation from the black-body law. Comparing eqs.~\eqref{eq:hotT2} and \eqref{eq:hotT3}, we see that eq.~\eqref{eq:RinComp} is satisfied provided we set 
\begin{equation}
\alpha = \frac{d}{d-1}\,\frac{a_d \sigma_d}{(2\pi)^{d-1}}\,.
\label{eq:tomatch}
\end{equation}

Given the constant prefactors $a_d \sigma_d$ listed in \cite{Takagi:1986kn}, employing the relation \eqref{eq:adstar} between $\alpha_0$ and $a_d^*$, and our identification $\alpha=\alpha_0$, this implies the following numerical values
\begin{equation}
    \alpha =
    \begin{cases}
        \frac{1}{6}, & d=2 \\
        \frac{9}{32\pi^{3}}\zeta(3), & d=3 \\
        \frac{1}{180\pi} & d=4\\
        \frac{15(\pi^2 \zeta(3)+15 \zeta(5)) }{1024\pi^6} & d=5\,.
    \end{cases}
    \qquad \qquad
    a_d^* =
    \begin{cases}
        \frac{1}{12}, & d=2 \\
        \frac{9}{32\pi^{3}}\zeta(3), & d=3 \\
        \frac{1}{360} & d=4\\
        \frac{5(\pi^2 \zeta(3)+15 \zeta(5)) }{512 \pi^5} & d=5\,.
    \end{cases}
    \label{eq:ours}
\end{equation}
We find agreement for $a_d^*$ with previous literature in even $d$, but not odd $d$. In particular,
we find agreement in $d=2$, since  we have\footnote{See discussion below eq.~\eqref{eq:thermalentropydensity}.} $\alpha = \frac{c}{6}$ and  $c=1$ for a massless free scalar.
In $d=4$, we find $a_d^*=\frac{1}{360}$ which corresponds to the $A$-anomaly coefficient  for a massless conformally coupled free scalar \cite{Birrell:1982ix,Duff:1977ay}.\footnote{The meticulous reader will find it useful to know that in \cite{Duff:1977ay}, $b'$ is related to $A$ via $A=-(4\pi)^2 b'$.}
However, in $d=3$ and $5$, we do not find agreement. For example, for a conformally-coupled free scalar field theory, $F=\frac{1}{8}\log 2 - \frac{3}{16\pi^2}\zeta(3)\approx 0.064$ in $d=3$ \cite{Nishioka:2021uef,Klebanov:2011gs,Hu:2024pen}, 
which does not match our result above, namely $F=2\pi a_3^*\approx 0.069$.
We speculate that this is related to regularization subtleties in odd $d$ such as those discussed around eq.~\eqref{stop2}.
It would be interesting to further test our proposal against the stress tensor for the Rindler vacuum in holographic CFTs \cite{Emparan:1999gf,Emparan:2023ypa}.

\vspace{.2cm} 

To close this discussion, we note that it is tempting to use eq.~\eqref{eq:UnruhEnergyMom} in evaluating
\begin{equation}
    \langle h \rangle =2\pi \int_{A} \rd\Sigma_{\mu} \langle T^{\mu}{}_{\nu}\rangle \xi^{\nu}\,.
    \label{help}
\end{equation}
Then following the standard relationship for a  finite-dimensional quantum system
\begin{equation}
    \langle h\rangle_{\rho} = -\Tr(\rho \log \rho) = S  \,,
    \label{eq:itish}
\end{equation}
one might expect that eq.~\reef{help} yields the entanglement entropy. However, examining $\langle T^{\mu}{}_{\nu}\rangle \xi^{\nu}$ in more detail, we find
\begin{equation}
    \langle T^{\mu}{}_{\nu}\rangle \xi^{\nu} = \frac{\alpha}{2\pi d\norm{\xi}^{d+2}} \left(d \xi^{\mu} \xi_{\nu} + \norm{\xi}^2 \delta^{\mu}_{\nu}\right) \xi^{\nu}
    = -\frac{d-1}{d}\alpha
    \frac{\xi^{\mu}}{2\pi \norm{\xi}^{d}} = \frac{d-1}{d}\,j_S^{\mu}\,,
    \label{housefire}
\end{equation}
where the entropy current is as in eq.~\eqref{eq:current}. Hence, we have an extra $\frac{d-1}{d}$ factor in eq.~\eqref{eq:itish}, which is the same factor that appears in eq.~\eqref{eq:actualTtt}.

In fact, there is no contradiction here, and this factor is easily accounted for. Indeed, eq.~\eqref{eq:localHam} and hence eq.~\eqref{eq:itish} do not properly account for the normalization of the density matrix. We correct eq.~\eqref{eq:itish} by writing\footnote{Alternatively, \cite{Casini:2011kv} adds a c-number to the right-hand side of eq.~\eqref{eq:localHam}. This does not affect our previous discussions, where we were primarily concerned with the operator $h$ and the geometric modular flows.}
\begin{equation}
    S= \langle h\rangle_{\rho} + \log Z \,, \label{help2}
\end{equation}
where $Z=\Tr e^{-h}$. 
With a standard vacuum-subtraction scheme, one has that $\langle h\rangle_{\rho}=0$ in the vacuum state, and thus the entanglement entropy is given entirely by this extra term, \ie $S=\log Z$. In eq.~\eqref{eq:disdiff}, we are changing the vacuum-subtraction scheme; this shifts some of the entanglement entropy from the $\log Z$ term to the expectation value. Implicitly, we are thus producing a c-number shift $h \to h-\langle h \rangle_{\Psi_{\mathrm{dis}}}$ and also redefining $Z=\Tr e^{-(h-\langle h \rangle_{\Psi_{\mathrm{dis}}})}$. Together these c-number shifts leave eq.~\reef{help2} unchanged, 
\begin{equation}
    S=\langle h-\langle h \rangle_{\Psi_{\mathrm{dis}}}\rangle_{\rho} + \log \Tr e^{-(h-\langle h \rangle_{\Psi_{\mathrm{dis}}})} 
    = \langle h\rangle_{\rho} 
    +
    \log \Tr e^{-h} 
    \,.
\end{equation}
Therefore, the difference between $\langle h- \langle h \rangle_{\Psi_{\mathrm{dis}}}\rangle_{\rho}$ as in eq.~\eqref{housefire} and $S$ as in eq.~\eqref{eq:TheEntropy} can be accounted for by the $Z$ term. It would be interesting in the future to determine if any vacuum subtraction scheme can fully transfer the entanglement entropy into the $\langle h\rangle_{\rho}$ term.

\section{Discussion}
\label{sec:discuss}

In this manuscript, we defined the concept of Unruh flows and constructed a state for every such flow in the Rindler wedge $\mathcal{R}$ of a 2$d$ CFT on Minkowski spacetime.
We have shown this for both the vacuum and the thermal sectors of the theory.
We also discussed how the existence of a state for any Unruh flow generalizes beyond the Rindler wedge to a wide range of causally complete subregions of a $2d$ conformally flat manifold.

We then derived general expressions for the entropy and energy densities of states with Unruh flows --- see eqs.~\eqref{eq:ENERGY} and \eqref{eq:Thermalenergydensity2}, respectively. We generalized these formulas to states with geometric modular flows in higher dimensional CFTs, and provided several consistency checks. We now discuss implications and extensions of our work. 

\subsection*{Multiple subregions}
\label{sec:multisubregion}

So far, we have studied modular flows of connected regions, e.g.,~the Rindler wedge or a causal diamond.
A generalization of our construction comes from considering subregions with multiple disconnected components.

Let $\mathcal{V}$ describe two spacelike-separated, finite regions, $\mathcal{V}=\mathcal{V}_1 \cup \mathcal{V}_2$, with a finite separation.
For a typical state, the modular Hamiltonian with respect to $\mathcal{V}$ contains nonlocal terms mixing $\mathcal{V}_1$ and $\mathcal{V}_2$ which represent entanglement between the two subregions (for example, see eq.~\eqref{eq:Hnonloc} in appendix \ref{App:multiple}).
However, we expect there exist special states for which $\mathcal{V}=\mathcal{V}_1 \cup \mathcal{V}_2$ and yet the modular Hamiltonian is local in $\mathcal{V}$, leading to vanishing entanglement between $\mathcal{V}_1$ and $\mathcal{V}_2$.

This is related to the following general statement \cite{Witten:2018zxz,Fewster:2019ixc,Roos:1970fm,Buchholz:1973bk,Doplicher:1984zz}: 
For well-behaved QFTs in Minkowski spacetime, given any two states $\ket{\Psi_1}$ and $\ket{\Psi_2}$ in a GNS Hilbert space $\mathcal{H}$, there exists a (non-unique) state $\ket{\Psi_p}\in \mathcal{H}$ 
which is indistinguishable from $\ket{\Psi_1}$ for measurements in $\mathcal{V}_1$ and indistinguishable from $\ket{\Psi_2}$ for measurements in $\mathcal{V}_2$.
That is, for $\a_1 \in \mathcal{A}_{\mathcal{V}_1}$ and $\a_2 \in \mathcal{A}_{\mathcal{V}_2}$,
\begin{equation}
    \langle \Psi_p| \a_1|\Psi_p\rangle
    =
    \langle \Psi_1| \a_1|\Psi_1\rangle
    \qquad
    \langle \Psi_p| \a_2|\Psi_p\rangle
    =
    \langle \Psi_2| \a_2|\Psi_2\rangle\,.
\end{equation}
Further, one can choose $|\Psi_p\rangle$ to be a product state, meaning
\begin{equation}
    \langle \Psi_p| \a_1 \a_2|\Psi_p\rangle
    =
    \langle \Psi_1| \a_1|\Psi_1\rangle\langle \Psi_2| \a_2|\Psi_2\rangle
    \label{eq:productstate}
\end{equation}
and there is no entanglement between $\mathcal{V}_1$ and $\mathcal{V}_2$ in the state $|\Psi_p\rangle$.
From \cite{Takesaki:ToOA}, the modular flow $U_{\Psi_p}^{\mathcal{V}}(s)$ then factorizes, i.e.~it acts as $U_{\Psi_1}^{\mathcal{V}_1}(s)$ on all operators in the $\mathcal{V}_1$ algebra, and $U_{\Psi_2}^{\mathcal{V}_2}(s)$ on all operators in the $\mathcal{V}_2$ algebra.

To see the relation with local modular Hamiltonians, consider the case $d=2$ and Unruh flows $\xi_1$ in $\mathcal{V}_1$ and $\xi_2$ in $\mathcal{V}_2$.
By our results in section \ref{sec:AStateForEveryUnruhFlow}, there exist vacuum-sector CFT$_2$ states $|\Psi_1\rangle$ and $|\Psi_2\rangle$ whose modular Hamiltonians generate $\xi_1$ in $\mathcal{V}_1$ and $\xi_2$ in $\mathcal{V}_2$, respectively.
Now using the result above, there exists a vacuum-sector state $\ket{\Psi_p}$ whose modular flow with respect to $\mathcal{V}$ factorizes, generating the local flows $\xi_1$ in $\mathcal{V}_1$ and $\xi_2$ in $\mathcal{V}_2$.

In summary, we can specify an arbitrary Unruh flow $\xi_1$ in $\mathcal{V}_1$ and an arbitrary Unruh flow $\xi_2$ in $\mathcal{V}_2$, and we expect there exists a state in the CFT$_2$ vacuum sector with the corresponding modular flow. Hence, our result on the existence of a state for every Unruh flow in Minkowski CFT$_2$ generalizes to multi-component subregions, and the states of interest are disentangled between the various subregions.\footnote{Unlike in section \ref{sec:disentangling}, there are no singularities associated with the disentangled state.
This is because the finite separation between $\mathcal{V}_1$ and $\mathcal{V}_2$ ensures the quantum correlations are UV-finite.}

We apply this result to a particular example in appendix \ref{App:multiple}.
There, we consider free fermion theory in $d=2$.
In this case, the one-sided modular Hamiltonian $h_{\Omega}^{\mathcal{V}}$ for the vacuum state with respect to $\mathcal{V}_1\cup \mathcal{V}_2$ is comprised of a local term as well as a nonlocal term mixing the two subregions \cite{Casini:2016fgb}.
Using the discussion above, there exists a state $\ket{\Psi_p}$ whose modular Hamiltonian is given by the local terms in $h_{\Omega}^{\mathcal{V}}$. 
This is useful because we expect our entanglement entropy formula eq.~\eqref{eq:TheEntropy} applies only when the modular Hamiltonian is local, hence for the subregion $\mathcal{V}$ it can be applied to $\ket{\Psi_p}$ but not $\ket{\Omega}$.
In appendix \ref{App:multiple}, we calculate the entanglement entropy of $\mathcal{V}$ in the state $\ket{\Psi_p}$ and show that it matches that of the vacuum state $\ket{\Omega}$, calculated in \cite{Casini:2016fgb}.
This implies that the nonlocal terms in $h_{\Omega}^{\mathcal{V}}$ do not actually contribute to the entanglement entropy, reinforcing the interpretation of the nonlocal terms as representing entanglement between $\mathcal{V}_1$ and $\mathcal{V}_2$, rather than entanglement between $\mathcal{V}$ and its complement.
Our analysis in appendix \ref{App:multiple} shows that this result easily extends to the case of $n>2$ subregions.

Having generalized our analysis of Unruh modular flows to the case of multi-component subregions, resulting in local expressions for certain corresponding entanglement entropies, we now explore further the relationship between local entropy densities and entanglement structures.

\subsection*{Entanglement structure}

Typically, entanglement entropy in QFT is  a complicated nonlocal quantity, as it encapsulates quantum correlations across extended regions. 
However, for CFT states with geometric modular flows, we found that it is given by a local entropy current $j_S$ (see eqs.~\eqref{eq:news} and \eqref{eq:current}), signaling the special and simple entanglement structure of these states. 

We can interpret the entropy current as a book-keeping device tracking the amount of entanglement we must add to obtain the state of interest from a corresponding disentangled state.
Heuristically, we can think of adding a CFT plasma with a local temperature which varies from point to point.
This temperature may even change from one moment to the next along the worldlines of observers following the modular flow.
The conservation of $j_S$ indicates that these changes in temperature precisely balance among the collection of observers such that the total entanglement entropy is unchanged with any choice of Cauchy slice for the subregion of interest.

The idea of local entanglement entropy densities has previously been studied independently of geometric modular flows. In particular, \cite{Vidal:2014aal} advocated for the construction of the ``entanglement contour,'' \ie an entropy density $s=s(x)$ whose integral across a given Cauchy surface  yields the entanglement entropy for the subregion of interest.\footnote{See related discussions in \cite{Botero:2004vpl,Coser:2017dtb}.}
The entanglement contour is defined to satisfy several axioms, such as positivity and invariance under the action of a local unitary. This latter requirement distinguishes entanglement contours from our entropy current, as a typical unitary is expected to disrupt the geometric nature of the modular flow, leaving us without a definition of the entropy current, see appendix \ref{app:connesappendix}. 
Further, in \cite{Vidal:2014aal} the contour axioms do not single out a unique $s=s(x)$, while our entropy current is uniquely determined be the modular flow given a state $\Psi$ and a subregion $\mathcal{V}$. 

Nevertheless, it would be interesting to see whether ideas from geometric modular flows may be useful in the study of entanglement contours.
In particular, as discussed further below, local temperatures can be defined even in states whose modular Hamiltonian is not of the form \eqref{eq:localHam} \cite{Arias:2016nip}.
It would therefore be interesting to understand whether a temperature-based entanglement current such as eq.~\eqref{eq:current} can be defined even in these states, providing a general realization of an entanglement density or contour. 
Indeed, this is suggested by the results of appendix \ref{App:multiple}, where we show that eq.~\eqref{eq:TheEntropy} gives the correct entanglement entropy for a state whose modular Hamiltonian contains not only local ``temperature'' terms but also additional nonlocal terms. 

The idea of local entanglement densities has also appeared in the context of AdS/CFT. 
In holographic CFTs, entanglement entropies can be calculated using the RT formula \cite{Ryu:2006bv}. 
The RT formula is indeed a local integral, but over an extremal surface in the dual bulk spacetime rather than over a Cauchy surface in the boundary. 
It is natural to wonder whether this integral in the bulk can be somehow projected to a local integral in the boundary. This idea is concretely realized with bit threads \cite{Freedman:2016zud,Headrick:2017ucz,Headrick:2022nbe}. This is most easily understood in static situations where one can restrict the analysis to a single bulk time slice. In this context, one studies particular vector fields $\chi^\mu$ on the bulk time slice describing an ``entanglement flux'' between a boundary subregion $A$ and its complement. The RT surface then represents a bottleneck for these conserved flows. Having optimized the flow with $\chi_{\max}^{\mu}$, the entanglement entropy can be expressed as a local integral $S = \int \rd\Sigma_{\mu} \chi_{\max}^{\mu}$ over the RT surface or the boundary subregion (or any other bulk surfaces which are homologous to the boundary subregion). 
In the language of the previous discussion, the flow  $\chi_{\max}^{\mu}$ defines a local entropy density on $A$. However, as emphasized in \cite{Freedman:2016zud}, the choice of $\chi_{\max}^{\mu}$ and hence the corresponding entropy density are not unique. In other words, there is no canonical projection from the RT surface to the boundary subregion.
Again, we expect there may be an interplay between ideas from our discussion of geometric modular flows and the bit thread framework, though this would require consideration of the covariant construction of the latter \cite{Headrick:2017ucz,Headrick:2022nbe}. 
At least, in the context of boundary states with a geometric modular flow, where $\xi^\mu$ uniquely defines the entropy current, there may be a preferred bit thread configuration which yields the same boundary entropy density.

In general, it would be fruitful to examine our results in a holographic context.
For instance, we can test our proposal \eqref{eq:adstar} that the constant prefactor $\alpha=\alpha_0$ discussed in sections \ref{sec:entanglementcurrents} and \ref{sec:energytomatch} is proportional to the $A$ anomaly in even $d$ by comparing eq.~\eqref{eq:UnruhEnergyMom} to the Rindler vacuum stress tensor computed holographically as in  \cite{Emparan:1999gf,Emparan:2023ypa}.

\subsection*{Primary states}

In sections \ref{sec:AStateForEveryUnruhFlow} to \ref{sec:beyondRindler}, we built excited states with geometric modular flows with respect to connected subregions, starting from either the vacuum state or the thermal state. 
In many cases, it is simple to generalize this construction to the case of primary states and/or multi-component subregions.

A primary state can be defined by acting with a local primary operator $V(x)$ on the vacuum state at the origin of a Euclidean CFT$_2$ on the plane,
\begin{equation}
    \ket{V} = V(0)\ket{\Omega}\,.
\end{equation}
Under radial quantization, this defines a state on circles centered at the origin.
With a suitable conformal mapping and Wick rotation, it defines a state on the Lorentzian cylinder.

In holography, an example of a primary state on the Lorentzian cylinder is the dual of the AdS$_3$ conical defect geometry \cite{Deser:1983nh,Deser:1983tn,Banerjee:2016qca}.
This geometry is obtained from empty AdS$_3$ by reducing the periodicity of the spatial $\varphi$-circle from $2\pi$ to $2\pi\lambda$, with $\lambda<1$. In this case, it can be argued using entanglement wedge reconstruction \cite{Dong:2016eik} 
that the modular flow associated with small causal diamonds in the CFT, whose entanglement wedge does not contain the defect, is a simple rescaling of the vacuum modular flow with $t\to \frac{t}{\lambda}$ and $\varphi\to \frac{\varphi}{\lambda}$.
On the other hand, the modular flow is non-geometric for a boundary subregion which is so large that its entanglement wedge contains the defect. Our result on the existence of a state for any one-sided Unruh flow straightforwardly carries over to the case of the small causal diamonds $\mathcal{D}_c$ in such primary states.
That is, a $\mathcal{D}_c$-preserving conformal transformation $U_{\varphi}$ maps $|V\rangle$ into a new state $U_{\varphi}|V\rangle$ with a different geometric modular flow in $\mathcal{D}_c$.
Of course, the flow will remain non-geometric in the complementary diamond.
It may be interesting to study how the modular flow changes as the size of $\mathcal{D}_c$ increases in this setting.

\subsection*{Beyond the Modular Wedge}
In section \ref{sec:beyondRindler}, we examined Unruh flows in causally complete regions of a conformally flat spacetime and argued that there is always a state in the CFT$_2$ with the corresponding modular flow. 
One may have noticed that in the figures describing the associated vector field, we have drawn the vectors over the entire spacetime, \ie beyond the causally complete region of interest, $\mathcal{V}$. That is, we extend the vector field to regions of the spacetime that naively are not accessible by starting in $\mathcal{V}$ and following the modular flow.\footnote{We thank the referee for suggesting that we clarify this point.}

For example, in figure \ref{fig:flows}, one might expect the modular flow with respect to the Rindler wedge $\mathcal{R}$ to act only on $\mathcal{R}$ and, by the two-sided property, on its complement $\mathcal{R}'$.
In this way one would be led to not
consider the causal future and past of the origin, \ie the regions where the associated boost vector \eqref{eq:boost} is spacelike. 

However, on the mathematical side, one can analyze the situation as follows: 
In a given Hilbert space sector $\mathcal{H}$ of a quantum field theory, to each causally complete region in spacetime one associates a von Neumann algebra of operators contained in $\mathcal{B}(\mathcal H)$.
Once we select a state in $\mathcal{H}$ which is cyclic and separating with respect to, say, the von Neumann algebra of the Rindler wedge, we can construct the associated modular Hamiltonian.
The modular Hamiltonian acts on $\mathcal{H}$ and thus can be used to Heisenberg-evolve operators in $\mathcal{B}(\mathcal{H})$.
In particular, it can be used to evolve operators that are supported neither in 
$\mathcal{R}$ nor in its causal complement $\mathcal{R}'$.
Since, in the specific case of the vacuum, the modular Hamiltonian associated to $\mathcal{R}$ is proportional to the boost generator, when it Heisenberg evolves an operator supported in, say, some subregion in the causal future of the origin, it will evolve it 
according to the boost vector field in that region. 

This discussion can be extended to all of the geometric modular flows constructed in this paper and hence explains the significance of the vector fields shown outside of $\cal{V}$ and its complement $\cal{V}'$.

From the physical point of view, this situation also reflects the well-known expectation that fields in a relativistic quantum field theory are determined by initial data on a Cauchy slice.\footnote{This statement is subtle at the quantum level. 
It holds precisely only as long as fields smeared on some codimension-one surface $\Sigma$ can make sense as (possibly unbounded) operators, which happens if and only if the OPE singularity arising from the multiplication of two point-like fields on $\Sigma$ is integrable when smeared over $\Sigma$. For non-conformal interacting quantum field theories, this is never the case, and one needs to ``fatten'' the Cauchy slice; the corresponding smearing in real time ensures everything is well-defined.
For conformal field theories in 2$d$, on the other hand, one can verify that smearing along some codimension-one spacelike surface is enough to turn any normal-ordered polynomial into a true operator \cite{Witten:2023qsv}.}
To elaborate, even if one prefers to think of the modular Hamiltonian as acting on fields on the $t=0$ slice, its action on a field smeared in some subregion in the causal future of the origin is uniquely fixed, since the latter is dynamically generated by fields smeared on the $t=0$ Cauchy slice.



Let us add here that as noted in footnote \ref{garmond}, the flow  determines the energy densities through eq.~\eqref{eq:manifest} and these expressions can be applied everywhere throughout the spacetime. In particular, these expressions still apply outside of the regions where the modular flow is timelike. 

\subsection*{Sectors}

Two sectors at different temperatures are unitarily inequivalent; see footnote \ref{foot:inequiv}. 
Consistency with this statement, in our framework, follows from the requirement that the conformal unitaries act trivially at spatial infinity, such that the vacuum or thermal sector falloffs for the modular flow in eqs.~\eqref{eq:vacFalloffsRindler} and \eqref{eq:thermalFalloffs} are preserved.
One may wonder how to extend our construction to allow for operators which change the temperature.
In this setting, it is important to understand the asymptotic behavior which characterizes the different sectors.

Our analysis indicates that vacuum sector states in CFT are characterized by a rapid asymptotic decay of the energy density, $\langle T_{tt}\rangle \sim \frac{1}{x^4}$, as $|x|\to\infty$.
This is consistent with the vacuum sector falloffs \eqref{eq:vacFalloffsRindler} for the Rindler wedge, and those for the diamond in eq.~\eqref{eq:psiDFalloffs}.
The situation is more complicated for the thermal sectors. As discussed in section \ref{sec:thermalProperties}, given two states with different temperatures, the differences of their energies and entanglement entropies are IR divergent, since changing the temperature changes the uniform energy and entropy densities. 
The relative entropy contains a similar IR divergence.
We interpret these divergences as indicating that these two sectors are unitarily inequivalent.

Divergences arising between states whose modular flows approach the same thermal flow $\frac{\beta}{2\pi} \partial_t$ at $x\to \infty$, but differ at ${\cal O}\left(\frac{1}{x}\right)$, are also addressed in section \ref{sec:thermalProperties}. In particular, both the total energy and entropy differences diverge logarithmically with the IR cutoff $L$.
This indicates an even finer grading of GNS sectors: we do not only have a GNS sector for each choice of asymptotic temperature, but possibly also GNS sectors associated with subleading behaviors of the modular flow. On the other hand, a cancellation in section \ref{sec:thermalProperties} yields a finite relative entropy between the thermal state and these states where the modular flows fall off more slowly to the thermal modular flow.
Since, in continuum QFT, only the relative entropy (and not the entanglement entropy) is well-defined, this supports the hypothesis that the two states belong to the same sector after all.\footnote{We expect similar questions may appear when comparing two thermal states related by a global boost, such that, for instance, one of the states corresponds to a thermal gas at rest while the other corresponds to a thermal gas with uniform constant velocity. 
These states have the same entanglement entropy difference due to the covariant form of eq.~\eqref{eq:covariantEntropy}.
On the other hand, the energy difference and relative entropy diverge linearly with the IR cutoff $L$ for nonzero boost parameter.}
Ironing out these subtleties may lead to interesting physical insights on the GNS construction in QFT.

\subsection*{Local temperatures}
\label{sec:localtemp}

In sections \ref{sec:vacuumProperties} and \ref{sec:derivation}, our entanglement entropy formula is derived by assigning a local modular temperature $\beta = 2\pi \norm{\xi}$ to Unruh observers following the geometric modular flow. 
We presented a heuristic picture that these observers see a CFT plasma carrying a local entropy density specified by this local temperature, as given in eq.~\eqref{eq:current}. This intuition was largely derived from the case of the vacuum flow in the Rindler wedge where $\xi$ generates boosts, as shown in eq.~\reef{eq:boost}. In this case, the corresponding observers follow trajectories of constant acceleration\footnote{For simplicity, we parameterize the acceleration by the position where the observers cross the $x$-axis.} $a(x)=\frac{1}{x}$ and the modular temperature,
\begin{equation}
  T=\frac{1}{2\pi\norm{\xi}}=\frac{1}{2\pi\,x} = \frac{a(x)}{2\pi}\,,
  \label{eq:acceler8}
\end{equation}
precisely matches the physical temperature  measured by the observer's detector. However, this is a special case; in general, $\beta = 2\pi\norm{\xi}$ is not necessarily identified with the physical inverse temperature that an observer following the modular flow experiences.

To illustrate this point, consider the example of the vacuum flow \eqref{eq:diamondflowVacuum} in a causal diamond. The integral curves corresponding to this flow can be obtained as in eq.~\eqref{eq:exponentiated},
\begin{equation}
u(s)=-R\,\frac{x_0-R\tanh\frac{s}{2}}{R-x_0\tanh\frac{s}{2}}\,,
\qquad
v(s)=R\,\frac{x_0+R\tanh\frac{s}{2}}{R+x_0\tanh\frac{s}{2}}\,,
    \label{eq:pop0}
\end{equation}
where we specify the curves by where they cross the $x$-axis (\ie $t=0$) with $v_0=x_0=-u_0$, which occurs at $s=0$. Now consider these curves \eqref{eq:pop0} to be the worldlines describing a family of Unruh observers traversing the diamond. 
Hence, we may wish to re-express them in terms of the proper time $\tau$ of the observers.
This is done using the relation\footnote{To derive this, integrate the relation $\frac{\rd \tau}{\rd s} = \norm{\xi}$ from footnote \ref{foot:dtauds}, using eqs.~\eqref{eq:pop0} and \eqref{eq:diamondflow}.
We have chosen the integration constant such that $\tau=0$ corresponds to $s=0$. Note that in traversing the entire diamond, $-\infty<s<\infty$ while the proper time $\tau$ spans a finite range $-\tau_{\rm max}<\tau<\tau_{\rm max}$ where at its maximum value one has $\cosh\frac{x_0\, \tau_{\rm max}}{R^2-x_0^2}=\frac{R}{\sqrt{R^2-x_0^2}}$. \label{foot:span}}
\begin{equation}
\tanh\frac{s}{2} = \frac{R}{x_0}\,\tanh\frac{x_0\,\tau}{R^2-x_0^2}\,.
    \label{eq:pop1}
\end{equation}
The resulting expressions $u=u(\tau), v=v(\tau)$ are periodic under the imaginary shift
\begin{equation}
    \tau\to \tau+i \beta_{\mathrm{phys}} \qquad {\rm with}\quad \beta_{\mathrm{phys}}=\pi\,\frac{R^2-x_0^2}{x_0}\,.
    \label{eq:pop3}
\end{equation}

Why do we label $\beta_{\mathrm{phys}}$ as a physical inverse temperature?
Imagine one of the observers carries an Unruh-DeWitt detector along her trajectory, which couples to a free massless scalar.\footnote{Or any scalar primary operator in the underlying CFT.} 
At leading order in the coupling between the detector and the scalar field, the response of the detector is fully specified by the Wightman function $W(\tau_1,\tau_2)$ of the scalar evaluated along the trajectory \cite{Birrell:1982ix,Unruh:1976db,DeWitt:1980hx,Crispino:2007eb}.
By Lorentz invariance, this correlator will depend only on the proper separation $\Delta s$ of the points along the trajectory, given by
\begin{equation}
  \Delta s^2=\Delta u\,\Delta v = (u(\tau_2)-u(\tau_1)) (v(\tau_2)-v(\tau_1))\,.
    \label{eq:pop4}
\end{equation}
Now given the periodicity in eq.~\eqref{eq:pop3} and the symmetry $\Delta s(\tau_1,\tau_2)=\Delta s(\tau_2,\tau_1)$, the Wightman function then satisfies $W(\tau_1+i\beta_{\rm phys},\tau_2)=W(\tau_2,\tau_1)$, which implies a thermal response of the detector at temperature $\beta_{\mathrm{phys}}$, as shown rigorously in \cite{Birrell:1982ix,DeBievre:2006pys}. 
That is, the observer experiences a thermal bath with temperature given by\footnote{One may be concerned that the detector will not reach thermal equilibrium because our observers traverse the causal diamond in a finite time --- see footnote \ref{foot:span}. But the observer can simply initialize the detector with its different energy levels populated in a thermal distribution determined by $\beta_{\rm phys}$. Then, apart from small fluctuations, the observer's detector remains in equilibrium as she crosses the diamond.}
\begin{equation}
T_{\mathrm{phys}}=\frac{1}{\beta_{\mathrm{phys}}}= \frac{x_0}{\pi\,(R^2-x_0^2)}\,.
    \label{eq:pop5}
\end{equation}

A few comments are in order here. For $x_0=0$, eq.~\eqref{eq:pop1} reduces to $\tanh{\frac{s}{2}}=\frac{\tau}{R}$ and the corresponding trajectory \eqref{eq:pop0} becomes $u(\tau)=v(\tau)=\tau$, hence describing a stationary observer with constant $x$-coordinate. In this case, $T_{\mathrm{phys}}$ vanishes, as to be expected for an inertial observer in the Minkowski vacuum state. On the other hand, as $x_0$ approaches the entangling surface at $x_0 = \pm R$, the modular flow approaches a boost flow and $T_{\mathrm{phys}}$ approaches the standard Unruh-Bisognano-Wichmann temperature $T\simeq \frac{1}{2\pi\,\delta x}$, where $\delta x=R-|x_0|$ is the distance to the entangling surface. 
More generally, we observe that the vacuum flow in the causal diamond has a special feature: the observers are following trajectories of constant acceleration $a(x_0) = 2x_0/(R^2-x_0^2)$.
Hence, eq.~\eqref{eq:pop5} has the form of the standard Unruh-Bisognano-Wichmann temperature, $T=\frac{a}{2\pi}$.\footnote{Similar statements apply for the geometric flow in the causal complement of the diamond, since the observers there also follow trajectories with constant acceleration.
\label{foot:cspan}}
Identical results hold for the double-cone region in $d>2$, since the vacuum modular flow \eqref{eq:diamondflow} for the double-cone takes the same form in lightcone coordinates in all dimensions.
Further, identical results hold in any theory where the corresponding Wightman functions depend only on the proper distance.

We now contrast the physical temperature $T_{\mathrm{phys}}$ experienced by observers following the diamond flow with the modular temperature $\beta = 2\pi \norm{\xi}$ appearing in our entanglement entropy formula.
From eq.~\eqref{eq:diamondflowVacuum}, $\beta = 2\pi \norm{\xi} = \frac{\pi}{R}\sqrt{(R^2-u^2)(R^2-v^2)}$, which yields 
\begin{equation}
    T    = \frac{x_0^2 R}{\pi (R^2-x_0^2)\left[R^2 -(R^2-x_0^2)\cosh^2\!{\frac{x_0\,\tau}{R^2-x_0^2}} \right]}\,.
\end{equation}
This does not match the physical temperature in eq.~\eqref{eq:pop5}. Indeed, it assigns a nonzero temperature for the inertial observer with $x_0 = 0$! From this expression, one may also be concerned that the modular temperature will become negative for large values of $\tau$.
However, the observers cross the causal diamond in finite proper time --- see footnote \ref{foot:span}. A more detailed examination shows that $T$ diverges if we follow the observers to either the future or past tips of the causal diamond, but $T$ never becomes negative.

Recall, the modular temperature can also be written as $\beta = 2\pi\frac{\rd\tau}{\rd s}$ (see footnote \ref{foot:dtauds}),
which is the temperature ascribed to observers in \cite{Martinetti:2002sz}.\footnote{See related discussions in \cite{Earman:2011zz}.} While the authors do not associate a physical meaning to this temperature, they propose it should be related to the fact that observers do not access all the degrees of freedom of the underlying quantum field theory. 
We see this intuition is correct in that the modular temperature provides a simple prescription \eqref{eq:TheEntropy} to evaluate the entanglement entropy for the region the observers can access.

When do the modular and physical temperatures coincide?
Consider parametrizing the Wightman function above in terms of the modular parameter.
Then, the KMS condition \eqref{eq:definingeq} yields an apparent thermal behaviour: $W(s_1+2\pi i,s_2)=W(s_2,s_1)$.
In general, this thermality is not physical, since the observer and her clock know nothing about the modular parameter $s$. There is, however, a special case where this behavior can be related to  physical thermality, namely when the proper time is proportional to the modular parameter, \ie when $\frac{\tau}{s}$ is a constant. Using eq.~\eqref{eq:ckv}, this arises precisely when $\xi^\mu$ is a Killing vector field,
\begin{equation}
    \frac{\rd^2\tau}{\rd s^2}=\partial_s \norm{\xi}
    = -\frac{1}{2\norm{\xi}}\xi^{\mu} \nabla_{\mu} \left(\xi^\nu\xi_\nu\right)
    = \frac{1}{d}\norm{\xi}\nabla_\nu \xi^\nu =0\,,
    \label{eq:divergence}
\end{equation}
where the final equality holds when $\xi$ is Killing. In this case, we then have $\beta=2\pi\frac{\rd\tau}{\rd s} = \beta_{\rm phys}$. That is, the modular temperature and the physical temperature experienced by the observer coincide, as with the vacuum Rindler flow. Conversely, when the geometric modular flow is described by a general conformal Killing vector, as is the case for the flow \eqref{eq:exponentiated} on the causal diamond, the divergence in eq.~\eqref{eq:divergence} is nonvanishing and hence $\beta \ne \beta_{\mathrm{phys}}$.

As an aside, recall that the vacuum flow in the causal diamond has a special feature: the observers are following trajectories of constant acceleration.  For general geometric flows in $2d$ CFTs, this is not the case, and there is no direct interpretation of the observer's experience in terms of a thermal bath at temperature $\frac{1}{\beta_{\mathrm{phys}}}$. One interesting situation is where $\nabla_\mu\xi^\mu$ is small and the observer's experience is approximately thermal.\footnote{In $d=2$, this means $|c'(u) + b'(v)|\ll 1$.
This condition holds, for example, for the thermal-state flow \eqref{eq:borchers1} in the Rindler wedge whenever $u$ is large and negative, and $v$ is large and positive (compared to $\beta$).}

Further, even when the observer experiences a thermal bath with $\beta=\beta_{\rm phys}$, our description of the entropy and energy, \eg in eqs.~\eqref{eq:TheEntropy} and \eqref{eq:UnruhEnergyMom}, is only heuristic, since the resulting Unruh temperatures do not describe a CFT plasma with a varying temperature which is in local thermal equilibrium.
In order to have local thermal equilibrium, the typical thermal wavelength should be smaller than the length scale over which the temperature is varying.
By contrast, for example, the vacuum modular flow in the Rindler wedge on the $t=0$ slice has typical thermal wavelength  $\lambda \simeq \beta = 2\pi x$, which is of the same order of magnitude of  the length scale over which the temperature varies, given by $x$.

Nevertheless, our thermodynamics-inspired entropy formula \eqref{eq:TheEntropy} for states with geometric modular flow $\xi$ is on good footing.
Recall $\xi$ is a future-directed conformal Killing vector in the subregion of interest, $\mathcal{V}$.
Using a singular Weyl transformation, $\xi\big|_{\mathcal{V}}$ be transformed into an ordinary Killing field which is future-directed and timelike everywhere; this happened, for example, in section \ref{sec:chm}, when conformally mapping the double cone to a hyperbolic spacetime.
In the new conformal frame where $\xi$ is Killing, the modular temperature equals the physical temperature, and this temperature is varying over very large length scales (or not at all, in the double-cone example).
Hence, the thermodynamic entropy formula \eqref{eq:thermalentropydensity} applies.
This validates the entanglement entropy formula \eqref{eq:TheEntropy} because by footnote \ref{foot:weyl}, eq.~\eqref{eq:TheEntropy} is Weyl-invariant, and hence it is also applicable for the original setting where $\xi$ is only \textit{conformally} Killing.

To close, we revisit the original question: In what sense does $\beta= 2\pi \norm{\xi}$ correspond to a temperature? 
A precise notion of local temperatures has been provided in \cite{Arias:2016nip,Arias:2017dda}. 
In these works, one excites the vacuum with a unitary operator localized at a spacetime point $\boldsymbol{w}$, yielding $\Delta S_{EE} = 0$ yet $\Delta \langle h\rangle \ne 0$.
In particular, $S_{rel}= \Delta \langle h\rangle$, and if the modular Hamiltonian has the form \eqref{eq:localHam}, one has $S_{rel}\approx 2\pi P_{\mu}\xi^{\mu}(\boldsymbol{w})$, where $P_{\mu}$ is the expectation value of the total energy-momentum of the excitation. The relative entropy $S_{rel}=S_{rel}(\rho_{\Psi_1}^{\cal V}||\rho_{\Psi_2}^{\cal V})$ measures how difficult it is to operationally distinguish $\Psi_1$ and $\Psi_2$ using measurements in $\mathcal{V}$.
That is, the probability of confounding the two states after $N$  measurements is $p\sim e^{-S_{rel} N}$.
If $\Psi_1$ differs from $\Psi_2$ by a localized unitary excitation such that $S_{rel}\approx 2\pi P_{\mu}\xi^{\mu}(\boldsymbol{w})$, then we find the excitation in the quantum fluctuations of $\Psi_1$ with the same probability as in a thermal state with ``covariant temperature'' $2\pi\xi^{\mu}(\mathbf{w})$, see \cite{Arias:2016nip,Arias:2017dda}. 
As an example, consider the Rindler wedge; setting $\boldsymbol{w}=(0,a)$  in $(t,x)$ coordinates, one finds $2\pi P_{\mu}\xi^{\mu}(\boldsymbol{w}) = 2\pi T_{00}\xi^{t}(\boldsymbol{w}) = 2\pi a E$, retrieving the Rindler result $T =\frac{1}{2\pi a}$.

There is therefore a precise sense in which $2\pi \xi^{\mu}$  can be interpreted as a covariant temperature and $\beta=2\pi \norm{\xi}$ is the inverse temperature of an observer along the modular flow.
While $\beta_{\mathrm{phys}}$ is a physical temperature measured by a detector, $\beta = 2\pi \norm{\xi}$ corresponds to an information-theoretic temperature related to the distinguishability and entanglement of states.

\acknowledgments

We thank Michele Bianchessi, Horacio Casini, Laurent Freidel, Samuel Goldman, Tom Hartman, Ted Jacobson, Marc Klinger, Rob Leigh, Caroline Lima, Tassos Petkou, Maria Preciado-Rivas, Pranav Pulakkat, Kasia Rejzner, Mykola Semenyakin, Antony Speranza, Eirini Telali, Marija Toma\v{s}evi\'c, Erik Tonni, Themistocles Zikopoulos for useful discussions and feedback. 
We especially thank Jon Sorce for key discussions and feedback on the draft.
Research at Perimeter Institute is supported in part by the Government of Canada through the Department of Innovation, Science and Economic Development Canada and by the Province of Ontario through the Ministry of Colleges and Universities. 
JC acknowledges the support of the Natural Sciences and Engineering Research Council of Canada (NSERC) [funding reference number 514036].
RCM is also
supported in part by a Discovery Grant from the Natural Sciences and Engineering
Research Council of Canada, and by funding from the BMO Financial Group.
FC is partially supported by University of Milan through the Scholarship for Master’s Degree Thesis Preparation Abroad, and by Professor Florian Girelli through the Graduate Research Studentship at University of Waterloo.

\appendix

\section{Flows under conformal maps}
\label{app:pushforward}

In section \ref{sec:vacuumsector}, we claimed that if $U_{\Omega}(s)$ is the modular flow unitary implementing $\xi_{\Omega}$, as in eqs.~\eqref{eq:localHam} and \eqref{eq:modularunitary}, then $U_{\Psi}(s)$ defined in eq.~\eqref{eq:PsiModularUnitary} implements $\xi_{\Psi}$, where $\xi_{\Psi}$ is defined to be the pushforward of $\xi_{\Omega}$ under the conformal map $\varphi$. 
In section \ref{sec:vacuumsector}, we were interested in the case where $\xi_{\Omega}$ is the vacuum modular flow with respect to $\mathcal{R}$, but the previous statement applies for any geometric modular flow with respect to any causally complete subregion $\mathcal{V}$.

To show the pushforward relationship of vector fields, it suffices to show that the integral curves of the two modular flows are simply related by pullback under $\varphi$.
That is, if $\gamma_s$ is an integral curve of $\xi_{\Omega}$ and $\lambda_s$ is an integral curve of $\xi_{\Psi}$ such that $\varphi(\gamma_0) := \lambda_0$, we must show  that for all $s$, $\varphi(\gamma_s) := \lambda_s$.

From eq.~\eqref{eq:primary}, we have, for a primary operator $O$,
\begin{equation}
    U_{\varphi}\, O(\gamma_0) \,U_{\varphi}^{\dag} = \Omega_{\varphi}(\gamma_0)^{-\Delta}\, O(\lambda_0) \,,
\end{equation}
or equivalently
\begin{equation}
    U_{\varphi}^{\dag}\, O(\lambda_0)\,U_{\varphi}  = \Omega_{\varphi}(\gamma_0)^{\Delta} \,O(\gamma_0)\,,
\end{equation}
where we introduce subscripts on the conformal factors $\Omega$ to keep track of the corresponding conformal mapping.
By definition of $U_{\Omega}(s)$ as the vacuum modular automorphism unitary (c.p.~eq.~\eqref{eq:rho}), we have
\begin{equation}
    U_{\Omega}(-s) \left[ U_{\varphi}^{\dag}\, O(\lambda_0)\,U_{\varphi} \right] U_{\Omega}(s) 
    =
    \Omega_{\varphi}(\gamma_0)^{\Delta} 
    \Omega_{\Omega,s}(\gamma_0)^{-\Delta} \,O(\gamma_s)\,.
\end{equation}
Conjugating again by $U_{\varphi}$, and using the definition \eqref{eq:PsiModularUnitary}, we find
\begin{equation}
    U_{\Psi}(-s)\, O(\lambda_0) \,U_{\Psi}(s) 
    = \Omega_{\varphi}(\gamma_0)^{\Delta} 
    \Omega_{\Omega,s}(\gamma_0)^{-\Delta} \Omega_{\varphi}(\gamma_s)^{-\Delta} \,O(\varphi(\gamma_s))  \,.
    \label{eq:postconj}
\end{equation}
By definition of $\lambda$, the left hand side is
\begin{equation}
    U_{\Psi}(-s)\, O(\lambda_0) \,U_{\Psi}(s) 
    =
    \Omega_{\Psi,s}(\lambda_0)^{-\Delta}O(\lambda_s)\,,
    \label{eq:getlambdas}
\end{equation}
suggesting $\varphi(\gamma_s) = \lambda_s$ for all $s$.
Indeed, the conformal factors in eqs.~\eqref{eq:postconj} and \eqref{eq:getlambdas} match;
\begin{equation}
    \Omega_{\Psi,s}(\lambda_0) = \Omega_{\varphi}(\gamma_0)^{-1}
    \Omega_{\Omega,s}(\gamma_0)
    \Omega_{\varphi}(\gamma_s)
    \label{eq:chain}
\end{equation}
thanks to the chain rule.\footnote{Explicitly, letting $\psi$ and $\omega$ denote the conformal mappings associated with $\Omega_{\Psi,s}$ and $\Omega_{\Omega,s}$, respectively, we have $\psi(\lambda_0)=\varphi(\omega(\varphi^{-1}(\lambda_0)))$.
Since, using the conventions in eq.~\eqref{eq:metric}, the conformal factors are given by products of first derivatives of the associated conformal map, we have from the chain rule
$\psi'(\lambda_0)=\varphi'(\gamma_s)\omega'(\gamma_0)(\varphi^{-1})'(\lambda_0)
=\varphi'(\gamma_0)^{-1}\omega'(\gamma_0)\varphi'(\gamma_s)$.}
We conclude $\xi_{\Psi}$ is simply the pushforward of $\xi_{\Omega}$ under $\varphi$.

In section \ref{sec:vacuumsector}, we also claimed that any $\xi_{\Psi}$ constructed via the above procedure from the vacuum modular flow in $\mathcal{R}$ is a vacuum-sector Unruh flow in both $\mathcal{R} = D(A)$ and $D(\bar{A})$.
The conformal property is automatic, since $U_{\Psi}$ is a composition of conformal unitaries. 
The future-directed and timelike condition in $\mathcal{R}$, and the past-directed and timelike condition in $D(\bar{A})$, are guaranteed by the preservation of causal structure under conformal mappings.
Let us now check the boost-like property at the entangling surface.
Since $\varphi$ preserves $\mathcal{R}$ (and we assume it is connected to the identity), we must have $u(0)=v(0)=0$, and therefore $u(\tilde{u})=\tilde{u}u'(0)+\mathcal{O}(\tilde{u}^2)$ near the entangling surface.
Note, $u'(0)$ is finite and nonzero because we assume conformal mappings like $\varphi$ are invertible.

The pushforward equation $\tilde{c}(\tilde{u}) = \frac{c(u(\tilde{u}))}{u'(\tilde{u})}$, evaluated near $\tilde{u}=0$, then gives
\begin{equation}
\tilde{c}(\tilde{u}) = -\frac{u(\tilde{u})}{u'(\tilde{u})}
=-\frac{\tilde{u}u'(0)+\mathcal{O}(\tilde{u}^2)}{u'(0)+ \mathcal{O}(\tilde{u})}
= - \tilde{u} + \mathcal{O}(\tilde{u}^2)\,,
\label{eq:stillBoost}
\end{equation}
and similarly for $\tilde{b}(\tilde{v})$, so indeed $\varphi$ preserves the boost-like property.
By performing an inversion or special conformal map and using smoothness and invertibility of $\varphi$ at infinity, the same procedure implies that $\varphi$ preserves the vacuum-sector falloffs \eqref{eq:vacFalloffsRindler}.

As a brief aside, let us explore what goes wrong for the boost-like property when $u(\tilde{u})$ fails to be invertible.
First, consider the case where $u'(0)=0$, or more generally, where $u(\tilde{u}) = \tilde{u}^n u^{(n)}_0/n! + {\cal O}(\tilde{u}^{n+1})$ with $n>1$, with $u^{(n)}_0$ shorthand for the $n$th derivative of $u(\tilde{u})$, evaluated at $\tilde{u}=0$.
Then, we have
\begin{equation}
\tilde{c}(\tilde{u}) = -\frac{u(\tilde{u})}{u'(\tilde{u})}
=-\frac{\tilde{u}^n u^{(n)}_0/n! + {\cal O}(\tilde{u}^{n+1})}{\tilde{u}^{n-1} u^{(n)}_0 /(n-1)! + {\cal O}(\tilde{u}^{n})}
= - \frac{\tilde{u} }{n} + \mathcal{O}(\tilde{u}^2)\,.
\label{eq:notBoost}
\end{equation}
We see that the boost behavior $c(u)\sim -u$ has been rescaled by a possibly very small overall coefficient of the form $1/n$.
The case $n=3$ is considered in section \ref{sec:exRindler} and corresponds to a super-entangling map.
Secondly, consider the case where $u'(0)$ diverges, or more generally, where $\tilde{u}(u) = u^N \tilde{u}^{(N)}_0/N! + {\cal O}(u^{N+1})$ with $N>1$.
Then, we have 
\begin{equation}
\begin{aligned}
\tilde{c}(\tilde{u}) 
= -N \tilde{u}+ {\cal O}(\tilde{u}^{1+\frac{1}{N}})\,,
\end{aligned}
\end{equation}
and we see the boost behavior $c(u)\sim -u$ has been rescaled by a possibly very large overall coefficient, $N$.
This case is useful in the context of disentangling maps (see section \ref{sec:disentangling}).

Another result we used in section \ref{sec:vacuumsector} is eq.~\eqref{eq:PsiModularUnitary}, namely $U_{\Psi}(s):=U_{\varphi}\, U_{\Omega}(s)\, U_{\varphi}^\dag$, and its generalization to the case of conformal maps $\varphi$ which do not preserve the subregion of interest but rather map it into a new region.
The appropriate generalization is
\begin{equation}
U_{\Psi}^{\mathcal{V}}(s):=U_{\varphi} \,U^{\varphi^{-1}(\mathcal{V})}_{\Omega}(s)\, U_{\varphi}^\dag \,,
\label{eq:PsiModularUnitaryNewRegion}
\end{equation}
where we allow $\varphi$ to move the arbitrary causally complete region $\mathcal{V}$.
To argue for eq.~\eqref{eq:PsiModularUnitaryNewRegion}, we will show $\sigma^{\Psi, \mathcal{V}}_s(\cdot) := U_{\Psi}^{\mathcal{V}}(\cdot) U_{\Psi}^{\mathcal{V}}\,^{\dag}$ satisfies the defining periodicity property \eqref{eq:definingeq} of modular flows.
To do this, consider $\a, \b \in \mathcal{A}_{\mathcal{V}}$ and $\mathsf{A}:= U_{\varphi}^{\dag}\, \a\, U_{\varphi}$, $\mathsf{B}:=U_{\varphi}^{\dag} \,\b \,U_{\varphi}\, \in \mathcal{A}_{\varphi^{-1}(\mathcal{V})}$.
Then,
\begin{equation}
    \langle \Psi | \sigma_{s-2\pi i}^{\Psi,\mathcal{V}}(\a)\, \b |\Psi\rangle
    = \langle \Omega | \sigma_{s-2\pi i}^{\Omega,\varphi^{-1}(\mathcal{V})}(\mathsf{A})\, \mathsf{B} |\Omega\rangle
    =\langle \Omega |  \mathsf{B} \,\sigma_{s}^{\Omega,\varphi^{-1}(\mathcal{V})}(\mathsf{A}) |\Omega\rangle
    =\langle \Psi | \b\, \sigma_{s}^{\Psi,\mathcal{V}}(\a)   |\Psi\rangle\,,
    \label{eq:movingregion}
\end{equation}
where we have used that property \eqref{eq:definingeq} holds for the vacuum modular flow, by definition.

\section{\texorpdfstring{Non-uniqueness of a state with modular flow in $\mathcal{V}$}{Non-uniqueness of a state with modular flow in V}}
\label{app:connesappendix}

In this section, we observe that one can disturb a state $|\Psi\rangle$ with geometric modular flow without changing the modular flow within the lightcone of the disturbance.
To see why this is true, let $|\Phi\rangle = U_\a|\Psi\rangle$, where $U_\a$ is a unitary operator localized in either $\mathcal{V}$ or its causal complement $\mathcal{V}'$, and let $|\Psi\rangle$ be a state with geometric modular flow in $\mathcal{V}$ and $\mathcal{V}'$.

Our goal is to show that for $\b$ spacelike separated from the disturbance $U_\a$, 
\begin{equation}
    U_{\Phi}(-s)\, \b\, U_{\Phi}(s)
    = U_{\Psi}(-s)\, \b\, U_{\Psi}(s)
    \,,
    \label{eq:sameModularFlow}
\end{equation}
for sufficiently small $s$, i.e.~as long as $\b_s:=U_{\Psi}(-s)\, \b\, U_{\Psi}(s)$ remains spacelike separated from $\a$. 
To do so, note from \cite{Lashkari:2018oke} (and our independent check above) that
\begin{equation}
    U_{\Phi}(s) = U_\a U_{\Psi}(s) U_\a^{\dag}.
    \label{eq:lashk}
\end{equation}
Therefore,
\begin{equation}
    U_{\Phi}(-s)\, \b\, U_{\Phi}(s)
    = U_\a U_{\Psi}(-s) U_a^{\dag}\, \b\, U_\a U_{\Psi}(s) U_\a^{\dag}
    = U_\a \b_s U_\a^{\dag}
    \,,
\end{equation}
where we have used that $\a$ and $\b$ commute.
If further $\b_s$ is spacelike separated from $a$ (this is always true for small enough $s$), then $U_\a \b_s U_\a^{\dag}=\b_s$ and eq.~\eqref{eq:sameModularFlow} holds.
We conclude that perturbing a state $|\Psi\rangle$ with geometric modular flow by a local unitary $U_\a$ only modifies the flow within the lightcone of $U_\a$.

It would be interesting to extend this result to any local invertible operator $I_{\a}$.
In \cite{Lashkari:2018oke}, it is shown that the class of states obtained by the action of invertible operators is dense in the Hilbert space, and the corresponding modular unitaries can be constructed explicitly.
Extending our result to invertible operators would thus strongly suggest that local perturbations never modify a state outside the lightcone of the perturbation.

\subsection*{Alternative approach: Connes' cocycles}

A similar result can be obtained by perturbing the state $|\Psi\rangle$ using Connes' cocycles instead of a local unitary operator.

The converse of Connes' cocycle derivative theorem \cite{connes1973theoreme} instructs us that if $h_{\Psi}$ is the modular Hamiltonian of a cyclic, separating state $\Psi$ with respect to the subregion algebra $\mathcal{A}_{\mathcal{V}}$,\footnote{For example, the vacuum or a local, unitary excitation of the vacuum.} then choosing any two Hermitian operators $\a_{\mathcal{R}} \in \mathcal{A}_{\mathcal{V}}$ and $\a_L \in \mathcal{A}_{\mathcal{V}}'$, the operator
\begin{equation}
    h_{\Phi} = \a_L + h_{\Psi} + \a_{\mathcal{R}}
\end{equation}
is the modular Hamiltonian for some other state $\Phi$ on $\mathcal{A}_{\mathcal{V}}$.

Specializing to the case where $\a_{\mathcal{R}}=0$ while $\a_L$ has finite support in $D(\bar{A})$, we would like to show that $\Psi$ and $\Phi$ have the same modular flow in $\mathcal{V}$.
To do so, let $\b\in\mathcal{A}_{\mathcal{V}}$; then,
\begin{equation}
\label{eq: CBH}
\begin{aligned}
    e^{i h_{\Phi} s}\, \b\, e^{-i h_{\Phi} s} 
    &= e^{i (h_{\Psi} + \a_L) s}\, \b\, e^{-i (h_{\Psi} + \a_L) s} \\
    &= \b + [i (h_{\Psi} + \a_L) s, \b] + \frac{1}{2!}[i (h_{\Psi} + \a_L) s, [i (h_{\Psi} + \a_L) s, \b]] +\ldots \\
    &= \b + (is)[h_{\Psi}, \b] + (is)^2 \frac{1}{2!}[h_{\Psi} + \a_L, [h_{\Psi}, \b]] +\ldots \\
    &= \b + (is)[h_{\Psi}, \b] + (is)^2 \frac{1}{2!}[h_{\Psi}, [h_{\Psi}, \b]] +\ldots  \\
    &= e^{-i h_{\Psi} s}\, \b\, e^{i h_{\Psi} s}
    \,,
\end{aligned}
\end{equation}
where to reach the second line, we apply the Baker-Campbell-Hausdorff formula; to reach the third line, we use that $\b$ commutes with $\a_L$; and to reach the fourth line, we use that $[h_{\Psi},\b]$, $[h_{\Psi},[h_{\Psi},\b]] $, as well as the higher commutators, commute with $\a_L$ because $h_{\Psi}$ generates an automorphism that preserves $\mathcal{A}_{\mathcal{V}}$.\footnote{See similar reasoning employed in appendix D of \cite{Jensen:2023yxy}.}
Hence we can simply drop all of the $\a_L$'s in the second line. We conclude that $\Psi$ can be perturbed by an operator associated with $\mathcal{V}$'s causal complement such that the resulting modular flow matches $\xi_{\Psi}$ in $\mathcal{V}$.\footnote{We might  generalize this result to the case of $\a=\a_L$ spacelike separated from $\b$ and not in the causal complement of $\mathcal{V}$. Assuming that $h_\Psi$ has geometric action in both $\mathcal{V}$ and $\mathcal{V}'$, the evolution of $\b$ under the perturbed modular Hamiltonian $h_\Phi$ is
\begin{equation}
\label{eq: CBH2}
\begin{aligned}
    e^{i h_{\Phi} s}\, \b\, e^{-i h_{\Phi} s} 
    = e^{is\,\mathrm{ad}_{h_\Phi}}\,\b 
    = \sum_{n=0}^\infty \frac{(is)^n}{n!}\mathrm{ad}^n_{h_\Phi}\b
    \,,
\end{aligned}
\end{equation}
where we introduce the standard notation $\mathrm{ad}_h(\cdot):= [h, \,\cdot\,]$.
From $h_\Phi = h_\Psi + \a$ and $[\a,\b]=0$, we have
\begin{equation}
\label{eq: comm_id}
\begin{aligned}
      \mathrm{ad}^n_{h_\Phi}\b &= \mathrm{ad}^n_{h_\Psi}\b + \sum_{k=1}^{n-1} \mathrm{ad}_{h_\Phi}^{n-1-k}\left( \left[\a, \mathrm{ad}_{h_\Psi}^k\b\right]\right). \\
\end{aligned}
\end{equation}
To show eq.~\eqref{eq:sameModularFlow} means showing the argument in the sum in \eqref{eq: comm_id} identically vanishes for each $k$. 
To do this, we define the operator-valued real function $\sf{f}$ by
$\mathsf{f}(t) := \left[ \a, e^{ish_\Psi} \b e^{-ish_\Psi} \right]$,
which tracks whether $\b_s$ remains spacelike separated from $\a$ under $s$-evolution.
Since, by assumption, $\b$ is contained in the interior of the causal complement of the support of $\a$ and the action of $h_\Psi$ is geometric everywhere, it follows that $\sf{f}$ identically vanishes for $s\in(-\epsilon, \epsilon)$ for $\epsilon$ sufficently small. Since $\mathsf{f}$ is an analytic function of $s$,
\begin{equation}
   \left[\a, \mathrm{ad}_{h_\Psi}^k\b\right] = \frac{i^k}{k!}\, \frac{\dd^k}{\dd{t}^k}\bigg|_{t=0} \mathsf{f} =0
\end{equation}
for all positive integers $k$. We are thus led to conclude that eq.~\eqref{eq:sameModularFlow} holds. 

However, in contrast to the case of the unitary disturbance $U_\a$ discussed above, eq.~\eqref{eq:sameModularFlow} here appears to hold not only for $s$ small enough such that $\b_s$ remains spacelike separated from $\a$, but indeed for all $s$. 
In particular, $\b_s$ can enter within the lightcone of $\a$ and eq~.\eqref{eq:sameModularFlow} holds.
Either there is a problem in the reasoning above, or we must conclude that perturbing the modular Hamiltonian via $h_{\Psi}\mapsto h_{\Phi} = h_{\Psi}+\a$ has very different causal consequences from directly perturbing the state via $|\Psi\rangle \mapsto |\Phi\rangle = U_\a|\Psi\rangle$.
Note, these are indeed two very different modifications.
Assuming $U_\a$ is of the form $U_\a:=e^{i\a}$ with $\a$ hermitian, we find by eq.~\eqref{eq:lashk} that $h_{\Phi}=e^{i\a}h_{\Psi}e^{-i\a}$, which in general differs from $h_{\Phi}=h_{\Psi}+\a$.}
Note, in this proof we did not actually need that $h_\Psi$ acts geometrically.

\section{Purity from smoothness}
\label{app:areabalance}

In this appendix, we specify necessary and sufficient conditions on $\tilde{c}(\tilde{u})$ and $\tilde{b}(\tilde{v})$ for the corresponding vacuum sector Unruh flow to be the pushforward of the vacuum flow under a conformal map.
By reasoning in the main text, this suggests the existence of a state with the corresponding modular flow.
We assume only that the flow is an Unruh flow in both $\mathcal{R}$ (the right Rindler wedge) and the left Rindler wedge.

We also briefly provide analogous results for the thermal sector case.

\subsection*{Vacuum sector}
In the main text, we constructed a conformal map \eqref{eq:uvsolutionRindler} sending the vacuum modular flow to an arbitrary Unruh flow with coefficient functions $\tilde{c}(\tilde{u})$ and $\tilde{b}(\tilde{v})$.
We checked various properties of the conformal map to ensure good behavior, including continuity and invertibility.
The property that remains to be explicitly checked is sufficient smoothness at $0$ and $\infty$, which requires suitable choice of constants $\alpha$ and $\alpha'$.

We first argue that letting $\alpha'+A' = \alpha+A$, or more explicitly,
\begin{equation}
    \alpha' = \alpha +  \fint_{-1} ^{1} \frac{\rd\tilde{u}}{\tilde{c}(\tilde{u})}:= \alpha +  \lim_{\epsilon \to 0^+} \left(\int_{-1}^{-\epsilon}\frac{\rd\tilde{u}}{\tilde{c}(\tilde{u})} + \int_{\epsilon}^{1}\frac{\rd\tilde{u}}{\tilde{c}(\tilde{u})}\right)
    \label{eq:betafromalpha}
\end{equation}
ensures smoothness at the origin.
Note, the above condition simply corresponds to demanding $f_{\alpha}(\tilde{u})$ and $g_{\alpha'}(\tilde{u})$ approach each other in their respective $\tilde{u}\to0$ limits.

To check smoothness, we will make use of the following lemma:\footnote{We thank Michele Bianchessi for discussions on this lemma.}

\paragraph{Lemma (``Log-removal lemma'').} Let $f(x)$ be a continuous function on $[0,1]$, and positive on $(0,1]$ such that $f(0) = 0$.
If, as $x \to 0^+$, $f(x) = x\, f'(0) + \cO(x^2)$, then
\begin{equation}
F(\epsilon):=\int_{\epsilon}^{1}\frac{\rd x}{f(x)}
\end{equation}
obeys, as $\epsilon \to 0^+$, 
\begin{equation}
F(\epsilon)= a \log\epsilon + b + \cO(\epsilon)\,,
\end{equation}
with finite $a$, $b$ defined by 
\begin{equation}
    a := -\frac{1}{f'(0)}
    \qquad 
    b := \int_0 ^{1}\rd x \left( \frac{1}{f(x)}+ \frac{a}{x}\right)\,.
\end{equation}
\paragraph{Proof}
L'H\^opital's rule gives
\begin{equation}
    \lim_{\epsilon\to 0^+} \frac{F(\epsilon)}{\log \epsilon} 
    = \lim_{\epsilon\to 0^+} \frac{F'(\epsilon)}{1/\epsilon} 
    = \lim_{\epsilon\to 0^+}  \left(-\frac{\epsilon}{f(\epsilon)}\right)
    = -\frac{1}{f'(0)}\,,
\end{equation}
which implies
\begin{equation}
     F(\epsilon) = a \log \epsilon + f_2(\epsilon)
     \qquad \textrm{such that}\qquad
     \lim_{\epsilon\to 0^+}\frac{f_2(\epsilon)}{\log \epsilon} = 0\,.
\end{equation}
It remains to show that $f_2(\epsilon)$ takes the form $b + \cO(\epsilon)$.
To show this, notice that
\begin{equation}
    f_2(\epsilon) = F(\epsilon) - a \log \epsilon
    = \int_{\epsilon}^{1}\frac{\rd x}{f(x)} - a \int_1^\epsilon \frac{\rd x}{x}
    = \int_{\epsilon}^{1}\rd x\left(\frac{1}{f(x)} +  \frac{a}{x}\right)
    =:  \int_{\epsilon}^{1}\rd x\, g(x) \,,
\end{equation}
where the integrand $g(x)$ is clearly continuous on $(0,1]$.
Furthermore, since
\begin{equation}
\begin{aligned}
   \lim_{x \rightarrow \epsilon^+}g(x) 
   &=  \lim_{x \rightarrow \epsilon^+}\frac{1}{x f'(0) + \cO(x^2)} +  \frac{a}{x} \\
   &=  \lim_{x \rightarrow \epsilon^+}\frac{1}{x f'(0)}\left(1 + \cO(x)\right) -  \frac{1}{x f'(0)} \\
   &=  \cO(1)\,,
\end{aligned}  
\end{equation}
we find $g(x)$ can be extended by continuity to $x=0$, and hence can be integrated in $[0,1]$.

We conclude, as claimed, that
\begin{equation}
\begin{aligned}
  b= \lim_{\epsilon\rightarrow 0^+}f_2(\epsilon) 
  = \int_0 ^{1}  \rd x\, g(x) 
    =  \int_0 ^{1}\rd x \left( \frac{1}{f(x)}+ \frac{a}{x}\right)\,.
\end{aligned}
\end{equation}

Equipped with the log-removal lemma, let us now check $u(\tilde{u})$ has the desired smoothness properties.

\paragraph{Differentiability at zero.}
The limit from the right gives
\begin{equation}
\begin{aligned}
\lim_{\epsilon \rightarrow 0^+}u'(\epsilon)= \lim_{\epsilon \rightarrow 0^+} f_{\alpha}'(\epsilon) e^{f_{\alpha}(\epsilon)} = \lim_{\epsilon\rightarrow 0^+} \frac{-e^{\alpha-\int^\epsilon_1\frac{\rd\tilde{u}}{\tilde{c}(\tilde{u})}}}{\tilde{c}(\epsilon)}  = \lim_{\epsilon\rightarrow 0^+} \frac{-e^{\alpha+ \log\epsilon + A}}{-\epsilon + \mathcal{O}(\epsilon^2)}  = e^{\alpha + A} \,,
\end{aligned}
\label{eq:uprimePositiveLimit}
\end{equation}
where $A$ is the finite constant provided in eq.~\eqref{eq:uprime0}, by the log-removal lemma.

Similarly, the limit from the left gives
\begin{equation}
\begin{aligned}
\lim_{\epsilon\rightarrow 0^-} u'(\epsilon) 
= \lim_{\epsilon\rightarrow 0^-}-g_{\alpha'}'(\epsilon) e^{g_{\alpha'}(\epsilon)} = \lim_{\epsilon\rightarrow 0^+} \frac{e^{g_{\alpha'}(-\epsilon)}}{\tilde{c}(-\epsilon)}  = \lim_{\epsilon\rightarrow 0^+} \frac{e^{\alpha'-\int^{-\epsilon}_{-1}\frac{\rd\tilde{u}}{\tilde{c}(\tilde{u})}}}{ \epsilon + \cO(\epsilon^2)}  =e^{\alpha'+ A'}
\end{aligned}
\end{equation}
where again $A'$ is a finite constant provided in eq.~\eqref{eq:uprime0}.
We conclude that differentiability requires precisely $\alpha + A = \alpha'+A'$, which is eq.~\eqref{eq:betafromalpha}.

\paragraph{2$\times$ and 3$\times$-differentiability at zero.}
The second derivative of $u(\tilde{u})$ is
\begin{equation}
u'' = \begin{cases}
(e^{f_{\alpha}})'' 
= f_{\alpha}' e^{f_{\alpha}}\left(\frac{f_{\alpha}''}{f_{\alpha}'} + f_{\alpha}'\right)\,, 
 &\tilde{u}>0 \\
(-e^{g_{\alpha'}})'' 
= -g_{\alpha'}' e^{g_{\alpha'}} \left(\frac{g_{\alpha'}''}{g_{\alpha'}'} + g_{\alpha'}'\right)\,, 
 &\tilde{u}<0\,.
\end{cases}
\end{equation}
Differentiability at zero implies $f_{\alpha}' e^{f_{\alpha}}$ varies continuously to $-g_{\alpha'}' e^{g_{\alpha'}}$ across $\tilde{u}=0$.
Further, in the $\tilde{u}\to0$ limit, l'H\^opital's rule gives
\begin{equation}
\begin{aligned}
\frac{f_{\alpha}''}{f_{\alpha}'} + f_{\alpha}' 
= -\frac{\tilde{c}'+1}{\tilde{c}} 
\to -\frac{\tilde{c}''}{\tilde{c}'} 
\to \tilde{c}'' \,,
\end{aligned}
\end{equation}
and similarly for $\frac{g_{\alpha'}''}{g_{\alpha'}'} + g_{\alpha'}'$, so that smoothness of $\tilde{c}(\tilde{u})$ directly implies twice-differentiability of $u(\tilde{u})$.
A similar trick gives thrice-differentiability.
We can explicitly demonstrate that many more derivatives are well-defined, and speculate that all of them are.

\paragraph{Differentiability at $\mathbf{\infty}$.}
We check this property by conjugating the conformal map $(u(\tilde{u}), v(\tilde{v}))$ by the inversion map, and then ensuring smoothness and invertibility at zero.
In $\tilde{u},\tilde{v}$ coordinates, the inversion map $I$ is $\tilde{u}\rightarrow -\frac{1}{\tilde{v}}$ and $\tilde{v}\rightarrow -\frac{1}{\tilde{u}}$.
Using $(u \circ I)(\tilde{u},\tilde{v}) = u(-\frac{1}{\tilde{v}},-\frac{1}{\tilde{u}}) = u(-\frac{1}{\tilde{v}})$, composition with $I$ gives
\begin{equation}
\begin{aligned}
    u \circ I = \begin{cases}
    e^{f_\alpha(-\frac{1}{\tilde{v}})}, &\tilde{v}<0\\
    -e^{g_{\alpha'}(-\frac{1}{\tilde{v}})}, &\tilde{v}>0\,,
    \end{cases}
\end{aligned}
\end{equation}
and $I\circ u \circ I$ is
\begin{equation}
\begin{aligned}
    I \circ u \circ I = \begin{cases}
    -e^{-f_\alpha(-\frac{1}{\tilde{v}})}, &\tilde{v}<0\\
    e^{-g_{\alpha'}(-\frac{1}{\tilde{v}})}, &\tilde{v}>0\,.
    \end{cases}
\end{aligned}
\end{equation}
Note, from this equation, continuity is automatic; as $\tilde{v}\rightarrow 0$ from either side, $f_{\alpha}$ or $g_{\alpha'}$ diverges to $+\infty$ such that $I \circ v \circ I \rightarrow 0$.
For differentiability, we have
\begin{equation}
\begin{aligned}
    \left(I \circ u \circ I\right)'(\tilde{v}) = \frac{1}{\tilde{v}^2 \tilde{c}\left(-\frac{1}{\tilde{v}}\right)}\begin{cases}
    -e^{-f_\alpha(-\frac{1}{\tilde{v}})}, &\tilde{v}<0\\
    e^{-g_{\alpha'}(-\frac{1}{\tilde{v}})}, &\tilde{v}>0\,.
    \end{cases}
\end{aligned}
\end{equation}
Using similar reasoning that led to eq.~\eqref{eq:betafromalpha}, $I\circ v\circ I$ is differentiable at zero if and only if
\begin{equation}
    \alpha' = \alpha -  \lim_{\epsilon \rightarrow 0^+}  \left(\int_{-\frac{1}{\epsilon}}^{-1}\frac{\rd\tilde{u}}{\tilde{c}(\tilde{u})}+ \int_1^{\frac{1}{\epsilon}}\frac{\rd\tilde{u}}{\tilde{c}(\tilde{u})} \right)\,.
    \label{eq:betafromalpha2}
\end{equation}
Consistency with eq.~\eqref{eq:betafromalpha} requires the purity 
condition \eqref{eq:purity}.

To explicitly calculate the slope at infinity, let $\tilde{v} = -\epsilon$, with $\epsilon>0$; then, we have
\begin{equation}
\begin{aligned}
    \left(I \circ u \circ I\right)'(0) 
    = \lim_{\epsilon \to 0^+}\frac{e^{-f_\alpha(\frac{1}{\epsilon})}}{-\epsilon^2 \tilde{c}\left(\frac{1}{\epsilon}\right)}
    = \lim_{\epsilon \to 0^+}\frac{e^{-\alpha+ \int_1^{\frac{1}{\epsilon}}\frac{\rd\tilde{u}}{\tilde{c}(\tilde{u})}}}{-\epsilon^2 \tilde{c}\left(\frac{1}{\epsilon}\right)}
    = \lim_{\epsilon \to 0^+}\frac{e^{-\alpha - A_{\infty} - \int_1^{\frac{1}{\epsilon}}\frac{\rd\tilde{u}}{\tilde{u}}}}{\epsilon + \cO(\epsilon^2)}
    = 
    e^{-\alpha-A_{\infty}}\,,
\end{aligned}
\end{equation}
justifying eq.~\eqref{eq:inftyslope}.
Note, finiteness of $A_{\infty}$ follows from a straightforward generalization of the log-removal lemma, and relies on the vacuum-sector falloffs \eqref{eq:vacFalloffsRindler}.

\paragraph{2$\times$ and 3$\times$-differentiability at $\mathbf{\infty}$.} This can be checked evaluating
\begin{equation}
\begin{aligned}
    \left(I \circ u \circ I\right)''(\tilde{v}) = \frac{-1 + 2 \tilde{v} \tilde{c}(-\frac{1}{\tilde{v}}) + 
    \tilde{c}'(-\frac{1}{\tilde{v}})}{\tilde{v}^4 \tilde{c}\left(-\frac{1}{\tilde{v}}\right)^2}
    \begin{cases}
    e^{-f_\alpha(-\frac{1}{\tilde{v}})}, 
    &\tilde{v}<0\\
    -e^{-g_{\alpha'}(-\frac{1}{\tilde{v}})}, &\tilde{v}>0\,.
    \end{cases}
\end{aligned}
\end{equation}
Vacuum-sector asymptotics implies we can expand $\tilde{c}(-\frac{1}{\tilde{v}})$ as $\frac{1}{\tilde{v}} + \alpha_1 + \alpha_2 \tilde{v} + \cO\left(\tilde{v}^2\right)$.
The assumption that $\tilde{c}(\tilde{u})$ is three-times differentiable at infinity guarantees the constants $\alpha_1$ and $\alpha_2$ in the $\tilde{v}\to 0^+$ expansion are the same as those in the $\tilde{v}\to 0^-$ expansion.
Using also eq.~\eqref{eq:betafromalpha2}, we find
\begin{equation}
    \left(I \circ u \circ I\right)''(0) = -\frac{\alpha_1}{\pi}e^{-\alpha - A_{\infty}}\,,
\end{equation}
guaranteeing twice-differentiability.
A similar calculation shows thrice-differentiability.

\section*{Thermal sector}
\label{app:thermalApp}

In this section, starting from eq.~\eqref{eq:uvsolutionThermal}, that is,
\begin{equation}
    u(\tilde{u}) = -\frac{\beta}{2\pi}
    \begin{cases}
        \log\left(1-e^{f_{\alpha}(\tilde{u})}\right),
        &\tilde{u}>0\\
        \log\left(1+ e^{g_{\alpha'}(\tilde{u})}\right),
        &\tilde{u}<0\,,
    \end{cases}
    \label{eq:uvsolutionThermal2}
\end{equation}
we show $\alpha$ and $\alpha'$ can be chosen to ensure $u(\tilde{u})$ behaves as described in section \ref{sec:thermalsector}.

\paragraph{Continuity at zero.}
Recall that eq.~\eqref{eq:uvsolutionRindler} continuously crosses zero as $\tilde{u}\to 0$ due to the boost-like property.
Similar reasoning applies here. 

\paragraph{Differentiability at zero.}
To check differentiability, note
\begin{equation}
    u'(\tilde{u}) = \frac{\beta}{2\pi \tilde{c}(\tilde{u})}
    \begin{cases}
        \frac{1}{1-e^{-f_{\alpha}(\tilde{u})}}\,, & \tilde{u}>0 \\
        \frac{1}{1+ e^{-g_{\alpha'}(\tilde{u})}}\,, & \tilde{u}<0 \,.
    \end{cases}
\end{equation}
Replicating eq.~\eqref{eq:uprimePositiveLimit}, we have in the $\tilde{u}\to 0^+$ limit
\begin{equation}
\begin{aligned}
\lim_{\epsilon \rightarrow 0^+}u'(\epsilon)
= \frac{\beta}{2\pi}\lim_{\epsilon \rightarrow 0^+}  \frac{1}{\tilde{c}(\epsilon)(1-e^{-f_{\alpha}})} 
= - \frac{\beta}{2\pi}\lim_{\epsilon \rightarrow 0^+}  \frac{e^{f_{\alpha}}}{\tilde{c}(\epsilon)}
= -\frac{\beta}{2\pi}\lim_{\epsilon \rightarrow 0^+}  \frac{e^{\alpha- \int_1^{\epsilon}\frac{\rd\tilde{u}}{\tilde{c}(\tilde{u})}}}{\tilde{c}(\epsilon)}\, .
\end{aligned}
\label{eq:uprimePositiveLimitThermal}
\end{equation}
Defining
\begin{equation}
    A = -\int_{1}^{0}\rd\tilde{u}\left(\frac{1}{\tilde{c}(\tilde{u})}-\frac{1}{c(\tilde{u})}\right),
\end{equation}
we can process this further and get
\begin{equation}
  \lim_{\epsilon \rightarrow 0^+}u'(\epsilon)  
  = -\frac{\beta}{2\pi}\lim_{\epsilon \rightarrow 0^+}  \frac{e^{\alpha+A -\int_1^{\epsilon}\frac{\rd\tilde{u}}{c(\tilde{u})}} }{\tilde{c}(\epsilon)}
= \frac{\beta}{2\pi} \lim_{\epsilon\rightarrow 0^+} \frac{e^{\alpha+  A 
+ \log \frac{2\pi \epsilon}{\beta} - \log\left(-1+e^{\frac{2\pi}{\beta}}\right)}}{\epsilon + \mathcal{O}(\epsilon^2)} = \frac{e^{\alpha + A}}{1-e^{-\frac{2\pi}{\beta}}}\,.
\label{eq:upNLT}
\end{equation}
In the $\tilde{u}\to 0^-$ limit, a similar computation leads to
\begin{equation}
\lim_{\epsilon \rightarrow 0^-}u'(\epsilon)= \frac{\beta}{2\pi}\lim_{\epsilon \rightarrow 0^+}\frac{1}{\tilde{c}(-\epsilon)(1+e^{-g_{\alpha'}(-\epsilon)})} 
= \frac{\beta}{2\pi}\lim_{\epsilon \rightarrow 0^+}  \frac{e^{\alpha'+A' -\int_{-1}^{-\epsilon}\frac{\rd\tilde{u}}{c(\tilde{u})}}}{\tilde{c}(-\epsilon)}
 = \frac{e^{\alpha' + A'}}{e^{\frac{2\pi}{\beta}}-1}\,,
\label{eq:uprimeNegativeLimitThermal}
\end{equation}
where we defined
\begin{equation}
    A' = -\int_{-1}^{0}\rd\tilde{u}\left(\frac{1}{\tilde{c}(\tilde{u})}-\frac{1}{c(\tilde{u})}\right)\,.
\end{equation}
Therefore, differentiability of $u(\tilde{u})$ at zero requires
\begin{equation}
    \alpha +A + \frac{2\pi}{\beta} = A'+\alpha'\,,
\end{equation}
as shown in the main text.
We conclude $u'(\tilde{u})$ is differentiable at $\tilde{u}=0$, with $u'(0) = \frac{e^{\alpha'+A'}}{e^{\frac{2\pi}{\beta}}-1}$.

\paragraph{2$\times$ differentiability at zero.}
We have
\begin{equation}
    u''(\tilde{u}) = \frac{\beta}{2\pi}
    \begin{cases}
        -\frac{\tilde{c}' - (1 + \tilde{c}')e^{-f_{\alpha}}}{\left(\tilde{c} -  \tilde{c}e^{-f_{\alpha}}\right)^2}, &\tilde{u}>0,  \\
        - \frac{\tilde{c}' +(1 + \tilde{c}')e^{-g_{\alpha'}}}{\left(\tilde{c} +  \tilde{c}e^{-g_{\alpha'}}\right)^2}, &\tilde{u}<0\,.     
    \end{cases}
\end{equation}
Here, it is useful to repackage eqs.~\eqref{eq:uprimeNegativeLimitThermal} and \eqref{eq:upNLT} into the following form:
\begin{equation}
    \begin{cases}
        \tilde{c}e^{-f} \to -\frac{\beta}{2\pi u'(0)},  &\tilde{u}\to 0^+ \\
        \tilde{c}e^{-g} \to +\frac{\beta}{2\pi u'(0)}, &\tilde{u}\to 0^-\,.
    \end{cases}
\end{equation}
Using this together with l'H\^opital's rule, we find that $u(\tilde{u})$ is twice-differentiable and
\begin{equation}
    u''(\tilde{u}) = u'(0) \left(\frac{2\pi}{\beta}u'(0)+\tilde{c}''\right)\,.
\label{eq:uprimeprimeNearZero}
\end{equation}

\paragraph{3$\times$ differentiability at zero.}
Using the methods described above, we find
\begin{equation}
    u'''(0) = \frac{1}{2}u'(0)
   \left[\left(\frac{4\pi}{\beta} u'(0)\right)^2 + \frac{12\pi}{\beta}u'(0) \tilde{c}'' +3\tilde{c}''^2 + \tilde{c}'''\right]\,.
\end{equation}
from both the left and the right.

\paragraph{Invertibility.}
The Unruh property of $\tilde{c}(\tilde{u})$ implies $u(\tilde{u})$ is strictly increasing on $(\infty,0) \cup (0, \infty)$; the positive slope at zero guarantees invertibility on $(-\infty, \infty)$.
We now check whether the inverse function is supported on the entire real line.
By the thermal sector property, $\lim_{\tilde{u}\rightarrow - \infty}\tilde{c}(\tilde{u}) = \frac{\beta}{2\pi}$ is a constant.
By an analogue of the log-removal lemma, $f_{\alpha}(\tilde{u})$ and $g_{\alpha'}(\tilde{u})$ behave in the $\tilde{u}\rightarrow \pm \infty$ limits as 
\begin{equation}
    f_{\alpha}(\tilde{u})
    \to \alpha + A_{\infty}  - \log(1-e^{-\frac{2\pi}{\beta}})
     \qquad
    g_{\alpha'}(\tilde{u})
    \to \alpha' + A_{-\infty} - \frac{2\pi}{\beta}\tilde{u} - \log(-1+e^{\frac{2\pi}{\beta}})\,,
    \label{eq:fgtou}
\end{equation}
respectively, where
\begin{equation}
    A_{\infty} = -\int_{1}^{\infty}\rd\tilde{u}\left(\frac{1}{\tilde{c}(\tilde{u})}-\frac{1}{c(\tilde{u})}\right)
    \qquad
    A_{-\infty} = -\int_{-1}^{-\infty}\rd\tilde{u}\left(\frac{1}{\tilde{c}(\tilde{u})}-\frac{1}{c(\tilde{u})}\right)\,,
\end{equation}
and the thermal sector falloffs \eqref{eq:thermalFalloffs} are essential to guarantee finiteness of $A_{-\infty}$.

It follows that, up to $\cO(\frac{1}{\tilde{u}})$,
\begin{equation}
u(\tilde{u}) \to 
\begin{cases}
    -\frac{\beta}{2\pi}\log\left(1-\frac{e^{\alpha+A_{\infty}}}{1-e^{-\frac{2\pi}{\beta}}}\right), &\tilde{u}\to \infty  \\
     \tilde{u}  +\frac{\beta}{2\pi}\left[-\alpha'- A_{-\infty} + \log(-1+e^{\frac{2\pi}{\beta}})\right] , &\tilde{u}\to -\infty \,,
\end{cases}
\label{eq:uThermalNearInf}
\end{equation}
and in particular, at leading order, $u(\tilde{u})\to - \infty$ as $\tilde{u} \rightarrow - \infty$, while $u(\tilde{u})$ saturates to a constant value for $\tilde{u}\to\infty$ unless $\alpha + A_{\infty} = \log(1-e^{-\frac{2\pi}{\beta}})$, which was the choice made in the main text (eq.~\eqref{eq:alphafromgen}).

Given this choice, the leading terms in the $f_{\alpha}$ falloffs \eqref{eq:fgtou} cancel, leaving the subleading term
\begin{equation}
    f_{\alpha}(\tilde{u})
    \to - e^{-\frac{2\pi \tilde{u}}{\beta}}\,,
    \label{eq:fgtou2}
\end{equation}
and hence eq.~\eqref{eq:uThermalNearInf} is modified to
\begin{equation}
u(\tilde{u}) \to 
\begin{cases}
     \tilde{u}\,, &\tilde{u}\to \infty  \\
     \tilde{u}  +\frac{\beta}{2\pi}\left[-\alpha'- A_{-\infty} + \log(-1+e^{\frac{2\pi}{\beta}})\right] , &\tilde{u}\to -\infty \,.
\end{cases}
\label{eq:inspire}
\end{equation}

\paragraph{Smoothness across infinity?}
From the preceding results, we see naively demanding smoothness at infinity would require
\begin{equation}
    \alpha'+ A_{-\infty} = \log(-1+e^{\frac{2\pi}{\beta}})\,.
\end{equation}
Consistency with eqs.~\eqref{eq:alphafromgen} and eq.~\eqref{eq:alpha2pibeta} requires $A- A_{\infty} = A'-A_{-\infty}$, which is equivalent to the vacuum-sector purity condition \eqref{eq:purity}.
Due to the discussion around eq.~\eqref{eq:notpure}, we do not expect this condition to hold for thermal-sector states.

As suggested in the main text, smoothness across infinity is not the right condition to impose for the thermal sector states.
Rather, we should impose smoothness across the two junctions at infinity between the original and the auxiliary theories in the thermofield double.

To do this, first solve the analog of eq.~\eqref{cases} in the thermofield copy, namely
\begin{equation}
    -\frac{2\pi}{\beta}\frac{\rd u}{1+e^{-\frac{2\pi u}{\beta}}} =
     \frac{\rd \tilde{u} }{\tilde{c}(\tilde{u})}
     \qquad
\end{equation}
where we have substituted $c(u)$ from eq.~\eqref{eq:copyflow}.
We will assume $\tilde{c}<0$ everywhere, in accordance with the flow being past-directed in the complement of $\mathcal{R}$.
The solution reads
\begin{equation}
    u(\tilde{u}) = \frac{\beta}{2\pi}\log\left(-1+ e^{h_{\alpha''}(\tilde{u})}\right)
\end{equation}
where
\begin{equation}
    h_{\alpha''}(\tilde{u}):= \alpha''- \int_{-\infty}^{\tilde{u}}\frac{\rd \tilde{u}^*}{\tilde{c}(\tilde{u}^*)}\,.
\end{equation}
Introducing
\begin{equation}
    A'' = -\int_{-\infty}^{\infty}\rd\tilde{u}
\left(\frac{1}{\tilde{c}(\tilde{u})}-\frac{1}{c(\tilde{u})}\right)\,,
\end{equation}
we find that as $\tilde{u}\to +\infty$:
\begin{equation}
    h_{\alpha''}\to \alpha'' +  A'' +\frac{2\pi}{\beta}\tilde{u}
\end{equation}
and as $\tilde{u}\to -\infty$,
\begin{equation}
    h_{\alpha''}\to \alpha''\,.
\end{equation}
Consequently,
\begin{equation}
    u(\tilde{u})\to \begin{cases}
        \tilde{u}+\frac{\beta}{2\pi}(\alpha''+A'')\, & \tilde{u}\to\infty \\
        \frac{\beta}{2\pi}\log(e^{\alpha''}-1)\,, & \tilde{u}\to -\infty\,.
    \end{cases}
\end{equation}
To ensure that $u(\tilde{u})$ spans the entire real line, we must set $\alpha''=0$.
In this case,
\begin{equation}
    u(\tilde{u})\to \begin{cases}
        \tilde{u}+\frac{\beta}{2\pi}A''\, & \tilde{u}\to\infty \\
        \tilde{u} \,, & \tilde{u}\to -\infty\,.
    \end{cases}
\end{equation}
Now, consulting eq.~\eqref{eq:inspire} we conclude we have the desired matching as long as
\begin{equation}
    A'' + (A-A_{\infty}) - (A'-A_{-\infty}) = 0\,,
\end{equation}
which translates to eq.~\eqref{eq:purityofpurification} in the main text.

\paragraph{Connectedness to identity.} This can be shown by employing similar steps to those discussed around eq.~\eqref{eq:connectedToId}.

\section{Entropy derivation from conformal mapping to annulus}
\label{app:annulus}

In this section, we confirm the entanglement entropy formula in eq.~\eqref{eq: EE Wong} with $\beta:=2\pi \xi_{\Psi}^t$ for our vacuum sector and thermal sector states satisfying $\tilde{b}(\tilde{v})=\tilde{c}(-\tilde{v})$.
To do this, we use the following entanglement entropy formula from \cite{Cardy:2016fqc}:
\begin{equation}
    S(\rho^{\mathcal{V}}) = \frac{c}{6}\,W + \cO(1)\,.
    \label{eq:CardyTonniFormula}
\end{equation}
This formula holds whenever $\rho^{\mathcal{V}}$ is prepared by a path integral on a Euclidean manifold conformally equivalent to the annulus, or more precisely, a rectangle periodically identified mod$(2\pi)$ in the Euclidean time direction, $\tau$.
Then, $W$ is the width of that annulus.
The $\cO(1)$ terms depend on boundary conditions in the regularization.

\section*{Vacuum sector}
\label{sec:annulusVacuum}

When $\mathcal{V}={\cal R}$ is the Rindler wedge and $\rho_{\Omega}$ is the Rindler density matrix associated with the Minkowski vacuum state, then the Euclidean state-preparation manifold is a (complex) plane parameterized by $z_{\mathcal{R}}:=x_{\mathcal{R}} + i\tau_{\mathcal{R}}$ with a slit running from $x_{\mathcal{R}} = 0$ to $x_{\mathcal{R}} = \infty$.
We are introducing the $\mathcal{R}$ subscript to emphasize that this is a coordinate on the Euclidean Rindler manifold, and not on the annulus.
Following \cite{Cardy:2016fqc}, introducing a UV cutoff corresponds to removing a circle of radius $r_{\min} = \epsilon$ centered at $z_{\mathcal{R}} = 0$; similarly, we introduce an IR cutoff by removing the exterior of the circle of radius $r_{\max} = L$.

The map from this Euclidean state-preparation manifold to the annulus is given by
\begin{equation}
    z_A = \log z_{\mathcal{R}}\,,
    \label{eq:RtoAcomplex}
\end{equation}
where $z_A:=x_A + i\tau_A$, and similarly for $z_{\mathcal{R}}$.
This map takes circles to the horizontal lines $\tau_A = \theta_{\mathcal{R}}$, and radial lines to vertical lines $x_A = \log r_{\mathcal{R}}$.
Hence, as desired, we obtain $2\pi$-periodicity in the $\tau_A$ direction and an annulus of width 
\begin{equation}
    W=\log r_{\max} - \log r_{\min}\,,
\end{equation}
yielding
\begin{equation}
S(\rho_{\Omega})=\frac{c}{6}\log\frac{L}{\epsilon}+ \cO(1)\,.    
\label{eq:beforeassumps}
\end{equation}
This quantity agrees with result presented in eq.~\eqref{eq: EE Wong}. 

To compute the entanglement entropy of the excited state $|\Psi \rangle$ associated with the conformal map $\varphi:(u,v) \to(\tilde{u},\tilde{v})$,
we will assume that:
\begin{itemize}
    \item $\varphi:(u,v) \to(\tilde{u},\tilde{v})$ admits a straightforward analytic extension via the formal substitution $u\to -z$ and $\tilde{u}\to -\tilde{z}$, denoted $\tilde{z}(z_{\mathcal{R}})$, 
    \item The inverse map $z_{\mathcal{R}}(\tilde{z})$ sends the Euclidean manifold\footnote{Note, this manifold will have a flat metric but possibly in curvilinear coordinates, see \eg eq.~\eqref{eq:metric}.}  preparing $\ket{\Psi}$ to a Rindler-vacuum-preparation manifold whose UV/IR boundaries we again label $r_{\min}$, $r_{\max}$.
\end{itemize}
If these assumptions are met, the composite map $z_A(\tilde{z}) := \log z_{\mathcal{R}}(\tilde{z})$ provides the desired map from the excited-state-preparation manifold to the annulus, and the entanglement entropy is thus again determined by $W = \log \frac{r_{\max}}{r_{\min}}$.

To solve for $r_{\max}$ and $r_{\min}$, first note that consistency demands the IR and UV regulator surfaces in the Euclidean manifold preparing $\rho^{\mathcal{R}}_{\Psi}$ again lie at $\tilde{r}_{\max} = L$ and $\tilde{r}_{\min} = \epsilon$.
These circles intersect $\tilde{\tau}=0$ at, for instance, $\tilde{z}= \tilde{z}^* = \tilde{r}_{\max}$, which in the Lorentzian setting corresponds to
\begin{equation}
-\tilde{u}=\tilde{v}
=\frac{1}{2}\left(\tilde{v}-\tilde{u}\right)
=\tilde{r}_{\max}\,.
\label{eq:uvrmax}
\end{equation}
If we assume that $z(\tilde{z})$ acts as a dilatation at very large and very small $\tilde{r}$, and that $u(\tilde{u})$ is odd, then 
\begin{equation}
    r_{\max} = \frac{1}{2} (v(\tilde{r}_{\max})-u(-\tilde{r}_{\max}))
    = \frac{1}{2} (v(\tilde{r}_{\max})+u(\tilde{r}_{\max}))
\end{equation}
and similarly for $r_{\min}$.
Thus,
\begin{equation}
    S(\rho^{\mathcal{R}}_{\Psi}) = \frac{c}{6}\log \frac{u(L)+v(L)}{u(\epsilon)+v(\epsilon)} \,.
    \label{eq:preEntropy}
\end{equation}

The assumption that $z_{\mathcal{R}}(\tilde{z})$ acts as a dilatation at large/small $\tilde{r}$ is partially justified by the boost property of the Unruh flow, which guarantees that $u(\tilde{u})$ becomes linear for large and small $|\tilde{u}|$, together with the assumption that $\tilde{u}(u)$ admits a straightforward analytic continuation via $\tilde{u}\to-\tilde{z}$, etc.
However, in general, the scaling factors at large and small $\tilde{r}$ can differ in the $\tilde{u}$ and $\tilde{v}$ directions.
Recalling from section \ref{sec:vacuumsector} that 
\begin{equation}
\begin{cases}
    u'(0) = e^{\alpha + A} \\
    u'(\pm\infty) = e^{\alpha+A_{\infty}}\end{cases}
    \qquad 
    \begin{cases}
    v'(0) = e^{\beta + B} \\
    v'(\pm\infty) = e^{\beta+B_{\infty}}\,,
    \end{cases}
\end{equation}
we see that our formula eq.~\eqref{eq:preEntropy} only applies for the subset of maps with 
\begin{equation}
    \alpha + A = \beta + B
    \qquad 
    \alpha + A_{\infty} = \beta + B_{\infty}\,.
    \label{eq:ABrelation}
\end{equation}
This is automatically true if $\tilde{b}(\tilde{v}) = \tilde{c}(-\tilde{v})$, as must also be assumed to use the formula eq.~\eqref{eq: EE Wong} from \cite{Wong:2013gua}.

Linearizing terms in eq.~\eqref{eq:preEntropy} via, e.g. $u(\epsilon) \approx \epsilon\, e^{\alpha + A}$, 
we obtain
\begin{equation}
    S(\rho^{\mathcal{R}}_{\Psi})= \frac{c}{6}\left[\log\frac{L}{\epsilon} + A_{\infty}-A\right]+ \cO(1) \,,
\end{equation}
in agreement with eq.~\eqref{eq:ABentropy}, which is equivalent to eq.~\eqref{eq:FINITE_ENTROPY}.

One final comment on this derivation is that in general, the surfaces used to regulate the Euclidean state-preparation manifolds only furnish proper cutoffs if they are tangent to the modular flow.
Otherwise, evolving along the modular flow would change the regularization scheme.
We expect tangency here because our starting point is the vacuum modular flow with concentric circles as the cutoffs.
These circles are tangent to the modular flow, since a Lorentzian boost is a Euclidean rotation.
The assumptions below eq.~\eqref{eq:uvrmax} guarantee that, from here, we are simply performing a Euclidean conformal map, which preserves the property that a surface is tangent to a vector field.

\section*{Thermal sector}
\label{sec:annulusThermal}

We now compute the entanglement entropy of the thermal state $|\beta\rangle$ using the Cardy-Tonni formula \eqref{eq:CardyTonniFormula}; later on, we consider the excited states $|\Psi\rangle = U_{\varphi}|\beta\rangle$.

\begin{figure}
    \centering
    \includegraphics[width=.7\linewidth]{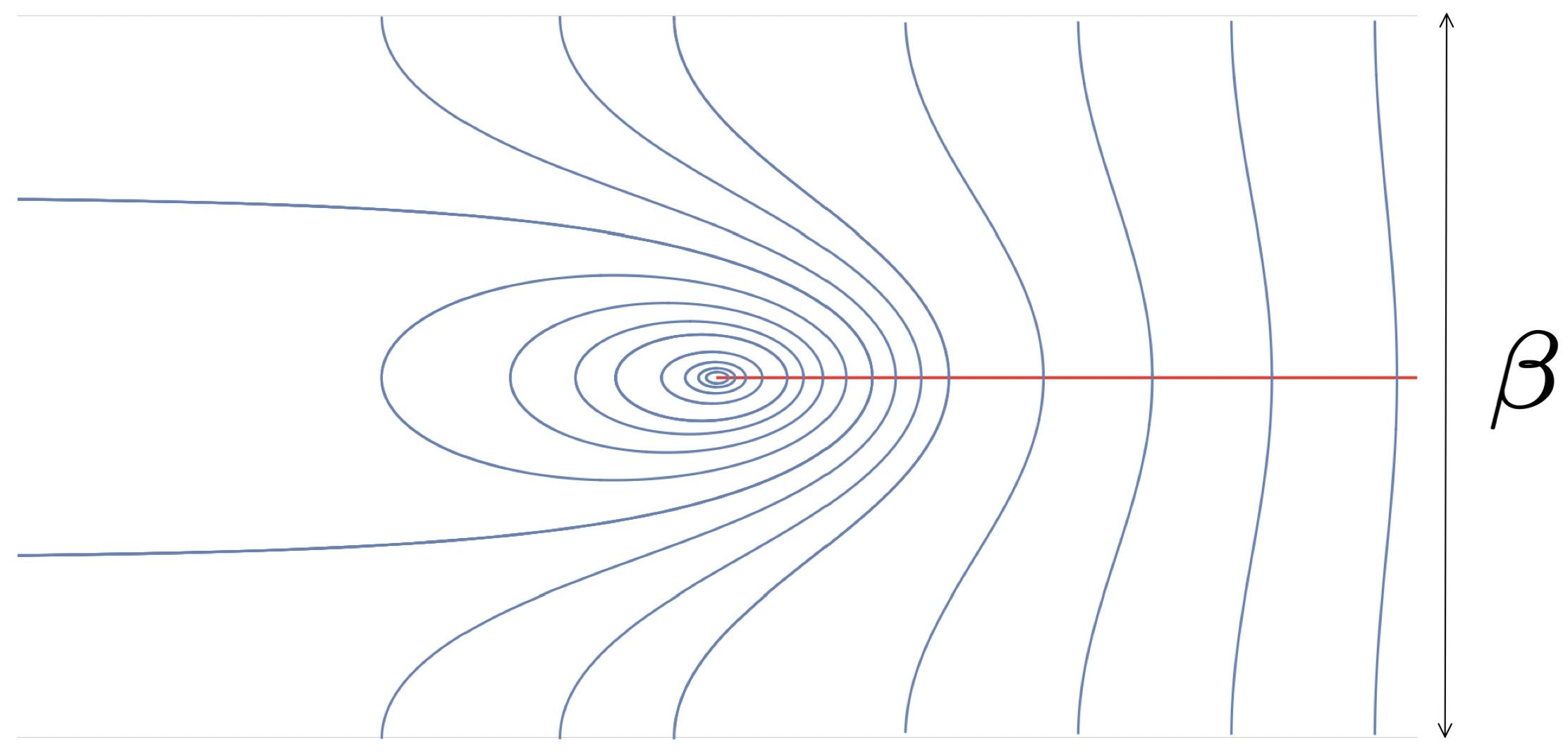}
    \caption{Euclidean state-preparation manifold for the thermal state density matrix on the $t=0$ slice of the Rindler wedge (red).
    Under the conformal map in eq.~\eqref{eq:confmap1}, constant Re$(z_A)$ slices correspond to the blue lines shown.}
    \label{fig:2confmaps}
\end{figure}

To use the formula, we must find a conformal mapping between the annulus and the Euclidean manifold preparing $\rho_\beta^\mathcal{R}$.
This density matrix is prepared by a path integral on a cylinder of circumference $\beta$, cut open along a semi-infinite line.
We define a complex coordinate $z_{\mathcal{R}} = x_{\mathcal{R}} + i \tau_{\mathcal{R}}$ on the cylinder, where $\tau_{\mathcal{R}}\sim \tau_{\mathcal{R}}+\beta$ runs around the cylinder, $x_{\mathcal{R}}$ runs along the cylinder, and the slit runs from $x_{\mathcal{R}}=0$ to $x_{\mathcal{R}}=\infty$ at $\tau_{\mathcal{R}} = 0$.
A mapping to the annulus is given by\footnote{This map can be obtained by taking an appropriate limit of eq.~(31) in \cite{Cardy:2016fqc}.}
\begin{equation}
    z_A(z_{\mathcal{R}})=\log ( e^{\frac{2\pi z_{\mathcal{R}}}{\beta}} - 1)\,,
    \label{eq:confmap1}
\end{equation}
and is illustrated in figure \ref{fig:2confmaps}.\footnote{See Figure 5 of \cite{Mintchev:2022fcp} for the corresponding flow for a finite interval in a thermal state.}

The image of the cut cylinder under eq.~\eqref{eq:confmap1} is an annulus of infinite width, corresponding to UV and IR divergences in the thermal entanglement entropy.
To regularize, we remove a disk of radius $\epsilon$ around $z_{\mathcal{R}} = 0$, and we truncate the Euclidean manifold computing $\rho_\beta^\mathcal{R}$ at a level curve of Re$(z_A)$ which intersects $\tau_{\mathcal{R}}=0$ at $x_{\mathcal{R}}=L$. 
As in the previous section, it can be seen that the modular flow runs tangent to these cutoff surfaces.\footnote{Explicitly, recall that the modular flow corresponds to time translations on the annulus. From figure \ref{fig:2confmaps}, the orbit of time translations on the annulus correspond to the blue lines shown, with respect to which the cutoff surfaces are defined. } The image of the regularized manifold is an annulus of width 
\begin{equation}
    \label{eq: width thermal annulus}
    W= z_A(L) - z_A(\epsilon) \simeq \frac{2\pi}{\beta}L-\log (\frac{2\pi \epsilon}{\beta}),
\end{equation}
yielding 
\begin{equation}
    \label{eq: EE thermal state, rindler, cardy}
    S(\rho_\beta^\mathcal{R})= \frac{c}{6}\left[\frac{2\pi L}{\beta}-\log (\frac{2\pi\epsilon}{\beta})\right]+\cO(1)\,,
\end{equation}
which agrees with eq.~\eqref{eq:thermalEntropyDiff} in the main text.

Consider now $\ket{\Psi}=U_\varphi \ket{\beta}$. 
By the assumptions under eq.~\eqref{eq:beforeassumps}, the Riemmanian manifold preparing $\rho_\Psi^\mathcal{R}$ is the image of the cut thermal cylinder under (the analytic continuation of) $\varphi$. We regularize the resulting manifold by removing a disk of radius $\epsilon$ at the origin and truncating at the level curve of Re$(z_A)$ which intersects $x_{\mathcal{R}}=L$. Note, a conformal mapping from this manifold to the annulus is given by $z_A \circ \varphi^{-1}$, with $z_A=z_A(z_{\mathcal{R}})$ one of the two maps provided above.

We can relate these quantities to the Lorentzian theory by
\begin{equation}
\label{eq: r_min, r_max, thermal}
\begin{aligned}
    r_{\mathrm{\min}}^\Psi = \frac{1}{2}\left(v(\epsilon)-u(-\epsilon)\right)
    \qquad
    r_{\mathrm{\max}}^\Psi = \frac{1}{2}\left(v\left( L \right)-u\left(-L\right)\right).
\end{aligned}
\end{equation}
Using the steps that led to eq.~\eqref{eq: EE thermal state, rindler, cardy}, this gives
\begin{equation}
    S(\rho_\Psi^\mathcal{R})= \frac{c}{6}\left[\frac{2\pi r^{\Psi}_{\max}}{\beta}-\log (\frac{2\pi r_{\min}^{\Psi}}{\beta})\right]+\cO(1).
\end{equation}
Recalling eq.~\eqref{eq:linearize},
and using the assumption in eq.~\eqref{eq:ABrelation}, one can linearize eq.~\eqref{eq: r_min, r_max, thermal} to obtain eq.~\eqref{eq:thermalEEAB} in the special case $A=B$, $A_{\infty}=B_{\infty}$.

We close by noting that eq.~\eqref{eq:confmap1} can be used to derive the Rindler modular flow in the thermal state, as provided in eq.~\eqref{eq:borchers1}.
As described in \cite{Cardy:2016fqc}, this is done by considering the push-forward of  Euclidean time-translation on the annulus, $\partial_{\tau_A}$, under the inverse of eq.~\eqref{eq:confmap1}.
By evaluating the flow at $\tau_R=0$ and Wick rotating to Lorentzian signature, we retrieve the modular flow \eqref{eq:borchers1}.
Furthermore, evaluating the flow at $\tau_R=\frac{\beta}{2}$ (or equivalently, $\tau_R=-\frac{\beta}{2}$) gives the action of this modular flow on the second copy of the theory in the thermofield double system.
After a suitable Wick rotation and $\mathsf{CPT}$ transformation, this flow is given by
\begin{equation}
    c(u) = -\frac{\beta}{2\pi}\left(1+e^{-\frac{2\pi u}{\beta}}\right)\qquad
    b(v) = -\frac{\beta}{2\pi}\left(1+e^{\frac{2\pi v}{\beta}}\right)\,,
    \label{eq:copyflow}
\end{equation}
and is plotted in figure \ref{fig:copy}.

\section{Nonlocal transformation rules for primary operators}
\label{app:nonlocal}

As explained in section \ref{sec:warmup}, subtleties arise in the study of special conformal transformations in Minkowski spacetime.
These subtleties are relevant when relating vacuum modular flows in the Rindler wedge $\mathcal{R}$ to vacuum modular flows in the diamond $\mathcal{D}$.

To simplify the discussion, it will actually be easier to study the conformal map of interest, eq.~\eqref{eq:DtoRmap}, in $d>2$. 
As in the $d=2$ case, this maps sends spatial infinity in the $(u_\cD,v_\cD)$ frame to a single point\footnote{Note, $i_0$ is always conformally mapped to a single point, not an extended surface, as explained in \cite{Hawking:1973uf}.} at finite distance in the $(u_{\mathcal{R}},v_{\mathcal{R}})$ frame, and the diamond (\ie double-cone) region $\mathcal{D}$ to the Rindler wedge $\mathcal{R}$.

However, unlike in the $d=2$ case, in the $\mathcal{D}$ frame we are now considering a spherically symmetric setup, hence
\begin{equation}
    u_\cD = t_\cD - r_\cD\qquad v_\cD = t_\cD + r_\cD\,,
\end{equation}
where the radial coordinate, $r_\cD$, obeys $r_\cD>0$.
On the other hand, in the $\mathcal{R}$ frame, we are considering the ordinary Rindler wedge, hence
\begin{equation}
    u_{\mathcal{R}} = t_{\mathcal{R}}-x_{\mathcal{R}}\qquad v_{\mathcal{R}} = t_{\mathcal{R}} +x_{\mathcal{R}}\,,
\end{equation}
with $x_{\mathcal{R}} \in (-\infty,\infty)$.

As in section \ref{sec:warmup}, we note that eq.~\eqref{eq:DtoRmap} is a composition of the special conformal transformation (SCT)
\begin{equation}
    u_{\mathcal{R}}(u)=2R\frac{u}{2R+u}
    \qquad
    v_{\mathcal{R}}(v) = 2R\frac{v}{2R-v}\,,
    \label{eq:baseSCTapp}
\end{equation}
with a shift $u= u_\cD-R$, $v= v_\cD+R$.

The representation theory of SCTs is subtle in Minkowski CFT because, as noted in \cite{Schroer:1974ay}, SCTs in Minkowski spacetime can send spacelike separated to timelike separated points, and this causes problems for the local transformation rule \eqref{eq:primary}.
To see why, let $O_1$ and $O_2$ be spacelike separated primary operators with scaling dimension $\Delta$, and let $U_0$ be the unitary representation \eqref{eq:primary} of an SCT such that $U_0O_1U_0^{\dag}$ and $U_0O_2U_0^{\dag}$ are timelike separated.
Since $[O_1,O_2]=0$ and $U_0$ is unitary operator, we have
$[U_0O_1U_0^{\dag},U_0O_2U_0^{\dag}]=0$, i.e.~commutation at timelike separation.
However, conformal primaries only commute at timelike separation if $\Delta$ is an integer \cite{Hartman:2015, Simmons-Duffin:2016gjk}.
This means the SCT transformation rule \eqref{eq:primary} cannot be naively applied for fields with non-integer $\Delta$, such as a free scalar in odd $d$.

In \cite{Swieca:1973gc}, this problem is resolved in the context of $d>2$ free scalar theory by replacing the transformation rule \eqref{eq:primary}
\begin{equation}
    U_0 O(x) U_0^{\dag} = \Omega(x)^{-\Delta}O(x_T)\,,
\end{equation}
with 
\begin{equation}
    \Omega(x) = 1-2b\cdot x + b^2 x^2\,,
    \label{eq:omegapmapp}
\end{equation}
by the following transformation rule:
\begin{equation}
    U_0 O^{\pm}(x) U_0^{\dag} = \Omega_{\pm}(x)^{-\Delta}O^{\pm}(x_T)\,,
    \label{eq:nonlocalapp}
\end{equation}
where $O^{\pm}(x)$ are the annihilation and creation parts, respectively, of the free field $O(x)$, and the subscript on $\Omega_{\pm}(x)$ indicates the following $i\epsilon$ prescription: 
\begin{equation}
    x^0 \rightarrow x^0 \pm i\epsilon\qquad b^0 \rightarrow b^0 \pm i\epsilon\,.
    \label{eq:iepsapp}
\end{equation}
These prescriptions instruct us which branch of the log to choose in $\Omega_{\pm}(x)^{-\Delta} = e^{-\Delta \log \Omega_{\pm}(x)}$.
That is, one plugs in the prescription \eqref{eq:iepsapp} into eq.~\eqref{eq:omegapmapp} and evaluates the principal branch of $\log$; schematically, $\Omega_\pm(x)=e^{-\Delta \mathrm{Log} (\Omega(x)\pm i\epsilon_{b,x})}$.
When the $i\epsilon$ and $-i\epsilon$ prescriptions give different answers, the transformation rule eq.~\eqref{eq:nonlocalapp} is nonlocal, since the positive and negative frequency parts of the field will acquire different phases under eq.~\eqref{eq:nonlocalapp}.

In the special case where $\Delta$ is an integer, the two prescriptions give the same answer and the transformation rule \eqref{eq:nonlocalapp} reduces to the usual local transformation rule \eqref{eq:primary}.
This is consistent with the fact that fields with integer $\Delta$ commute at timelike separation.
Eq.~\eqref{eq:nonlocalapp} reduces to the local transformation rule \eqref{eq:primary}
also when 
$\Omega(x)$ in eq.~\eqref{eq:omegapmapp} is positive.
The sign of $\Omega(x)$ roughly tracks whether or not the conformal map of interest changes $x$ from being spacelike to timelike separated with respect to another point, $y$. 

Focusing on the special conformal transformation eq.~\eqref{eq:baseSCTapp}, we find
eq.~\eqref{eq:omegapmapp} reads 
\begin{equation}
    \Omega_{\pm}(x) = \frac{1}{4R^2}\left(2R+u\right)\left(2R-v\right)
    =\left(\frac{1}{4R^2}(2R-u_{\cal R})(2R+v_{\cal R})\right)^{-1} = \Omega_{\pm}(x_{\cal R})^{-1}\,,
    \label{eq:OmegaForDtoRapp}
\end{equation}
and thus throughout the diamond region $\mathcal{D}$, given in these coordinates by $u\in(-2R,0)$, $v\in(0,2R)$, we have $\Omega_{\pm}(x)>0$, and eq.~\eqref{eq:nonlocalapp} reduces to the standard local transformation rule \eqref{eq:primary}.
Further, $\Omega_{\pm}(x)<0$ only at timelike separation of the boundary of the sphere.

To obtain the vacuum modular flow unitaries $U_\cD(s)$, we first define $U_{\mathcal{R}}(s)$ to implement the vacuum modular flow in the Rindler wedge, i.e.~eq.~\eqref{eq:vacflow}.
We also denote the unitary which sends $\mathcal{R}$ to $\mathcal{D}$ as $U_0$, and propose based on appendix \ref{app:pushforward} that
\begin{equation}
    U_\cD(s) = U_0 U_{\mathcal{R}}(s) U_0^\dag\,.
\end{equation}
Now, consider spacetime points $w_\cD(s)$ and $w_\cD(0)=:w_\cD$ related to $w_{\mathcal{R}}(s)$ and $w_{\mathcal{R}}$ via eq.~\eqref{eq:DtoRmap}. Letting $O^{\pm}(w_{\mathcal{R}}(s)):=U_{\mathcal{R}}(-s) O^{\pm}(w_{\mathcal{R}})  U_{\mathcal{R}}(s)$, we have
\begin{equation}
\begin{aligned}
    U_\cD(-s)O^{\pm}(w_\cD)U_\cD(s) &= \Omega_{\pm}(w_\cD)^{-\Delta} U_0 U_{\mathcal{R}}(-s) O^{\pm}(w_{\mathcal{R}})  U_{\mathcal{R}}(s) U_0^\dag\\
    &= \Omega_{\pm}(w_\cD)^{-\Delta} U_0  O^{\pm}(w_{\mathcal{R}}(s))  U_0^\dag\\
    &= \Omega_{\pm}(w_\cD)^{-\Delta}\Omega_{\pm}(w_{\mathcal{R}}(s))^{-\Delta} O^{\pm}(w_\cD(s))\\
    &= \frac{\Omega_{\pm}(w_\cD)^{-\Delta}}{\Omega_{\pm}(w_\cD(s))^{-\Delta}} O^{\pm}(w_\cD(s))\,.
\end{aligned}
\label{eq:savingSCTapp}
\end{equation}
If $\Omega_{\pm}(w_\cD)$ and $\Omega_{\pm}(w_\cD(s))$ are of the same sign, then the phases coming from the $i\epsilon$ prescription will cancel and altogether the transformation rule will be local.
This is the case whenever $w_\cD$ and $w_\cD(s)$ are connected by a line which never crosses the lightsheets emanating from the boundary of the sphere.
On the other hand, if $w_\cD$ and $w_\cD(s)$ are separated by one of the lightsheets, then $\Omega_{\pm}(w_\cD)$ and $\Omega_{\pm}(w_\cD(s))$ have different sign, and the transformation rule is nonlocal.

The vacuum modular flow for $\mathcal{D}$, as described in 
eq.~\eqref{eq:diamondflowVacuum}, runs tangent to these lightsheets.
Nevertheless, it is possible that points can jump across the lightsheets by passing through infinity, as discussed below 
eq.~\eqref{eq:exponentiated}.
Therefore, the local transformation rule eq.~\eqref{eq:primary} cannot naively be applied to operators of non-integer scaling dimension which pass through infinity via eq.~\eqref{eq:exponentiated}. The transformation rule eq.~\eqref{eq:savingSCTapp} guarantees that such operators actually transform nonlocally, since $\Omega_{\pm}$ has opposite sign in $\mathcal{D}$ and $\mathcal{D}_F$.
Thus, generalizing the local transformation rule eq.~\eqref{eq:primary} to eq.~\eqref{eq:nonlocalapp} ensures that SCTs are consistent with microcausality.

We expect similar arguments apply in $d=2$.
In particular, the special conformal transformation rule \eqref{eq:nonlocalapp} is generalized to certain $2d$ theories in \cite{Schroer:1974ay}, and is conjectured to hold for a general interacting CFT$_2$.
The key feature of $d=2$ is that  the radial coordinate $r_\cD$ is replaced by a spatial coordinate $x_\cD$ which can be negative.

It would be interesting to investigate whether, under the modular flow, a local operator $O(w_{\mathcal{D}})$ which is initially spacelike separated from $\mathcal{D}$ continues to commute with operators in $\mathcal{D}$ after crossing infinity.

\section{Multiple intervals}\label{App:multiple}

In this appendix, we apply our entropy formula eq.~\eqref{eq:TheEntropy} to an example not considered in the main text: the case of two intervals in $d=2$ free fermion theory.

In this case, the modular Hamiltonian contains both local and non-local terms \cite{Casini:2009vk}.
In the case of two symmetric intervals $x\in(-R, -r)\cup(r,R)$ with $r \in (0,R)$, on the $t=0$ slice, the local part reads \cite{Casini:2016fgb}
\begin{equation}
    h_{\mathrm{loc}} = 2\pi \int_A \rd x\, \frac{(x^2-r^2)(R^2-x^2)}{2(R-r)(x^2 + rR)}T_{tt}(x)\,,
    \label{eq:Hloc}
\end{equation}
and it generates a conformal diffeomorphism of the spacetime.
The nonlocal part reads
\begin{equation}
    h_{\mathrm{nonloc}} = i \pi \int_A \rd x\, \psi^{\dag}(x) \frac{rR (x^2-r^2)(R^2-x^2)}{(R-r)x(x^2 + rR)^2}\psi(\bar{x})\,, \qquad
    \bar{x}:=-\frac{rR}{x}\,.
    \label{eq:Hnonloc}
\end{equation}
Since this term is not of the form \eqref{eq:localHam}, the vacuum modular flow with respect to two intervals is non-geometric.
One expects our entropy formula \eqref{eq:TheEntropy} to apply only for states with geometric modular flows.
Fortunately, from the discussion in section \ref{sec:discuss}, we know there exists another state $\ket{\Psi_p}$ whose modular Hamiltonian with respect to the two intervals is given simply by $h_{\mathrm{loc}}$, and we can apply the entanglement entropy formula to this state instead.
We will show that $\ket{\Psi_p}$ and $\ket{\Omega}$ have the same entanglement entropy.

To do so, first note that $h_{\mathrm{loc}}$ can be generalized beyond the case of symmetric intervals in the $t=0$ slice.
The more general operator reads \cite{Arias:2016nip}
\begin{equation}
    h_{\mathrm{loc}} = 2\pi \int_{I_v} \rd v \,b(v)  T_{vv}(v) 
    + 2\pi\int_{I_u} \rd u \,c(u)  T_{uu}(u)\,,
    \label{eq:fermionHam}
\end{equation}
where $I_v=I_{\Delta v_1}\cup I_{\Delta v_2} = [X_1^v,Y_1^v]\cup[X_2^v,Y_2^v]$ and similarly for $I_u$, denoting the null coordinate ranges of the two intervals.

Further,
\begin{equation}
    \frac{1}{b(v)} := \sum_{i=1}^{2} \frac{1}{b_i(v)}\,,
\end{equation}
where 
\begin{equation}
    b_i(v) = \left(\frac{1}{v-X_i^v}+\frac{1}{Y_i^v-v}\right)^{-1}
\end{equation}
is the vacuum modular flow associated with the $i^{th}$ region, considered in isolation.
This last statement holds setting $X_i^v=-R$ and $Y_i^v = R$ and comparing to eq.~\eqref{eq:diamondflow}.
Similarly, $\frac{1}{c(u)}=\sum_{i=1}^2 \frac{1}{c_i(u)}$.

By eq.~\eqref{eq:TheEntropy}, the two intervals in the state $\ket{\Psi_p}$ have an entanglement entropy of
\begin{equation}
    S_{12} = \frac{c}{12}\left(\int_{I_v}\frac{\rd v}{b(v)}+\int_{I_u}\frac{\rd u}{c(u)}\right)\,,
\end{equation}
which can be expanded into four terms, for each of the intervals $I_{\Delta v_1}$, $I_{\Delta v_2}$,$I_{\Delta u_1}$, and $I_{\Delta u_2}$.
Focusing on the $I_{\Delta v_1}$ term, we have
\begin{equation}
    \frac{c}{12}\int_{I_{\Delta v_1}}\frac{\rd v}{b(v)}
    = \frac{c}{12}\int_{X_1^v}^{Y_1^v}\rd v\left(\frac{1}{b_1(v)}+\frac{1}{b_2(v)}\right)
    = \frac{c}{6}\log\frac{\Delta v_1 }{\epsilon} + \cO(1)
    \,,
\end{equation}
where we have introduced $\Delta v_i:=Y_i^v-X_i^v$.
The $b_2(v)$ term does not contribute at leading order in $\epsilon$ because $\frac{1}{Y_2}$ is well-behaved within $I_v^1$.
The subleading term reads 
\begin{equation}
    \cO(1) \rightarrow \frac{c}{12}\log\frac{(X_2^v-Y_1^v)(Y_2^v-X_1^v)}{(Y_2^v-Y_1^v)(X_2^v-X_1^v)}\,,
    \label{eq:subl}
\end{equation}
and is symmetric under $1\leftrightarrow 2$.

Applying similar logic for each of the four terms, we obtain
\begin{equation}
    S\left(\rho_{\tilde{\Omega}}^{\mathcal{V}_1,\mathcal{V}_2}\right)
    = \frac{c}{3}\left(\log\frac{L_1}{\epsilon}+\log\frac{L_2}{\epsilon}\right)
    +
    \frac{c}{3}\log\frac{|X_2-Y_1||X_1-Y_2|}{|X_2 -X_1||Y_2-Y_1|}
    \label{eq:2intervalEntropy}
    \,,
\end{equation}
where $L_i = |X_i-Y_i|=\sqrt{-\Delta u_i \Delta v_i}$ is the proper length of region $i$.

This result straightforwardly generalizes to the case of more than two intervals, and matches the vacuum state entanglement entropy calculated in \cite{Casini:2009sr} for $p$ subregions.\footnote{Note that in \cite{Casini:2009sr}, the points $X_i$ and $Y_i$ lie on the same $t=0$ slice, whereas this is not required here.}
The interesting point here is that we recovered the correct result for the vacuum entanglement entropy using only the local term \eqref{eq:fermionHam} in the modular Hamiltonian, suggesting the nonlocal term does not actually contribute.
Of course, we do not expect to recover the correct mutual information, since $\ket{\Psi_p}$ has no entanglement between the two subregions, and in particular the mutual information vanishes.
On the other hand, the mutual information between two subregions in the vacuum state is nonvanishing.

\bibliographystyle{uiuchept}
\bibliography{file.bib}

\end{document}